\documentclass[12pt,axodraw]{article}
\usepackage{epsfig}
\usepackage{amsmath}
\usepackage{amsfonts}
\usepackage{amssymb}
\usepackage{graphicx}
\providecommand{\U}[1]{\protect\rule{.1in}{.1in}}
\begin{document}

\begin{center}
{\Large {\bf The Minimal Supersymmetric Standard Model (MSSM) and General Singlet Extensions of the MSSM (GSEMSSM), a short review}}\\
M. C. Rodriguez   \\
{\it Grupo de F\'\i sica Te\'orica e Matem\'atica F\'\i sica \\
Departamento de F\'\i sica \\
Universidade Federal Rural do Rio de Janeiro - UFRRJ \\
BR 465 Km 7, 23890-000 \\
Serop\'edica, RJ, Brazil, \\
email: marcoscrodriguez@ufrrj.br \\} 
\end{center}


\begin{abstract}
In this lectures, we give a review about the Minimal Supersymmetric Standard Model (MSSM) and the General Singlet Extensions of the MSSM (GSEMSSM). 
We, first introduce the minimal set of fields to built both models. Then we introduce their superfields and using them we build the lagrangian of those 
models in the superspace formalism. We show how to get the mass spectrum of those model in the $R$-parity scenarios and we also show how to 
get some Feynman Rules with the Gauge Bosons. The second part of this review was presented at Dark Mattter Workshop – Early Universe 
Cosmology, Baryogenesis and Dark Matter held in Instituto de F\'\i sica Te\'oria (IFT-Unesp) from 21 to 25 of October 2019.
\end{abstract}

PACS number(s): 12.60. Jv

Keywords: Supersymmetric Models



\section{Introduction}

Although the Standard Model (SM) describes the observed properties
of charged leptons and quarks. The success of the SM has been astonishing. Nevertheless, the SM is 
not considered as the ultimate theory since neither the fundamental 
parameters, masses and couplings, nor the symmetry pattern are 
predicted. 

It is commonly agreed that the Higgs sector of the SM is
unsatisfactory. One of this reason is the hierarchy problem. When we define the Higgs potential, we 
need to define the squared mass parameter $\mu^{2}$, this is a free parameter and it is not fixed by the theory. 
The Higgs potential is defined as 
\begin{equation}
V(H)=-\mu^{2} |H|^{2}+ \lambda \left( |H|^{2}\right)^{2}.
\label{higgspotsm}
\end{equation}
In the SM there is only one physical neutral Higgs scalar in the spectrum, and its mass is given by \cite{Gunion:1989we}
\begin{equation}
M_{H}= \sqrt{2}\mu .
\label{higgsmasssm}
\end{equation}

The SM is renormalizable, wchich means that finite results are obtained for all higher-order (loop)
corrections even if we extend the virtual momenta in the loop integral all
the way to infinity. The 4-boson self interaction in Eq.(\ref{higgspotsm}) 
generate, at one loop order, the self-energy, wchich is proportional to \cite{Gunion:1989we}
\begin{equation}
\lambda \int^{\Lambda}d^{4}k \frac{1}{k^{2}-M^{2}_{H}}.
\end{equation}
The coefficient $\mu$ is then replaced by the one loop-corrected ` `physical" value $\mu_{\mbox{ phys}}$
where
\begin{equation}
\mu^{2}_{\mbox{ phys}}= \mu^{2} - \lambda \Lambda^{2} ,
\end{equation}
The parameter $\Lambda$ is the energy scale wehere we expect to find the physics beyond the SM, in GUT scenarios we expect 
$\Lambda \sim M_{Pl}\approx 10^{19}$ GeV, the one loop correction
is then vastly greater than $M_{W}\approx \left( 100 \mbox{GeV} \right)^{2}$ and
it is known as hierachy problem. It means, the hierarchy problem \cite{Susskind:1978ms} is to
understand why $M_{H}$, given at Eq.(\ref{higgsmasssm}), is much less then the Planck mass Scale $M_{Pl}$.  

It is clearly prudent to explore the implication of more complicated Higgs
Models, both in the context of SM and in extended theories. Although the
minimal Higgs satisfy \cite{Gunion:1989we}
\begin{equation}
\rho = \frac{M_{W}}{M_{Z} \cos \theta_{W}} \approx 1,
\end{equation}
so does any version of the SM with any number of the Higgs doublets, one
example is the Two Higgs Doublet Model (THM) \cite{Gunion:1989we,Deshpande:1977rw,Georgi:1978wr,Donoghue:1978cj,Abbott:1979dt,McWilliams:1980kj,Haber:1978jt,Gunion:1984yn}, 
(and singlets \cite{Hooper:2009gm,Wang:2012ry}), for more details about this interesting problem see 
\cite{Gunion:1989we,Langacker:1984nf,Ellis:1986ij}. But unfortunatelly the THM 
can not solve the hierachy problem, for more details about this fact see \cite{Gunion:1989we}.

One intersting class of models to be the physics beyond the SM is 
Supersymmetry, it is more known as SUSY \cite{Salam:1974ig}. The hierachy can be solved in Supersymmetric Model, known as SUSY by short 
\cite{ogievetskivi,wb,MullerKirsten:1986cw}.  The supersymmetry automatically cancels all
quadratic corrections in all orders of perturbation theory due to
the contributions of superpartners of the ordinary particles.

The most intensively studied
model is the Minimal Supersymmetric Standard Model (MSSM) \cite{R,ssm,grav},
which is the most economical realization of SUSY. Control of the radiative
corrections so that a small value of $\mu^{2}$ (and hence $M^{2}_{H}$ and 
$M^{2}_{W}$) becomes natural was a primary motivation of the MSSM where we show the exact cancelation to 
the Higgs masses coming from fermion contribution. A light
Higgs sector ($M^{2}_{H}\ll M^{2}_{Pl}$) is natural in a softly broken SUSY
scenarios where the quadratically divergent loop contribution cancel leaving
a finite correction of the form 
\begin{equation}
\delta M^{2}_{H}= {\cal O}\left( \frac{\alpha}{\pi} \right) \left(
M^{2}_{B}- M^{2}_{F}\right) ,
\end{equation}
where $M_{B,F}$ are the masses of the bosonic and fermionic partner
particles circulating in the loops. When we consider 
scenarios with soft SUSY breakings terms we have the following constraints
\begin{equation}
M^{2}_{B}-M^{2}_{F}\simeq {\cal O}(M_{W})= M_{SUSY}^{2},
\end{equation}
which  should not be very large ($\leq$ 1 TeV) to make the fine-tuning natural. 
Therefore, it provides a solution to the hierarchy 
problem by protecting the eletroweak scale from large radiative corrections 
\cite{INO82a,INO82b}. However, the origin of the hierarchy is  the other part of the problem. 

SUSY arose in theoretical papers more than 30 years ago independently and the 
originals works were made by the following set of authors:
\begin{itemize}
\item[1-)] Golfand and Likhtman \cite{gl};
\item[2-)] Volkov and Akulov \cite{va}; 
\item[3-)] Wess and Zumino \cite{wz,Ferrara:1974ac}.
\end{itemize}
Some very nices reviews about the begining of SUSY can be found in the following references \cite{volkov1,shifman,shifman1}. 

SUSY, is basically a symmetry between bosons (particles with integer
spin) and fermions (particles with half-integer spin) \cite{ogievetskivi,wb,MullerKirsten:1986cw}. 
Since that time there appeared thousands of papers and the reason for this remarkable is due the fact that 
there are a number of theoretical and phenomenological issues that the 
SM fails to address adequately \cite{Chung:2003fi,dress,Baer:2006rs,Aitchison:2005cf,mssm,Simonsen:1995cf,Kuroda:1999ks}: 
\begin{itemize}
\item Unification with gravity; The point is that SUSY algebra being a generalization of
Poincar\'e algebra \cite{wb,dress,Baer:2006rs,Aitchison:2005cf}
\begin{equation}
\{Q_\alpha, \bar{Q}_{\dot{\alpha}}\}=2\sigma_{\alpha,\dot{\alpha}}^{m}P_{m}.
\end{equation}
Therefore, when we make SUSY local, one obtains supergravity~\cite{sugra}.
\item Unification of Gauge Couplings; According to hypothesis of Grand Unification Theory (GUT) all gauge couplings change with energy. In the 
MSSM the slopes of Renormalization Group Equation (RGE) curves about the running of behaviour of 
gauge couplings can achieve perfect unification \cite{ABF}.
\item Hierarchy problem (discussed above);  
\item Electroweak symmetry breaking (EWSB); The ``running" of the Higgs masses, using the RGE of MSSM, leads to the phenomenon known as 
radiative electroweak symmetry breaking. Indeed, the mass parameters from the Higgs
potential $m_{1}^{2}$ and $m_{2}^{2}$ (or one of them) decrease while
running from the GUT scale to the scale $M_{Z}$ may even change the
sign.  Thus the breaking of the electroweak symmetry is not introduced by brute force as in 
the SM, but appears naturally from the radiative corrections \cite{running}.
\end{itemize}

The main sucess of SUSY  is to solve all the problems listed above. However, SUSY has also 
made several correct predictions \cite{Chung:2003fi}:
\begin{itemize}
\item SUSY predicted in the early 1980s that the top quark would be heavy \cite{Ibanez:wd,Pendleton:as}, its experimental value is given by 
$m_{t}=172.25\pm 0.08\mathrm{(stat.)} \pm 0.62\mathrm{(syst.)}~\mathrm{GeV}$ \cite{Sirunyan:2018gqx};
\item SUSY GUT theories with a high fundamental scale accurately predicted the present experimental value of $\sin^{2} \theta_{W}$ 
before it was mesured \cite{Dimopoulos:1981yj,Dimopoulos:1981zb,Ibanez:yh,Einhorn:1981sx};
\item SUSY requires a light Higgs boson to exist \cite{INO82a,INO82b,Kane:1992kq,Espinosa:1992hp,haber2,Djouadi:2005gj,Casas:1994us}, its 
experimental value is $M_{H} = 125.09\pm0.21\,\mathrm{(stat.)}\pm0.11\,\mathrm{(syst.)}~\mathrm{GeV}$ \cite{Aad:2015zhl}.
\end{itemize}
Together these success provide powerful indirect evidence that low energy SUSY is indeed part 
of correct description of nature. The MSSM 
\cite{R,ssm,grav,dress,Baer:2006rs,Aitchison:2005cf,mssm,Simonsen:1995cf,Kuroda:1999ks}\footnote{About the history of MSSM, see e. g. 
\cite{Fayet:2001xk,Rodriguez:2009cd}}  have, also viable candidates to 
be the Dark Matter on this case it can be the lightest neutralino, 
lighest sneutrino\footnote{Unfortunatelly this particle have been ruled out by the combination of collider experiment as LEP and direct 
searches for cosmological relics as discussed at \cite{hillwalkers}.} and the gravitino, more details about SUSY Dark Matter Candidates 
is presented in nice way in \cite{Ellis:2010kf}.

However, the MSSM suffers from the $\mu$-problem. The $\mu$-parameter is
the only dimensional parameter in the SUSY conserving sector and one would
expect the $\mu$ to be either zero or at the Planck scale. A simple solution
is to promote $\mu$ to a dynamical field in extensions of the MSSM that
contain an additional singlet superfield $\hat{S}$ which does not interact
with the MSSM fields other than the two Higgs doublets, it is a singlet. An effective $\mu$
can be reasonably got at Electro-Weak scale when $\hat{S}$ denotes the vacuum
expectation value (VEV) of the singlet scalar field. Among these extension models
the Next-to-Minimal Supersymmetric Model (NMSSM) 
\cite{R,nmssm,barr,nilsredwy,derendinger,Drees:1988fc,Ellis:1988er,Ananthanarayan:1996zv,Ellwanger:1998jk,Maniatis:2009re,Ellwanger:2009dp}.

If introduce a singlet superfield to the MSSM, the Higgs sector will have
one more CP even component and one more CP odd component, and the neutralino
sector will have one more singlino component. These singlet multiplets
compose a ``singlet sector'' of the MSSM, the nearly Minimal Supersymmetric Model (nMSM) \cite{xnMSSM1,xnMSSM2} 
and the General Singlet Extensions of the MSSM (GSEMSSM). It can make the phenomenologies of
SUSY dark matter and Higgs different from the MSSM, and it can get some
interesting cosmological consequences as discussed recently \cite{Hooper:2009gm,Wang:2012ry}. 

As in our opinion, the MSSM and GSEMSSM are good candidates to be the extension of the 
SM we will on this review present those models in some details inclunding some numerical analyses. 

\section{Superfields}

The superfield formalism was formulated in independent way by
\begin{itemize}
\item A. Salam and J. Strathdee \cite{superspcae1};
\item S. Ferrara, J. Wess and B. Zumino \cite{superspcae2}.
\end{itemize} 

The most general superfield is given by \cite{wb,MullerKirsten:1986cw,dress,Baer:2006rs,Aitchison:2005cf,Lykken:1996xt}:
\begin{eqnarray}
{\cal F}(x, \theta , \bar{\theta})&=&f(x)+ \left( \theta \phi(x) \right) +
\left( \bar{\theta} \bar{\chi}(x) \right) + \left( \theta \theta \right) m(x)+ \left( \bar{\theta}\bar{\theta}\right) n(x)+
\left( \theta \sigma^{m} \bar{\theta}\right) v_{m}(x)  \nonumber \\ &+& 
\left( \theta \theta \right) \left( \bar{\theta} \bar{\lambda}(x) \right) + 
\left( \bar{\theta} \bar{\theta} \right) \left( \theta \psi(x) \right) +
\left( \theta \theta \right) \left( \bar{\theta} \bar{\theta} \right) d(x),
\label{supercampodefinition}
\end{eqnarray}
where
\begin{eqnarray}
&&f(x), \,\ m(x), \,\ n(x), \,\ d(x), \,\ {\mbox scalars}, 
\nonumber \\
&&\phi(x), \,\ \psi(x), \,\ \bar{\chi}(x), \,\ \bar{\lambda}(x), \,\ {\mbox spinors}, \nonumber \\
&&v_{m}(x), \,\ {\mbox vector}.
\end{eqnarray}

The infinitesimal variation of the general superfield under supersymmetric transformation, are given by:
\begin{eqnarray}
\delta_{ \xi}{\cal F} (x, \theta, \bar{\theta})&=& \delta_{ \xi}f(x)+ \left( \theta \delta_{ \xi} \phi(x) \right) +
\left( \bar{\theta} \delta_{ \xi}\bar{\chi}(x) \right) + 
\left( \theta \theta \right) \delta_{ \xi}m(x)+ \left( \bar{\theta}\bar{\theta} \right) \delta_{ \xi}n(x)+
\left( \theta \sigma^{m} \bar{\theta} \right) \delta_{ \xi}v_{m}(x)  \nonumber \\ &+& 
\left( \theta \theta \right) \left( \bar{\theta} \delta_{ \xi}\bar{\lambda}(x) \right) +
\left( \bar{\theta} \bar{\theta} \right) \left( \theta \delta_{ \xi}\psi(x) \right) +
\left( \theta \theta \right) \left( \bar{\theta} \bar{\theta} \right) \delta_{ \xi}d(x) \nonumber \\
&\equiv& ( \xi Q+ \bar{ \xi} \bar{Q}){\cal F}(x, \theta, \bar{\theta})
\quad .\nonumber
\label{eq:biglongtransf}
\end{eqnarray}
Using the Fierz identities, we then have that
the component fields of
${\cal F}$ transform as follows:
\begin{eqnarray}
\delta_{ \xi} f &=& \xi\phi + \bar{\xi}\bar{\chi} \quad ,\nonumber\\
\delta_{ \xi} \phi _{\alpha} &=& 2\xi_{\alpha}m + \sigma^{m}_{\alpha\dot{\beta}}
\bar{\xi}^{\dot{\beta}} \left[ \imath \left( \partial _{m}f \right) + v_{m}  \right] \quad ,\nonumber\\
\delta_{ \xi} \bar{\chi}^{\dot{\alpha}} &=& 2\bar{\xi}^{\dot{\alpha}}n +
\xi^{\beta}\sigma^{m}_{\beta\dot{\gamma}}\epsilon^{\dot{\gamma}\dot{\alpha}}
\left[ \imath \left( \partial _{m}f \right) - v_{m} \right]\quad ,\nonumber\\
\delta_{ \xi} m &=& \bar{\xi}\bar{\lambda} - \frac{\imath}{2}
\left[ \left( \partial_{m}\phi \right) \sigma^{m}\bar{\xi} \right]
\quad ,\nonumber\\
\delta_{ \xi} n &=& \xi\psi + \frac{\imath}{2}\left[ \xi\sigma^{m}\left( \partial_{m}\bar{\chi} \right) \right] \quad ,
\label{eq:stgenscal}\\
\delta_{ \xi} v_{m} &=& \left( \xi\sigma_{m}\bar{\lambda} \right) + \left( \psi\sigma_{m}\bar{\xi} \right) +
\frac{\imath}{2}\left[ \xi \left( \partial_{m} \phi \right) \right] - 
\frac{\imath}{2}\left[ \left( \partial_{m}\bar{\chi} \right) \bar{\xi} \right]
\quad ,\nonumber\\
\delta_{ \xi} \bar{\lambda}^{\dot{\alpha}} &=& 2\bar{\xi}^{\dot{\alpha}} d
+ \frac{\imath}{2}\bar{\xi}^{\dot{\alpha}} \left( \partial^{m}v_{m} \right)
+ \imath (\xi\sigma^{m}\epsilon )^{\dot{\alpha}} \left( \partial_{m} m \right)
\quad ,\nonumber\\
\delta_{ \xi} \psi _{\alpha} &=& 2\xi _{\alpha} d -
\frac{\imath}{2}\xi _{\alpha} \left( \partial^{m}v_{m} \right) + \imath (\sigma^{m}\bar{\xi})_{\alpha}
\left( \partial_{m} n \right) \quad ,\nonumber\\
\delta_{ \xi} d &=& \frac{\imath}{2}\partial_{m} \left[
\left( \psi\sigma^{m}\bar{\xi} \right) + \left( \xi\sigma^{m}\bar{\lambda} \right) \right]\quad .\nonumber
\end{eqnarray}
Note the important fact that the complex scalar component field
$d(x)$ transforms by a total derivative.

The covariant derivative and it is given by\cite{wb,MullerKirsten:1986cw}
\begin{eqnarray}
D_{\alpha} &=& \frac{\partial}{\partial \theta^{\alpha}}+i 
\sigma^{m}_{\alpha \dot{\alpha}}\bar{\theta}^{\dot{\alpha}}\frac{\partial}{\partial x^{m}} \,\ \nonumber \\ 
\bar{D}_{\dot{\alpha}}&=&- \frac{\partial}{\partial \bar{\theta}^{\dot{\alpha}}}-i 
\theta^{\alpha}\sigma^{m}_{\alpha \dot{\alpha}}\frac{\partial}{\partial x^{m}} \,\ .
\label{The Non-Abelian Fieldstrength prop 4}
\end{eqnarray}
and $\alpha = 1,2$ is a spinorial index.

We know that the chiral superfield\footnote{This superfield satisfy $\bar{D}_{ \dot{ \alpha}} \Phi = 0$, where $D_{ \alpha}$ is defined 
at Eq.(\ref{The Non-Abelian Fieldstrength prop 4}) and it is similar to Dirac equation to chiral fermion.}, denoted as ($\Phi$), and the anti-chiral, given by ($\bar{\Phi}$), has the following expansion\footnote{We want to emphasize that a chiral superfield $\Phi$ is called as ``left" chiral superfield because it cointains a left-handed fermion while 
the anti-chiral $\bar{\Phi}\equiv \Phi^{\dagger}$ is ``right" chiral superfield due the fact it has a right-handed fermion \cite{dress,Baer:2006rs,Aitchison:2005cf}.}\cite{wb,MullerKirsten:1986cw}
\begin{eqnarray}
\Phi (y, \theta )&=&A(y)+ \sqrt{2}\left( \theta \psi (y) \right) + \left( \theta \theta \right) F(y) , \,\ 
\left( \theta \psi \right) \equiv \theta^{\alpha} \psi_{\alpha}, \,\ \alpha =1,2,   \nonumber \\
\bar{\Phi} ( \bar{y}, \bar{\theta} )&=&A^{\dagger}( \bar{y})+ \sqrt{2}\left( \bar{\theta}\bar{\psi} ( \bar{y} ) \right) + 
\left( \bar{\theta} \bar{\theta}\right) F^{\dagger}( \bar{y}), \,\ 
\left( \bar{\theta}\bar{\psi} \right) \equiv \bar{\theta}_{\dot{\alpha}}\bar{\psi}^{\dot{\alpha}}, \,\ \dot{\alpha}=1,2,
\label{chiralsuperfield}
\end{eqnarray} 
respectivelly and $A$ is a scalar field, $\psi$ is a Weyl-van der Waerden fermions field \cite{wb,MullerKirsten:1986cw,vdWaerden1,Willenbrock:2004hu}
and $F$ is an scalar necessary to close the SUSY algebra \cite{wb,MullerKirsten:1986cw}. The new coordinate $y$ is defined as
\begin{eqnarray}
y^{m}&=&x^{m}+ \imath \left( \theta \sigma^{m} \bar{\theta} \right) ,
\label{fermioniccoordinate}
\end{eqnarray}
and $\bar{y}$ is the conjugate complex of $y$ defined above.

The superpotential $W$ is defined in general in the following way
\begin{equation}
W \equiv \lambda_{i}\Phi_{i}+\mu_{ij}\Phi_{i}\Phi_{j}+ f_{ijk}\Phi_{i}\Phi_{j}\Phi_{k},
\end{equation}
in terms of the fields components, see Eq.(\ref{chiralsuperfield}), can be written as \cite{wb,MullerKirsten:1986cw}
\begin{eqnarray}
\Phi_{i}(y, \theta) \Phi_{j}(y, \theta) &=& A_{i}(y)A_{j}(y) + \sqrt{2} \theta 
\left[ \psi_{i}(y)A_{j}(y)+A_{i}(y) \psi_{j}(y) \right] \nonumber \\
&+& \theta \theta \left[ A_{i}(y)F_{j}(y)+F_{i}(y)A_{j}(y)- \psi_{i}(y) \psi_{j}(y) 
\right] \nonumber \\
\Phi_{i}(y, \theta) \Phi_{j}(y, \theta) \Phi_{k}(y, \theta) &=& A_{i}(y)A_{j}(y)A_{k}(y)
\nonumber \\
&+& 
\sqrt{2} \theta 
\left[ \psi_{i}(y)A_{j}(y)A_{k}(y)+ A_{i}(y) \psi_{j}(y)A_{k}(y) 
+A_{i}(y)A_{j}(y) \psi_{k}(y) \right] \nonumber \\
&+& \theta \theta \left[ F_{i}(y)A_{j}(y)A_{k}(y)+ A_{i}(y)F_{j}(y)A_{k}(y) + 
A_{i}(y)A_{j}(y)F_{k}(y) \right. \nonumber \\
&-& \left. \psi_{i}(y) \psi_{j}(y)A_{k}(y)- A_{i}(y) \psi_{j}(y) \psi_{k}(y) 
- \psi_{i}(y)A_{j}(y) \psi_{k}(y) \right] \,\ .
\nonumber \\
\label{compsuperpotential}
\end{eqnarray}

The degrees of freedom are grouped in gauge superfields for gauge bosons and their 
supersymmetric partner, the gauginos, are put in a vector superfield
\footnote{It is also known as Real Superfield due the fact that this superfield satisfy the following constraint 
$V(x, \theta, \bar{ \theta})=V^{\dagger}(x, \theta, \bar{ \theta})$ \cite{wb,MullerKirsten:1986cw}.} 
in the Wess-Zumino gauge is written as \cite{wb,MullerKirsten:1986cw}
\begin{equation}
V_{WZ}(x, \theta, \bar{ \theta}) = - \left( \theta \sigma^{m} \bar{ \theta}\right) v_{m}(x) +
\imath \left( \theta \theta \right) \left( \bar{ \theta} \bar{ \lambda}(x) \right) 
- \imath \left( \bar{ \theta} \bar{ \theta} \right) \left( \theta \lambda(x) \right)
+ \frac{1}{2} \left( \theta \theta \right) \left( \bar{ \theta} \bar{ \theta} \right) D(x).
\label{eq:vwzdef}
\end{equation}
$v_{m}$ is a gauge boson, $\lambda$ is their superpartner known as gauginos\footnote{They satisfy the constraint $\lambda \equiv \bar{\lambda}$ and therefore they are Majorana fermions \cite{Majorana}, in the SM all the fermions are Dirac ones \cite{Dirac}.}, and $D$ again is a scalar necessary to close SUSY algebra \cite{wb,MullerKirsten:1986cw}.

The K\"ahler potential is defined as 
\begin{equation}
K( \bar{\Phi}, \Phi ) \equiv \bar{\Phi}\Phi .
\end{equation} 
It is defined in terms of the 
fields components as \cite{wb,MullerKirsten:1986cw,dress,Baer:2006rs,Aitchison:2005cf}
\begin{eqnarray}
\int d^{4}\theta K \left( \hat{\bar{\Phi}}e^{gT^{a}\hat{V}^{a}}, \hat{\Phi} \right) &\equiv& 
\int d^{4}\theta \left( \hat{\bar{\Phi}}e^{gT^{a}\hat{V}^{a}}\hat{\Phi} \right) =- 
\left( {\cal D}_{m}A \right)^{\dagger}\left( {\cal D}^{m}A \right)- 
\imath \left( \bar{\psi}\bar{\sigma}^{m} {\cal D}_{m}\psi \right) 
\nonumber \\ &+&
F^{\dagger}F + 
\imath \sqrt{2}g \left[ \left( A^{\dagger}T^{a}\lambda^{a}\right) \psi- 
\overline{\psi} \left( \bar{\lambda}^{a}\bar{T}^{a}A \right) \right] 
+g \left( A^{\dagger}T^{a}A \right) D^{a}, \nonumber \\
{\cal D}_{m}\psi&=&\partial_{m} \psi + \imath g \left( T^{a}v^{a}_{m} \right) \psi , \nonumber \\
\frac{1}{4}\left(  \int d^{2} \theta \,\ W^{\alpha a}W^{a}_{\alpha} 
+ \int d^{2} \bar{\theta} \bar{W}^{a}_{ \dot{ \alpha}} \bar{W}^{ \dot{ \alpha}a} \right) 
&=& -\frac{1}{4} F^{a}_{mn}F^{mna}
- \imath \left( \lambda^{a} \sigma^{m} D_{m} \bar{ \lambda}^{a} \right) + \frac{1}{2} \left( D^{a}\right)^{2}, \nonumber \\
F^{a}_{mn}&=& \partial_{m} v^{a}_{n}- \partial_{n} v^{a}_{m}-gf^{abc}v^{b}_{m}v^{c}_{n}, \nonumber \\
D_{m} \bar{\lambda}^{\dot{\alpha}a}&=& \partial_{m} \bar{\lambda}^{\dot{\alpha}a}-
gf^{abc}v^{b}_{m} \bar{\lambda}^{\dot{\alpha}c},
\label{kahlerpotential}
\end{eqnarray}
where $f^{abc}$ are the totally antisymmetric struture constant of some gauge group.

\section{$R$ Symmetry}
\label{apen:rsymmetry}

The R-symmetry was introduced in 1975 by A. Salam and J. Strathdee
\cite{r1} and in an independent way by P. Fayet \cite{R} to avoid
the interactions that violate either lepton number or baryon number.

$R$-symmetry is better understood with the superspace
formalism. It  is a continuous $U(1)$ 
symmetry acting on the supersymmetry generator, parametrized by $\alpha$. The corresponding
operator will be denoted as ${\bf R}$. $R$-symmetry 
acts on the superspace coordinate $\theta$, $\bar{\theta}$ as
follows \cite{wb} 
\begin{eqnarray}  
{\bf R} \theta &=& e^{-i \alpha} \theta, \nonumber \\ 
{\bf R} \bar{\theta} &=& e^{i \alpha} \bar{\theta}. 
\label{The R-Invariance prop 1} 
\end{eqnarray} 
$\theta$ has $R$-charge $\mathrm{R}( \theta )= -1$, while
$\bar{\theta}$ has $\mathrm{R}( \bar{\theta} )=1$.

The operator ${\bf R}$ acts on left-handed chiral superfields
$\Phi(x,\theta,\bar{\theta})$ and (right-handed) 
anti-chiral ones
$\bar{\Phi}(x,\theta,\bar{\theta})$ in the
following way~\cite{wb} 
\begin{eqnarray}
{\bf R} \Phi(x,\theta,\bar{\theta})&=& e^{i n_{\Phi}\alpha} \Phi(x, e^{-i\alpha}\theta ,e^{i\alpha}\bar{\theta} ),\label{eq3}\\
{\bf R} \bar{ \Phi}(x,\theta,\bar{\theta})&=&e^{- i n_{\Phi}\alpha}\bar{ \Phi}(x, e^{-i\alpha}\theta , e^{i\alpha}\bar{\theta} ),
          \label{The R-Invariance prop 2}
\end{eqnarray} 
where $n_{ \Phi}$ is the $R$-charge of the chiral superfield. We get the transformations for the field components:
\begin{eqnarray} \left.
\begin{array}{lcr}
A(x)    &\stackrel{{\bf R}}{\longmapsto}&       e^{in_{\Phi}\alpha} A(x) \\
\psi (x) &\stackrel{{\bf R}}{\longmapsto}&e^{i
\left( n_{\Phi}-1 \right) \alpha}\psi (x) \\
F(x)    &\stackrel{{\bf R}}{\longmapsto}&       e^{i \left( n_{\Phi}-1
\right)\alpha} F(x)
\end{array} \right\}.
\label{The R-Invariance prop 4a}
\end{eqnarray}

The products of chiral superfields\footnote{The superpotential}, 
\begin{eqnarray}
{\bf R} \prod_{a}\;\Phi_{a}(x,\theta ,\bar{\theta}) &=& e^{i\sum_{a}n_{a}\alpha}\,\prod_{a}\Phi_{a}(x, e^{-i\alpha}\theta ,e^{i\alpha}\bar{\theta} ).
\end{eqnarray}
is invariant under $R$ symmetry only if
\begin{eqnarray}
\sum_{a}n_{a}=0.
\end{eqnarray}   

The $R$ symmetry acts on vectorial (gauge) superfields, and by definition this superfield is invariant by this transformation, it means
\begin{eqnarray}
   {\bf R} V(x,\theta,\bar{\theta}) &=&
V(x, e^{-i\alpha}\theta , e^{i\alpha}\bar{\theta} ),
         \label{The R-Invariance prop 3}
\end{eqnarray}
the  field components in the vector superfield transform as 
\begin{eqnarray}
   \left.  \begin{array}{lcr}
A_{m}(x) &\stackrel{{\bf R}}{\longmapsto}&       A_{m}(x) \\
\lambda (x)     &\stackrel{{\bf R}}{\longmapsto}&
    e^{i\alpha} \lambda (x) \\
\bar{\lambda}(x)     &\stackrel{{\bf R}}{\longmapsto}&
    e^{-i\alpha} \bar{\lambda}(x) \\
D(x)       &\stackrel{{\bf R}}{\longmapsto}&       D(x)
          \end{array} \right\}.
          \label{The R-Invariance prop 5}
\end{eqnarray} 

The $R$ symmetry as defined above can avoid the proton decay in the Minimal Supersymmetric Standard Model. The we can ask, 
Why we need to introduce $R$ parity? The main reason to do it is the following: an unbroken continuous $R$ symmetry acting 
chirally on gauginos, and gluinos in particular, would maintain the gauginos massless, because the gaugino's mass term is given by \cite{10} 
\begin{equation} 
m_{\lambda} \left( \lambda \lambda + \bar{\lambda} \bar{\lambda} \right), 
\label{gaugino mass term} 
\end{equation} 
which, under the $R$-symmetry, see Eq.(\ref{The R-Invariance prop 5}), transforms into
\begin{equation} 
m_{\lambda} \left( e^{2i \alpha}\lambda \lambda + e^{-2i \alpha}\bar{\lambda} \bar{\lambda} \right) ,
\label{forbidgaugino mass term} 
\end{equation}
so that the mass term, given at Eq.(\ref{gaugino mass term}), is not invariant under $R$ symmetry. 

This forces us to abandon the continuous $R$-symmetry, in favour of its discrete version
called $R$ parity, we will discuss it at Sec.(\ref{subsec:r-parity}). This one allows for gluinos and other gauginos to acquire masses.
Moving from $R$ symmetry to $R$ parity is in any case necessary within supergravity, so that the spin-$\frac{3}{2}\,$ (Majorana) gravitino can 
acquire a mass $\,m_{3/2}$, which does also violate the continuous $R$ symmetry \cite{grav}. 

\section{Review of the MSSM.}
\label{sec:mssm}

The Minimal Supersymmetric Standard Model (MSSM) is the supersymmetric extension of the SM that contains a minimal number of states and interactions \cite{dress,Baer:2006rs,Aitchison:2005cf,mssm,Simonsen:1995cf,Kuroda:1999ks,kraml}. The model has the
gauge symmetry $SU(3)_{C} \otimes SU(2)_{L} \otimes U(1)_{Y}$ extended by the supersymmetry to include the supersymmetric partners of the SM fields which have spins that differ by $+1/2$ as required by the supersymmetric algebra. Since the SM fermions are left-handed and right-handed and they 
transform differently under $SU(3)_{C}$, $SU(2)_{L}$ and $U(1)_{Y}$  groups. The leptons and the Higgs must belong to chiral or anti-chiral supermultiplets.

The chiral supermultiplet \cite{dress,Baer:2006rs,Aitchison:2005cf} contains three families of left-handed quarks $Q_{iL}$, three families of leptons 
$L_{iL}$ plus the Higgs fields $H_{1}$ \cite{dress,Baer:2006rs,Aitchison:2005cf} and the particle content of each chiral 
superfield, given above, is presented in the Tab.(\ref{lfermionnmssm}). The anti-chiral supermultiplet \cite{dress,Baer:2006rs,Aitchison:2005cf} 
contains three families of right-handed quarks, given by  
$(u_{iR}$, $d_{iR}$)\footnote{Remember that in the field theory is hold the followings relations 
$u_{iR}=(\bar{u}_{iL})^{c}$ and $d_{R}=(\bar{d}_{iL})^{c}$and at this point they are, still Weyl-van der Waerden fermions~\cite{wb,MullerKirsten:1986cw,vdWaerden1}.}, three families of 
right-handed leptons ($l_{iR}$) and another Higgs fields $H_{2}$ \cite{dress,Baer:2006rs,Aitchison:2005cf} and the particle content of each 
anti-chiral superfield is presented in the Tab.(\ref{rfermionnmssm}). 
\begin{table}[h]
\begin{center}
\begin{tabular}{|c|c|c|}
\hline 
$\mbox{ Chiral Superfield} $ & $\mbox{ Fermion} $ & $\mbox{ Scalar} $ \\
\hline
$\hat{L}_{iL}=( \hat{\nu}_{i}, \hat{l}_{i})^{T}_{L}\sim({\bf 1},{\bf2},-1)$ & 
$L_{iL}=(\nu_{i},l_{i})^{T}_{L}\sim({\bf 1},{\bf2},-1)$ & 
$\tilde{L}_{iL}=( \tilde{\nu}_{i}, \tilde{l}_{i})^{T}_{L}\sim({\bf 1},{\bf2},-1)$ \\
\hline
$\hat{Q}_{iL}=(\hat{u}_{i}, \hat{d}_{i})^{T}_{L}\sim({\bf 3},{\bf2},1/3)$ & 
$Q_{iL}=(u_{i},d_{i})^{T}_{L}\sim({\bf 3},{\bf2},1/3)$ & 
$\tilde{Q}_{iL}=(\tilde{u}_{i}, \tilde{d}_{i})^{T}_{L}\sim({\bf 3},{\bf2},1/3)$ \\ 
\hline
$\hat{H}_{1}=( \hat{h}^{0}_{1}, \hat{h}^{-}_{1})^{T}\sim({\bf 1},{\bf2},-1)$ & 
$\tilde{H}_{1}=( \tilde{h}^{0}_{1}, \tilde{h}^{-}_{1})^{T}\sim({\bf 1},{\bf2},-1)$ & 
$H_{1}=(h^{0}_{1},h^{-}_{1})^{T}\sim({\bf 1},{\bf2},-1)$  \\
\hline
\end{tabular}
\end{center}
\caption{\small Particle content in the left-chiral superfields in MSSM, the numbers in parenthesis refers to the 
$(SU(3)_{C}, SU(2)_{L}, U(1)_{Y}$) quantum numbers, respectively and $i=1,2,3$ refers to the generation index (or flavor indices) 
and we neglected the color indices.}
\label{lfermionnmssm}
\end{table}

There are at least three reasons for introduce $\hat{H}_{2}$, they the following \cite{dress}
\begin{itemize}
\item Cancel chiral anomaly;
\item Give masses to all quarks in the model.
\end{itemize}

\begin{table}[h]
\begin{center}
\begin{tabular}{|c|c|c|}
\hline 
$\mbox{ Anti-Chiral Superfield} $ & $\mbox{ Fermion} $ & $\mbox{ Scalar} $ \\
\hline
$\hat{l}^{c}_{iL}\sim({\bf 1},{\bf1},2)$ & $l^{c}_{iL}\equiv \bar{l}_{iR}\sim({\bf 1},{\bf1},2)$ & 
$\tilde{l}^{c}_{iL}\sim({\bf 1},{\bf1},2)$ \\ 
\hline
$\hat{u}^{c}_{iL}\sim({\bf \bar{3}},{\bf1},-4/3)$ & $u^{c}_{iL}\equiv \bar{u}_{iR}\sim({\bf \bar{3}},{\bf1},-4/3)$ & 
$\tilde{u}^{c}_{iL}\sim({\bf \bar{3}},{\bf1},-4/3)$ \\ 
\hline
$\hat{d}^{c}_{iL}\sim({\bf \bar{3}},{\bf1},2/3))$ & $d^{c}_{iL}\equiv \bar{d}_{iR}\sim({\bf \bar{3}},{\bf1},2/3))$ & 
$\tilde{d}^{c}_{iL}\sim({\bf \bar{3}},{\bf1},2/3))$ \\ 
\hline
$\hat{H}_{2}=( \hat{h}^{+}_{2}, \hat{h}^{0}_{2})^{T}\sim({\bf 1},{\bf \bar{2}},1)$ & 
$\tilde{H}_{2}=( \tilde{h}^{+}_{2}, \tilde{h}^{0}_{2})^{T}\sim({\bf 1},{\bf \bar{2}},1)$ & 
$H_{2}=(h^{+}_{2},h^{0}_{2})^{T}\sim({\bf 1},{\bf \bar{2}},1)$   \\
\hline
\end{tabular}
\end{center}
\caption{\small Particle content in the right-anti-chiral superfields in MSSM, the quantum number is the same meaning as presented at Tab.(\ref{lfermionnmssm}).}
\label{rfermionnmssm}
\end{table}

In the MSSM we need to introduce the following three vector superfields $\hat{V}^{a}_{C}\sim({\bf 8},{\bf 1}, 0)$
\footnote{The gluinos are the superpartner of gluons, and therefore they are in the adjoint representation of $SU(3)$, wchich is real.}, 
where $a=1,2, \ldots ,8$, 
$\hat{V}^{i}\sim({\bf 1},{\bf 3}, 0)$, with $i=1,2,3$, and 
$\hat{V}^{\prime}\sim({\bf 1},{\bf 1}, 0)$. The particle content 
in each vector superfield is presented in the Tab.(\ref{gaugemssm}).
\begin{table}[h]
\begin{center}
\begin{tabular}{|c|c|c|c|}
\hline 
${\rm{Vector \,\ Superfield}}$ & ${\rm{Gauge \,\ Bosons}}$ & ${\rm{Gaugino}}$ & ${\rm Gauge \,\ constant}$ \\
\hline 
$\hat{V}^{a}_{C}\sim({\bf 8},{\bf 1}, 0)$ & $g^{a}_{m}\sim({\bf 8},{\bf 1}, 0)$ & $\lambda _{C}^{a}\sim({\bf 8},{\bf 1}, 0)$ & $g_{s}$ \\
\hline 
$\hat{V}^{i}\sim({\bf 1},{\bf 3}, 0)$ & $V^{i}_{m}\sim({\bf 1},{\bf 3}, 0)$ & $\lambda^{i}\sim({\bf 1},{\bf 3}, 0)$ & $g$ \\
\hline
$\hat{V}^{\prime}\sim({\bf 1},{\bf 1}, 0)$ & $V^{\prime}_{m}\sim({\bf 1},{\bf 1}, 0)$ & $\lambda \sim({\bf 1},{\bf 1}, 0)$ & $g^{\prime}$ \\
\hline
\end{tabular}
\end{center}
\caption{\small Particle content in the vector superfields in MSSM.}
\label{gaugemssm}
\end{table}

The supersymetric Lagrangian of the MSSM is given by
\begin{equation}
{\cal L}_{SUSY} = {\cal L}^{chiral}_{SUSY} + {\cal L}^{Gauge}_{SUSY} .
\label{SUSY-Lagrangian1}
\end{equation}
The Lagrangian defined in the equation (\ref{SUSY-Lagrangian1}) contains contributions from all sectors of the model
\begin{equation}
{\cal L}^{chiral}_{SUSY} = 
{\cal L}_{leptons} + {\cal L}_{quarks} + {\cal L}_{Higgs}+ {\cal L}_{sup} ,
\label{SUSY-Lagrangian}
\end{equation}
and thoae terms have the following explicit form 
\begin{eqnarray}
{\cal L}_{lepton}&=& \int d^{4}\theta\;\sum_{i=1}^{3}\left[\,
K \left( \hat{ \bar{L}}_{i}e^{2g\hat{V}+g^{\prime} \left( - \frac{1}{2}\right) \hat{V}^{\prime}}, \hat{L}_{i} \right) +
K \left( \hat{ \bar{l^{c}}}_{i}e^{g^{\prime}\hat{V}^{\prime}}, \hat{l}^{c}_{i} \right) \,\right], \nonumber \\
{\cal L}_{quarks}&=&\hspace{-2mm} \int d^{4}\theta \sum_{i=1}^{3}\left[
K \left( \hat{\bar{Q}}_{i}e^{2g_{s}\hat{V}_{c}+2g\hat{V}+g^{\prime} \left( \frac{1}{6} \right) \hat{V}^{\prime}}, \hat{Q}_{i} \right) +
K \left( \hat{ \bar{u^{c}}}_{i}e^{2g_{s}\hat{V}_{c}+ g^{\prime} \left( - \frac{2}{2}\right) \hat{V}^{\prime}}, \hat{u}^{c}_{i} \right) 
\right. \nonumber \\ &+& \left.
K \left( \hat{ \bar{d^{c}}}_{i}e^{2g_{s}\hat{V}_{c}+g^{\prime}\left( \frac{1}{3}\right) \hat{V}^{\prime}}, \hat{d}^{c}_{i} \right) 
\right], \nonumber \\ 
{\cal L}_{Higgs}&=&  \int d^{4}\theta\;\left[\,
K \left( \hat{ \bar{H}}_{1}e^{2g\hat{V}+g^{\prime}\left( - \frac{1}{2}\right) \hat{V}^{\prime}}, \hat{H}_{1} \right) +
K \left( \hat{ \bar{H}}_{2}e^{2g\hat{V}+g^{\prime}\left( \frac{1}{2}\right) \hat{V}^{\prime}}, \hat{H}_{2} \right) \right], \nonumber \\
{\cal L}_{sup}&=& \int d^{2}\theta\; W+ \int d^{2}\bar{\theta}\;\bar{W} . \nonumber \\ 
\label{allsusyterms}     
\end{eqnarray}
here, $\hat{V}_{c} \equiv T^{a}\hat{V}^{a}_{c}$ and $T^{a}=\lambda^{a}/2$ (with $a=1,\cdots,8$) are the generators of $SU(3)_{C}$ and $\hat{V}=T^{i}\hat{V}^{i}$ where $T^{i} \equiv \sigma^{i}/2$ (with $i=1,2,3$) are the
generators of $SU(2)_{L}$. As usual, $g_{s}$, $g$ and $g^{\prime}$ are the gauge couplings for the $SU(3)$, $SU(2)$ and $U(1)$ groups, respectively, as shown in the Table \ref{gaugemssm}. 

The superpotential\footnote{To get renormalizable interactions the superpotential has $[W] \leq 3$ it is because 
$\Phi_{1}\Phi_{2}\Phi_{3}\Phi_{4} \propto A_{1}A_{2}A_{3}A_{4}$ and they are no renormalizable at 
two loops \cite{dress,Baer:2006rs,Aitchison:2005cf}.} that conserve $R$-parity is given by \cite{dress,Baer:2006rs,Aitchison:2005cf,Dong:2006vk,barbier} 
\begin{eqnarray}
W_{MSSM}&=&W^{MSSM}_{2RC}+W^{MSSM}_{3RC},   \\
W^{MSSM}_{2RC}&=&\mu\; \left( \hat{H}_{1}\hat{H}_{2} \right),   \\
W^{MSSM}_{3RC}&=& \sum_{i,j=1}^{3}\left[\, f^{l}_{ij}\left( \hat{H}_{1}\hat{L}_{i}\right) \hat{l}^{c}_{j}+
f^{d}_{ij}\left( \hat{H}_{1}\hat{Q}_{i}\right) \hat{d}^{c}_{j}+ 
f^{u}_{ij}\left( \hat{H}_{2}\hat{Q}_{i}\right) \hat{u}^{c}_{j}\,\right],
\label{suppotMSSM}
\end{eqnarray}
where $\left( \hat{H}_{1}\hat{H}_{2} \right) \equiv \epsilon_{\alpha \beta} \hat{H}_{1}^{\alpha} \hat{H}_{2}^{\beta}$. 

The free parameter $\mu$ is a complex number. In general the parameters $f$\footnote{This parameter is call Yukawa term}  
are complex numbers; they are symmetric in $ij$ exchange; they are 
dimensionless~\cite{dress,Baer:2006rs,Aitchison:2005cf}. It is one of the necessary 
conditions in order to implement CP violatiog in this model. Moreover, $f^{d}$ and $f^{u}$ 
account for the mixing between the quark current eigenstates as described by the Cabibbo-Kobayashi-Maskawa (CKM) matrix and 
we also can explain the mass hierarchy in the charged fermion masses as showed in 
\cite{cmmc,cmmc1}. The color indices on the triplet (antitriplet) 
superfield $\hat{Q}$ $( \hat{u}^{c}, \hat{d}^{c})$ contract trivially, and have been suppressed. 

The terms that break $R$-parity are given by \cite{dress,Baer:2006rs,Aitchison:2005cf,Dong:2006vk,barbier}
\begin{eqnarray}
W_{2RV}&=&\sum_{i=1}^{3}\mu_{0i} \left( \hat{L}_{i}\hat{H}_{2} \right),\nonumber \\
W_{3RV}&=&\sum_{i,j,k=1}^{3} \left[  
\lambda_{ijk} \left( \hat{L}_{i}\hat{L}_{j}\right) \hat{l}^{c}_{k}+
\lambda^{\prime}_{ijk}\left( \hat{L}_{i}\hat{Q}_{j}\right) \hat{d}^{c}_{k}+ 
\lambda^{\prime\prime}_{ijk}\hat{u}^{c}_{i}\hat{d}^{c}_{j}\hat{d}^{c}_{k} \right] .
\label{mssmrpv}
\end{eqnarray}
The mass matrix of neutrinos arise when we allow a mixing between the usual leptons with the higgsinos and its mixings is generated by
\begin{equation} 
\left( \hat{L}_{i}\hat{H}_{2}\right) \subset \left( L_{i}\tilde{H}_{2}\right) 
=l_{i}\tilde{h}^{+}_{2}- \nu_{i}\tilde{h}^{0}_{2},
\end{equation}
it is the mechanism to generate masses to two neutrinos at tree level and one neutrino get mass at one loop level as 
discuss at \cite{banks,hall,rv1,fb,rnm,rv2,marta,Montero:2001ch}.

\subsection{$R$ Parity}
\label{subsec:r-parity}

The discrete $R$-Parity, denoted by ${\bf R}_{d}$, which is able to
solve the above problem can be obtained by putting $\alpha = \pi$ at Eq.(\ref{The R-Invariance prop 1}). It means that
\begin{eqnarray}
{\bf R}_{d} \theta &\stackrel{{\bf R}_{d}}{\longmapsto}&- \theta , \nonumber \\
{\bf R}_{d} \bar{\theta} &\stackrel{{\bf R}_{d}}{\longmapsto}& - \bar{\theta}. 
\label{invrparitydiscreteattheta} 
\end{eqnarray}
Taking this value into account on Eqs.(\ref{The R-Invariance prop 1}, \ref{eq3}), \ref{The R-Invariance prop 2}) and 
Eq. (\ref{The R-Invariance prop 3}) we get the following transformations 
\begin{eqnarray}
{\bf R}_{d} \Phi(x,\theta,\bar{\theta}) &\stackrel{{\bf R}_{d}}{ \longmapsto}&
e^{2 i n_{\Phi}\pi}\Phi(x,- \theta ,- \bar{\theta} ), \nonumber \\
{\bf R}_{d} \bar{ \Phi}(x,\theta,\bar{\theta}) & \stackrel{{\bf R}_{d}}{\longmapsto}& e^{-2 i n_{\Phi}\pi} \bar{ \Phi}(x,- \theta
,- \bar{\theta} ), \nonumber \\ 
{\bf R}_{d} V(x,\theta,\bar{\theta}) &\stackrel{{\bf R}_{d}}{\longmapsto} & V(x,- \theta ,- \bar{\theta} ). 
\label{invrparitydiscrete} 
\end{eqnarray}
It is worth
emphasizing that, under this (discrete) transformation law, the
terms $\theta \theta$ and $\theta \theta \bar{\theta}
\bar{\theta}$ are invariants which is very helpful in further
analysis.

Now, under the discrete symmetry, the components of the
superfields transform as: 
\begin{eqnarray}
\left.  
\begin{array}{lcr}
A(x) &\stackrel{{\bf R}_{d}}{\longmapsto}&       e^{2in_{\Phi}\pi} A(x) \\
\psi (x) &\stackrel{{\bf R}_{d}}{\longmapsto}& e^{2i
\left( n_{\Phi}-\frac{1}{2} \right) \pi}\psi (x)\\
F(x) &\stackrel{{\bf R}_{d}}{\longmapsto}& e^{2i \left( n_{\Phi}-1
\right) \pi} F(x)
\end{array} 
\right\},
\label{The R-discrete-parity}\\
\left.  \begin{array}{lcr}
A_{m}(x) &\stackrel{{\bf R}_{d}}{\longmapsto}&       A_{m}(x) \\
\lambda (x)     &\stackrel{{\bf R}_{d}}{\longmapsto}&- \lambda (x) \\
\bar{\lambda}(x)     &\stackrel{{\bf R}_{d}}{\longmapsto}&- \bar{\lambda}(x) \\
D(x)       &\stackrel{{\bf R}_{d}}{\longmapsto}&       D (x)
\end{array} \right\}.
\label{The R-Invariance prop 5a}
\end{eqnarray} 
From (\ref{The R-Invariance prop 5a}), we see that
(\ref{gaugino mass term}) is invariant under this discrete symmetry if
\begin{eqnarray} 
\sum_{a} n_{a} = 0,2. 
\label{invrparitydiscreta} 
\end{eqnarray}

We defined at Tab.(\ref{allrpqchargesinMSSM}) the $R$-charges of the superfields in the MSSM.  
\begin{table}[h]
\begin{center}
\begin{tabular}{|c|c|c|}
\hline  
$\mbox{ Superfield}$ & $R$-charge & $(B-L)$-charge \\
\hline
$\hat{L}_{i}=( \hat{\nu}_{i}, \hat{l}_{i})^{T}$ & $n_{L}=+ \left( \frac{1}{2} \right)$ & $- \left( 1 \right)$ \\ 
\hline
$\hat{Q}_{i}=(\hat{u}_{i}, \hat{d}_{i})^{T}$ & $n_{Q}=+ \left( \frac{1}{2} \right)$ & $ \left( \frac{1}{3} \right)$ \\ 
\hline
$\hat{H}_{1}=( \hat{h}^{0}_{1}, \hat{h}^{-}_{1})^{T}$ & $n_{H_{1}}=0$ & $0$  \\ 
\hline
$\hat{H}_{2}=( \hat{h}^{+}_{2}, \hat{h}^{0}_{2})^{T}$ & $n_{H_{2}}= 0$ & $0$ \\ 
\hline
$\hat{l}^{c}_{i}$ & $n_{l^{c}}=- \left( \frac{1}{2} \right)$ & $+ \left( 1 \right)$  \\ 
\hline
$\hat{u}^{c}_{i}$ & $n_{u^{c}}=- \left( \frac{1}{2} \right)$ & $- \left( \frac{1}{3} \right)$ \\ 
\hline
$\hat{d}^{c}_{i}$ & $n_{d^{c}}=- \left( \frac{1}{2} \right)$ & $- \left( \frac{1}{3} \right)$  \\ 
\hline
\end{tabular}
\end{center}
\caption{\small $R$-charge and $(B-L)$-charge assignment to all superfields in the MSSM.}
\label{allrpqchargesinMSSM}
\end{table}
Taking this values we can show
\begin{equation}
\begin{array}{rcl}
H_{1,2}(x) &\stackrel{{\bf R}_{d}}{\longmapsto}& H_{1,2}(x), \\
\tilde{H}_{1,2}(x) &\stackrel{{\bf R}_{d}}{\longmapsto}&- \tilde{H}_{1,2}(x), \\
\tilde{f}(x) &\stackrel{{\bf R}_{d}}{\longmapsto}& - \tilde {f}(x), \\
\Psi(x) &\stackrel{{\bf R}_{d}}{\longmapsto}&  \Psi(x).
\end{array}
\label{Rpa2} 
\end{equation}
Here $\tilde{f}$ is a sfermion while $\Psi$ is a fermion.

The second terms of the Lagrangian defined by the Eq.(\ref{SUSY-Lagrangian1}) is given by the following equation
\begin{eqnarray}
{\cal L}^{Gauge}_{SUSY}&=&  \frac{1}{4} \left\{ \int  d^{2}\theta\;
\left[ \sum_{a=1}^{8} W^{a \alpha}_{s}W_{s \alpha}^{a}+ \sum_{i=1}^{3} W^{i \alpha}W_{ \alpha}^{i}+
W^{ \prime \alpha}W_{ \alpha}^{ \prime}\, \right] \right\} + hc  \,\ . \nonumber 
\end{eqnarray}
The gauge superfields have the following explicit form\cite{wb,MullerKirsten:1986cw}
\begin{eqnarray}
W^{a}_{s \alpha}&=&-\frac{1}{8g_{s}}\, \left( \bar{D}\bar{D} \right)
e^{-2g_{s}\hat{V}^{a}_{c}}D_{\alpha}e^{2g_{s}\hat{V}^{a}_{c}} 
\,\ , \nonumber \\
W^{i}_{\alpha}&=&-\frac{1}{8g}\, \left( \bar{D}\bar{D} \right)
e^{-2g\hat{V}^{i}}D_{\alpha}e^{2g\hat{V}^{i}} 
\,\ , \nonumber \\
W_{\alpha}^{\prime}&=&-\frac{1}{4}\, \left( \bar{D}\bar{D}\right) D_{\alpha} \hat{V}^{\prime} \,\ ,
\label{W-a}
\end{eqnarray}
where $D_{\alpha}$ is defined at Eq.(\ref{The Non-Abelian Fieldstrength prop 4}).

\section{Soft SUSY breaking Terms}

The experimental evidence suggests that the supersymmetry is not an exact symmetry. Therefore, supersymmetry 
breaking terms should be added to the Lagrangian defined by the Eq.(\ref{SUSY-Lagrangian}). The most general soft supersymmetry 
breaking terms, which do not induce quadratic divergence, where described by Girardello and Grisaru \cite{10}. They 
found that the allowed terms can be categorized as follows: a scalar field $A$ with mass terms
\begin{equation}
{\cal L}_{SMT}=- A^{\dagger}_{i}m^{2}_{ij}A_{j},
\end{equation}
 a fermion field gaugino  $\lambda$ with mass  terms
\begin{equation}
{\cal L}_{GMT}=- \frac{1}{2} (M_{ \lambda} \lambda^{a} \lambda^{a}+hc)
\end{equation}
and finally trilinear scalar interaction terms
\begin{equation}
{\cal L}_{INT}= B_{ij}\mu_{ij}A_{i}A_{j}+A_{ijk}f_{ijk}A_{i}A_{j}A_{k}+hc \,\ .
\end{equation}
The terms in this case are similar with the terms allowed in the superpotential of the model we are going to consider next.

Taken all this information into account, we can add the following soft supersymmetry breaking terms to the MSSM
\begin{eqnarray}
{\cal L}^{MSSM}_{Soft} &=& {\cal L}^{MSSM}_{SMT} + {\cal L}^{MSSM}_{GMT}+ {\cal L}^{MSSM}_{INT} \,\ ,
\label{The Soft SUSY-Breaking Term prop 2aaa}
\end{eqnarray}
where the scalar mass term ${\cal L}_{SMT}$ is given by the following relation
\begin{eqnarray}
{\cal L}^{MSSM}_{SMT} &=& - \sum_{i,j=1}^{3} \left[\,
\left( M_{L}^{2}\right)_{ij}\;\tilde{L}^{\dagger}_{i}\tilde{L}_{j}+ 
\left( M^{2}_{l}\right)_{ij} \tilde{l^{c}}^{\dagger}_{i}\tilde{l^{c}}_{j}+ 
\left( M_{Q}^{2}\right)_{ij}\;\tilde{Q}^{\dagger}_{i}\tilde{Q}_{j}
\right. \nonumber \\  
\hspace{1.7cm} &+& \left.
\left( M^{2}_{u}\right)_{ij} \tilde{u^{c}}^{\dagger}_{i}\tilde{u^{c}}_{j}+ 
\left( M^{2}_{d}\right)_{ij} \tilde{d^{c}}^{\dagger}_{i}\tilde{d^{c}}_{j}
+ M_{1}^{2} H^{\dagger}_{1}H_{1} +
M_{2}^{2} H^{\dagger}_{2}H_{2}
\right] \,\ ,
\label{burro}
\end{eqnarray}
The $3 \times 3$ matrices $M_{L}^{2},M^{2}_{l},M_{Q}^{2},M^{2}_{u}$ and $M^{2}_{d}$ are hermitian and $M_{1}^{2}$ and $M_{2}^{2}$ are real. The gaugino mass term is written as
\begin{eqnarray}
{\cal L}^{MSSM}_{GMT} &=&- \frac{1}{2}  \left[
\left(\,M_{3}\; \sum_{a=1}^{8} \lambda^{a}_{C} \lambda^{a}_{C}
+ M\; \sum_{i=1}^{3}\; \lambda^{i} \lambda^{i}
+ M^{\prime} \;   \lambda \lambda \,\right)
+ hc \right] \,\ .
\label{The Soft SUSY-Breaking Term prop 3}
\end{eqnarray}
Here, $M_{3},M$ and $M^{\prime}$ are complex. Finally, there is an interaction term ${\cal L}_{INT}$, see the equation (\ref{mssmrpv}), of the form
\begin{eqnarray}
{\cal L}^{MSSM}_{INT} &=&- M_{12}^{2} \left( H_{1}H_{2} \right) 
+  \sum_{i,j,k=1}^{3} \left[ 
A^{E}_{ij}f^{l}_{ij} \left( H_{1}\tilde{L}_{i}\right) \tilde{l}^{c}_{j}+
A^{D}_{ij}f^{d}_{ij} \left( H_{1}\tilde{Q}_{i}\right) \tilde{d}^{c}_{j}+
A^{U}_{ij}f^{u}_{ij} \left( H_{2}\tilde{Q}_{i}\right) \tilde{u}^{c}_{j}
\right] +hc  \,\ .
\label{burroint}
\end{eqnarray}
The parameter $M_{12}$, sometimes is written as $B \mu$ \cite{dress}, is 
in general complex. The $3 \times 3$ matrices $A$ are complex.

The total Lagrangian of the MSSM is obtained by adding all Lagrangians above
\begin{equation}
\mathcal{L}^{MSSM} = \mathcal{L}_{SUSY} + \mathcal{L}^{MSSM}_{soft},
\label{L-total}
\end{equation}
see the Eq.(\ref{SUSY-Lagrangian},\ref{The Soft SUSY-Breaking Term prop 2aaa}). 

\subsection{Simple Way to Broke Supersymmetry}

One of the most intrigatin problem is Supersymmetric theory is the way we broke Supersymmetry. 
However, we can parameterize all these term, via the spurion field
\footnote{Spurion is a fictious auxiliar field that can be used to parameterize any Symmetry 
breaking and to determine all the operators invariant under this Symmetry. It has only a 
non-vanishing $F$-component (equal to the Supersymmetry breaking parameter), is not a dynamical 
superfield.}\cite{dress,Baer:2006rs,Srednicki:2004hg}. On this case we add 
to our model a constant chiral superfield of the form\footnote{Some works they use the generic 
expanion $\hat{{\cal Z}}=Z+( \theta \theta Z^{F}+hc)+ \theta \theta \bar{\theta}\bar{\theta}Z^{D}$, 
where $Z,Z^{F}$ and $Z^{D}$ are scalars fields.}
\begin{equation}
\hat{{\cal Z}}=m^{2}_{S}\theta \theta ,
\end{equation}
where $m_{S}$ is the supersymmetry breaking scale and to a scalar field $\varphi$ we have \cite{wb}
\begin{equation}
m^{2}_{M}\left[ \int d^{4} \theta \left( \hat{{\cal Z}}^{\dagger}\hat{{\cal Z}} \right) \right] C^{\varphi} \varphi^{\dagger}\varphi,
\end{equation}
where the parameter $m_{M}$ is the messenger scale. We can generate the gaugino masses as
\begin{equation}
m^{-1}_{M}C^{\lambda} \left[ \int d^{2} \theta \hat{{\cal Z}} \left( WW \right) \right] +hc
\end{equation}
where 
\begin{equation}
WW \equiv Tr \left(W^{a \alpha}W^{a}_{\alpha} \right) ,
\end{equation} 
is the supersymmetric stregth field defined at Eq.(\ref{W-a}). The last term is generated in 
the following way
\begin{equation}
m^{-1}_{M}C^{\prime \varphi} \left[ \int d^{2} \theta \hat{{\cal Z}} 
\left( {\cal W}( \varphi ) \right) \right] +hc
\end{equation}
where $W(\varphi)$ is the superpotential of the model. 

\section{Parameter Space of the MSSM}
\label{sec:freeparameters}

It is well known that SM has 19 free parameters. The MSSM contains 124 free 
parameters and the symmetry breaking parameters are completely arbitrary \cite{dress,Baer:2006rs,Aitchison:2005cf}. The main goal in the 
SUSY phenomenology is to find some approximation about the way we can break SUSY in order 
to have a drastic reduction in the number of these parameters\footnote{Different assumptions result in different version of the 
Constrained Minimal Supersymmetric Model (CMSSM).}. 

Many phenomenological analyses adopt the universality hypothesis at the scale $Q \simeq M_{GUT}\simeq 2 \times 10^{16}$ GeV:
\begin{eqnarray}
g_{s}&=&g=g^{\prime} \equiv g_{GUT}, \nonumber \\
M_{3}&=&M=M^{\prime}\equiv m_{1/2}, \nonumber \\
M_{L}^{2}&=&M^{2}_{l}=M_{Q}^{2}=M^{2}_{u}=M^{2}_{d}=M_{1}^{2}=M_{2}^{2}\equiv m^{2}_{0}, \nonumber \\
A^{E}&=&A^{D}=A^{U}\equiv A_{0}.
\label{msugra}
\end{eqnarray}
The assumptions that the MSSM is valid between the weak scale and GUT scale, and that the "boundary conditions", 
defined by  the Eq.(\ref{msugra}) hold, are often referred to as mSUGRA, or minimal supergravity model. The mSUGRA 
model is completely specified by the parameter set \cite{dress,Baer:2006rs,Aitchison:2005cf}
\begin{eqnarray}
m_{0}, \,\ m_{1/2}, \,\ A_{0}, \,\ \tan \beta , \,\ \mbox{sign}( \mu ) .
\end{eqnarray}
The new free parameter $\beta$ is defined in the following way
\begin{eqnarray}
\tan \beta \equiv \frac{v_{2}}{v_{1}},
\label{defbetapar}
\end{eqnarray}
where $v_{2}$ is the vev of $H_{2}$ while $v_{1}$ is the vev of the 
$H_{1}$. Due the fact that $v_{1}$ and $v_{2}$ are both positive, it imples that 
\begin{equation}
0 \leq \beta \leq (\pi/2) \,\ \mbox{rad}.
\label{constraintinbeta}
\end{equation} 
Before we present the discussion about the mass spectrum on this model, 
we want to say once known some masses is possible to get the parameters 
in the Lagrangian, see for example \cite{moultaka}.
Now we are ready to present the mass spectrum of this model.

\section{Masses of all the particles of this model}
\label{sec:masses}

We already introduced all the particles of this model. Until now, they are symmetries 
eigenstates. It means they are not the physical ones we can observe in 
collider experiments and we will consider the $R$-Parity conservation. Then we need to get the mass eigenstates, we will 
perform it on this section, they represent the real particle we can 
measure at laboratories.

\subsection{gluinos}

The gluinos $\tilde{g}$ are the fermionic partner of the gluons\footnote{They are Majorana Fermions \cite{Majorana}, and see the comments 
below Eq.(\ref{eq:vwzdef}).} and its defined as 
\begin{equation}
\tilde{g}^{a}=\left( 
\begin{array}{c}
- \imath \lambda _{C}^{a} \\ 
\imath \overline{\lambda _{C}^{a}}
\end{array}
\right) \,\ ,\hspace{1cm}a=1,\ldots ,8, 
\end{equation}
is the Majorana four-spinor defining the physical gluinos states. Therefore 
they are the only color octet fermion and therefore they carry color charge as we discuss at \cite{Espindola:2011nb}.

Since the $SU(3)_{C}$ symmetry is not broken, the gluino cannot mix with any other fermion, and must be a 
mass eigenstate. Its mass term then arises just from the soft 
supersymmetry breaking, given at 
Eq.(\ref{The Soft SUSY-Breaking Term prop 3}), so that its mass at tree 
level is simply
\begin{equation}
M_{\tilde{g}}=|M_{3}|e^{\imath \phi_{\tilde{g}}}.
\end{equation}
The real parameter $M_{3}$ can be both positive or negative. Due this fact, we can define the gluino field, in the following way
\begin{eqnarray}
\tilde{g}\rightarrow \left( - \imath \gamma_{5} \right)^{\theta}\tilde{g},
\end{eqnarray}
where $\theta$ is defined as
\begin{equation}
\theta =\left\{ 
\begin{array}{c}
0, \mbox{for} M_{3}>0, \\
1, \mbox{for} M_{3}<0,
\end{array}
\right. 
\end{equation}
and the chiral Dirac matrix $\gamma_{5}$ is presented at Eq.(\ref{gamma5}), when we will derive some Feynman Rules of 
this model at Sec.(\ref{sec:feynmanrulesinMSSM}). The Feynman rules to the gluinos are presented at \cite{dress,Baer:2006rs,Aitchison:2005cf} and 
you can found the coupling of gluinos was explicity derived besides the differential cross section and the total cross section to gluinos production 
is presented at \cite{Mariotto:2008zt}.

The gluinos are expected to be one of the most massive sparticles which constitute the MSSM and therefore their production is only feasible 
at a very energetic machine such as the Large Hadron Collider (LHC). The gluino production in nuclear collisions was presented at 
\cite{Baer:2006rs,Haber:1994pe,dreiner1,Dawson,Espindola:2011nb,BrennerMariotto:2011wm,Espindola:2010zz}.Therefore one of the sources of CP violation in the 
MSSM arises from the gluinos sector. Some phenomenological studies about this case see \cite{Heinemeyer:2011ab}.

\subsection{Fermion Masses}

The fermion mass comes from the following terms, see the last line at Eq.(\ref{compsuperpotential}), of the superpotential 
defined at Eq.(\ref{mssmrpv}):
\begin{equation}
W=-\left(  f_{ij}^{l}L_{i}H_{1}l_{j}^{c}+y_{ij}^{d}Q_{i}H_{1}d_{j}^{c}+y_{ij}^{u}Q_{i}H_{2}u_{j}^{c}+hc\right) ,
\end{equation}
where $f_{ij}^{l}$, $f_{ij}^{d}$ and $f_{ij}^{u}$ are the yukawa couplings of
Higgs with leptons families, ``down" sector quarks and ``up" sector quarks
respectively. 

The fact that $m_{u},m_{d},m_{s}$ and $m_{e}$ are many orders of magnitude 
smaller than the masses of others fermions may well be indicative of a 
radiative mechanism at work for these masses as considered at 
\cite{banks,ma}. We can explain the mass hierarchy in the charged 
fermion masses as showed in \cite{cmmc,cmmc1}. We will review this topic below 

The key feature of this kind of mechanism is to allow only the quarks $c,b,t,$
and the leptons $\mu$ and $\tau$ have Yukawa couplings to the Higgs bosons. It
means to prevent $u,d,s$ and $e$ from picking up tree-level masses.

Following we calculated the masses of the $u,d$ quarks and the electron, and theis expression are given by:
\begin{eqnarray}
m_{u}  & \propto& \frac{\alpha_{s}\sin(2\theta_{\tilde{u}})}{\pi}m_{\tilde{g}%
}\left[  \frac{M_{\tilde{u_{1}}}^{2}}{M_{\tilde{u_{1}}}^{2}-m_{\tilde{g}}^{2}%
}\ln\left(  \frac{M_{\tilde{u_{1}}}^{2}}{m_{\tilde{g}}^{2}}\right)  \right.
\nonumber\\
& -& \left.  \frac{M_{\tilde{u_{2}}}^{2}}{M_{\tilde{u_{2}}}^{2}-m_{\tilde{g}%
}^{2}}\ln\left(  \frac{M_{\tilde{u_{2}}}^{2}}{m_{\tilde{g}}^{2}}\right)
\right]  \,\ ,\nonumber\\
m_{d}  & \propto& \frac{\alpha_{s}\sin(2\theta_{\tilde{d}})}{\pi}m_{\tilde{g}%
}\left[  \frac{M_{\tilde{d_{1}}}^{2}}{M_{\tilde{d_{1}}}^{2}-m_{\tilde{g}}^{2}%
}\ln\left(  \frac{M_{\tilde{d_{1}}}^{2}}{m_{\tilde{g}}^{2}}\right)  \right.
\nonumber\\
& -& \left.  \frac{M_{\tilde{d_{2}}}^{2}}{M_{\tilde{d_{2}}}^{2}-m_{\tilde{g}%
}^{2}}\ln\left(  \frac{M_{\tilde{d_{2}}}^{2}}{m_{\tilde{g}}^{2}}\right)
\right]  \,\ ,\nonumber\\
m_{e}  & \propto& \frac{\alpha_{U(1)}\sin(2\theta_{\tilde{e}})}{\pi}m^{\prime
}\left[  \frac{M_{\tilde{e_{1}}}^{2}}{M_{\tilde{e_{1}}}^{2}-m^{\prime2}}%
\ln\left(  \frac{M_{\tilde{e_{1}}}^{2}}{m^{\prime2}}\right)  \right.
\nonumber\\
& -& \left.  \frac{M_{\tilde{e_{2}}}^{2}}{M_{\tilde{e_{2}}}^{2}-m^{\prime2}}%
\ln\left(  \frac{M_{\tilde{e_{2}}}^{2}}{m^{\prime2}}\right)  \right]  \,\ .
\end{eqnarray}

\begin{figure}[ptb]
\parbox{14cm}{
\epsfxsize=13cm
\epsfbox{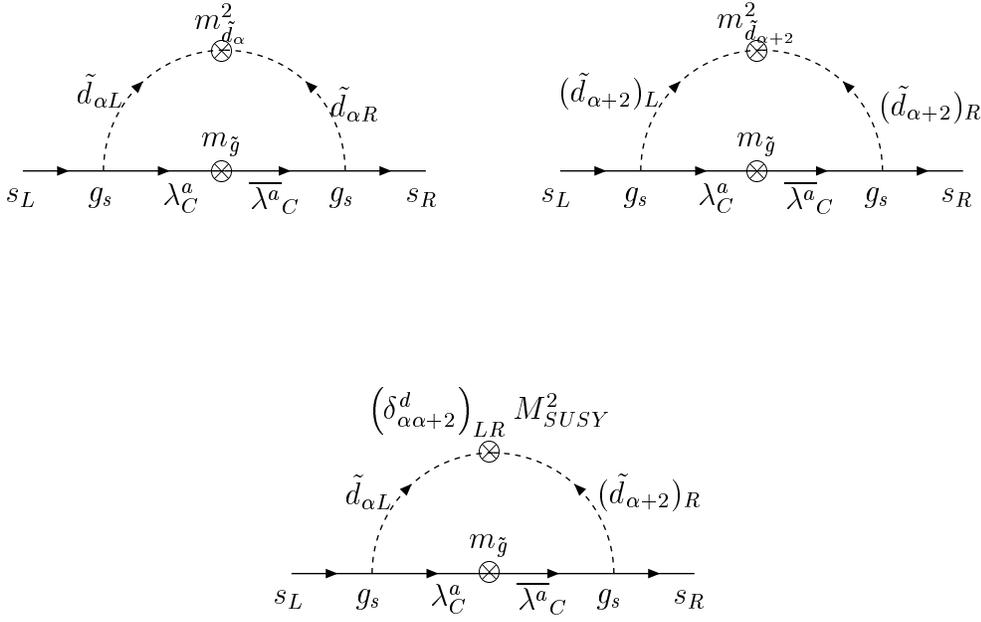}
}\newline
\caption{The diagram which gives mass to quark $s$, $\lambda^{a}_{C}$ is the gluino while $\tilde {s}
_{i}$ and $\tilde{b}_{i}$, $i=1,2$, are the squarks s-squark and sbottom,
respectively.}
\label{mass4}
\end{figure}

The expression for the mass of $s$ quark has a more complicated integral to
be solved, see Fig.(\ref{mass4}), turns into the following \cite{cmmc1}: 
\begin{eqnarray}
M_{s} &=&\frac{g_{s}^{2}m_{\tilde{g}}}{16\pi ^{4}}
\sum_{\alpha =1}^{2}\left\{ R_{1\alpha }^{(d)}R_{2\alpha }^{(d)}
\frac{m_{\tilde{g}}^{2}}{(m_{\tilde{g}}^{2}-m_{\tilde{d}_{\alpha }}^{2})}\ln \left( 
\frac{m_{\tilde{g}}^{2}}{m_{\tilde{d}_{\alpha }}^{2}}\right) +R_{1\alpha +2}^{(d)}
R_{2\alpha +2}^{(d)}\frac{m_{\tilde{g}}^{2}}{(m_{\tilde{g}}^{2}-
m_{\tilde{d}_{\alpha +2}}^{2})}\ln \left( \frac{m_{\tilde{g}}^{2}}{m_{\tilde{d}_{\alpha +2}}^{2}}
\right) \right.   \nonumber \\
&+&\left. \frac{R_{1\alpha }^{(d)}R_{2\alpha +2}^{(d)}}{(
m_{\tilde{d}_{\alpha }}^{2}-m_{\tilde{d}_{\alpha +2}}^{2})(m_{\tilde{g}}^{2}-
m_{\tilde{d}_{\alpha }}^{2})(m_{\tilde{d}_{\alpha +2}}^{2}-m_{\tilde{g}}^{2})}\left(
\delta _{\alpha \alpha +2}^{d}\right) _{LR}M_{SUSY}^{2}\left[ 
m_{\tilde{d}_{\alpha }}^{2}m_{\tilde{d}_{\alpha +2}}^{2}\ln \left( 
\frac{m_{\tilde{d}_{\alpha }}^{2}}{m_{\tilde{d}_{\alpha +2}}^{2}}\right) \right. \right.  
\nonumber \\
&+&\left. \left. m_{\tilde{d}_{\alpha }}^{2}m_{\tilde{g}}^{2}\ln \left( 
\frac{m_{\tilde{g}}^{2}}{m_{\tilde{d}_{\alpha }}^{2}}\right) +
m_{\tilde{d}_{\alpha +2}}^{2}m_{\tilde{g}}^{2}\ln \left( 
\frac{m_{\tilde{d}_{\alpha +2}}^{2}}{m_{\tilde{g}}^{2}}\right) \right] \right\} .
\end{eqnarray}

\subsection{Bosons Masses}

On the other hand, ${\cal L}_{Higgs}$ give mass to the gauge bosons, throught the following expression: 
\begin{eqnarray}
\left( {\cal D}_{m}H_{1}\right)^{\dagger}\left( {\cal D}_{m}H_{1}\right) + 
\left( {\cal D}_{m}H_{2}\right)^{\dagger}\left( {\cal D}_{m}H_{2}\right) ,
\label{originmassgaugebosons}
\end{eqnarray}
where ${\cal D}_{m}$ is covariant derivates of the SM given by:
\begin{eqnarray}
{\cal D}_{m}H_{1}\equiv \partial_{m}H_{1}+ \imath g \left( \frac{\sigma^{i}}{2}W^{i}_{m} \right) H_{1}+ \imath
g^{\prime} \left( \frac{Y_{H_{1}}}{2} b^{\prime}_{m} \right)H_{1}, \nonumber \\
{\cal D}_{m}H_{2}\equiv \partial_{m}H_{2}+ \imath g \left( \frac{\sigma^{i}}{2}W^{i}_{m} \right) H_{2}+ \imath
g^{\prime} \left( \frac{Y_{H_{2}}}{2} b^{\prime}_{m} \right)H_{2}.
\end{eqnarray}
From the Eq.(\ref{originmassgaugebosons}) beyond the masses of the Gauge bosons we also 
get the interactions between the usual scalars with the Gauge boson to see the Feynman Rules on this case see 
\cite{dress,Baer:2006rs,Aitchison:2005cf}.

After some simple calculatio, we get the following expression to the masses of the charged ones 
\begin{equation}
W^{\pm}_{m}= \frac{1}{\sqrt{2}}\left( V^{1}_{m} \mp \imath V^{2}_{m}\right)
\label{wdef}
\end{equation} 
get the following mass
\begin{eqnarray}
M^{2}_{W}= \frac{g^{2}}{4}(v_{1}^{2}+v_{2}^{2})=
\frac{g^{2}v_{1}^{2}}{4}(1+\tan^{2} \beta)= 
\frac{g^{2}v^{2}_{1}}{4 \cos^{2} \beta}=\frac{g^{2}v^{2}_{1}}{4}\sec^{2} \beta \,\ ,
\label{wmass}
\end{eqnarray}
where $\tan \beta$ is defined at Eq.(\ref{defbetapar}). 
Due the fact that both $v_{1}$ and $v_{2}$ are real positive number with this in mind we can justify Eq.(\ref{constraintinbeta}). 

In the MSSM, the masses of the charged boson has two free parameters: $v_{1}$, as in the SM, plus the new parameter $\beta$. In this model we 
recover the results given at SM when we fix 
$\beta =0$ rad and the $W$ mass considerating 
several values of $\beta$ parameter is showing at Fig.(\ref{fig1}), 
the black line represent the experimental values of 
$M_{W}= \left( 80.399\pm 0.023 \right) \,\ {\mbox GeV}$. On this figure 
we also show the behaviour of $W$ mass in terms of $\beta$ parameter and we see when $v_{1}>180$ GeV, we can consider any $\beta$ parameter to explain the 
$W$ mass. 

The next plot is given at Fig.(\ref{fig2}), 
where we show the W mass in function of $\beta$ parameter when we take several values to $v_{1}$ parameter. 
When we taken into account all the figures, we conclude that for the case of 
$v_{1} \geq 174 {\mbox GeV}$ we can fix the W mass in concordance with the experimental data. 

\begin{figure}[ht]
\begin{center}
\vglue -0.009cm
\mbox{\epsfig{file=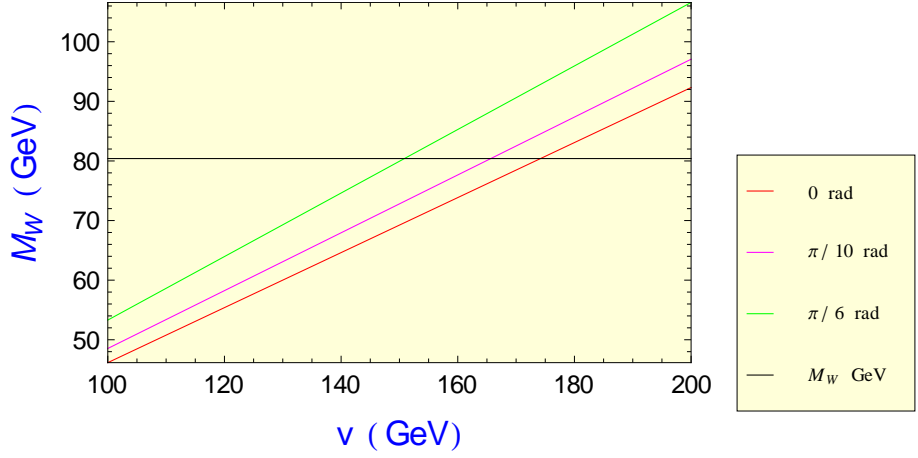,width=0.7\textwidth,angle=0}}       
\end{center}
\caption{The masses of $W$ to several values of the $\beta$ parameter 
in therms of the vev of $H_{1}$, the black line means the experimental 
values of $M_{W}$.}
\label{fig1}
\end{figure}

\begin{figure}[ht]
\begin{center}
\vglue -0.009cm
\mbox{\epsfig{file=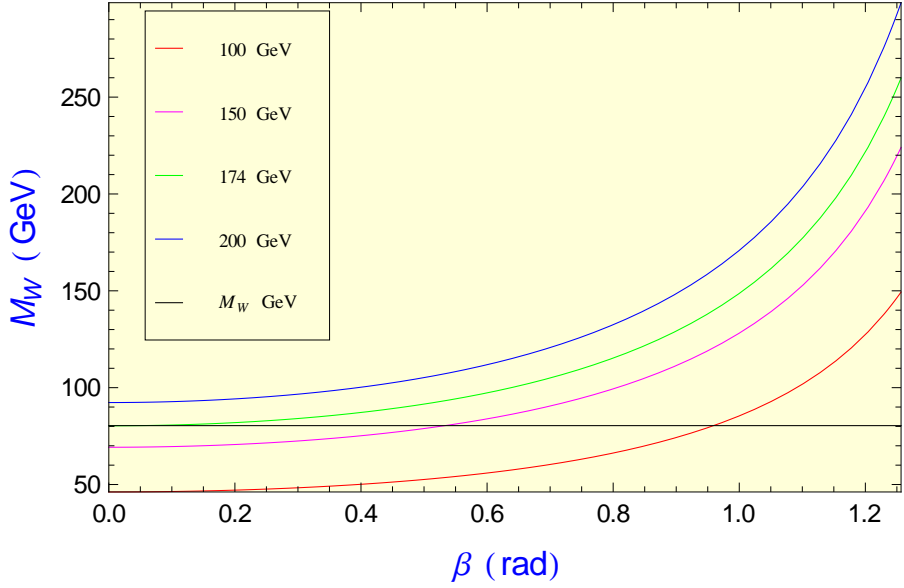,width=0.7\textwidth,angle=0}}       
\end{center}
\caption{The masses of $W$ in terms of $\beta$ parameter considerating 
several values of $v_{1}$, again the black line is the experimental value 
of $M_{W}$.}
\label{fig2}
\end{figure}

The neutral massive gauge boson ($Z^{0}$) get the following mass
\begin{eqnarray}
M^{2}_{Z}&=& \left( \frac{g^{2}+g^{\prime 2}}{4}\right) (v_{1}^{2}+v_{2}^{2})=
\frac{g^{2}}{4 \cos^{2} \theta_{W}} (v_{1}^{2}+v_{2}^{2})
= \frac{M^{2}_{W}}{ \cos^{2} \theta_{W}}, 
\label{z-mass}
\end{eqnarray}
where $\theta_{W}$ is the Weinberg angle and it is defined as
\begin{eqnarray}
e=g \sin \theta_{W}=g^{\prime}\cos \theta_{W},
\label{weinbergangledefinition}
\end{eqnarray}
and get a massless foton $A_{m}$. The experimental values are 
\begin{eqnarray}
M_{Z}&=&\left( 91.1888 \pm 0.0044 \right)\,\mbox{\rm GeV}, \nonumber \\
\sin^{2}\theta_{W}&=&1- \frac{M^{2}_{W}}{M^{2}_{Z}}=0.2320\pm 0.0004,
\label{expzmass}
\end{eqnarray}
The rotation in this case is
\begin{eqnarray}
\left( \begin{array}{c} A_{m} \\ Z_{m} \end{array} \right)= 
\left(\begin{array}{cc}
\sin \theta_{W} & \cos \theta_{W} \\
\cos \theta_{W} & - \sin \theta_{W}
\end{array} \right)
\left( \begin{array}{c} V^{3}_{m} \\ V_{m} \end{array} \right) \,\ ,
\label{boson5}
\end{eqnarray} 
it is the exact expression we get in the SM. Therefore the neutral 
boson gauge sector is exact the same as in the SM.

\subsubsection{Photino is not a mass eigenstate}
\label{sec:mixphotino-zino}

Before we present the neutralinos, we want to discuss first, that we learnt the photon and $Z^{0}$ gauge boson diagonalize the neutral 
boson sector of the  SM, throught the rotation defined at Eq.(\ref{boson5}). As we are dealing with supersymmetry, we can ask, 
are the photino ($\tilde{\gamma}$) and 
the zino ($\tilde{Z}^{0}$), the superpartnes of photons and $Z^{0}$ gauge boson respectivelly, are masses eigenstates? The answer 
to this question is no and we will show this results below, this material is a review of the results presented at \cite{mssm}. 

In order to show it let us first define their four-component spinors to the 
photino ($\tilde{\gamma}$) and the zino ($\tilde{Z}$)
\footnote{They are Majorana Fermions \cite{Majorana}, and see the comments below Eq.(\ref{eq:vwzdef}).} are 
\begin{eqnarray}
\tilde{\gamma}= \left( \begin{array}{r}  
- \imath \lambda_{\gamma}(x) \\
\imath \bar{\lambda}_{\gamma}(x)   
\end{array}   \right) \,\ , \,\
\tilde{Z}= \left( \begin{array}{r}   
- \imath \lambda_{Z} \\
\imath \bar{\lambda}_{Z}   
\end{array}   \right) \,\ .
\end{eqnarray}
By another we can define the winos ($\tilde{W}$) in the following way
\begin{eqnarray}
\tilde{W}(x) \;=\; \left( 
\begin{array}{r}   
- \imath  \lambda^{+}(x) \\
\imath  \bar{\lambda}^{-}(x)   
\end{array}   \right) \,\ , 
\end{eqnarray} 
Where $\lambda^{\pm}$ is defined at Eq.(\ref{wino2comp}), while $\lambda_{\gamma}$ and 
$\lambda_{Z}$ can be defined as function of $\lambda^{3}$ and $\lambda$ in the following way:
\begin{eqnarray}
\lambda_{\gamma}&=& \cos \theta_{W}\lambda^{3}- \sin \theta_{W}\lambda 
, \nonumber \\
\lambda_{Z}&=&\sin \theta_{W}\lambda^{3}+ \cos \theta_{W}\lambda .
\end{eqnarray}
$\theta_{W}$ is the Weinberg angle defined at Eq.(\ref{weinbergangledefinition}). 

The best way to see this result is taking into account 
Eq.(\ref{The Soft SUSY-Breaking Term prop 3}) we can write \cite{mssm} 
\begin{eqnarray}
&-& \frac{1}{2} M\left( 
\lambda^{1}\lambda^{1}+\lambda^{2}\lambda^{2}+\lambda^{3}\lambda^{3}+ 
\bar{\lambda}^{1}\bar{\lambda}^{1}+\bar{\lambda}^{2}\bar{\lambda}^{2}+
\bar{\lambda}^{3}\bar{\lambda}^{3}\right)
- \frac{1}{2} M^{\prime}\left(\lambda \lambda + \bar{\lambda}\bar{\lambda}\right) \nonumber \\
&=&- M \left(\lambda^{-}\lambda^{+}+\bar{\lambda}^{-}\bar{\lambda}^{+}\right) 
- \frac{1}{2} \left( M \sin^{2}\theta_{\mbox{w}}+ M^{\prime} \cos^{2}\theta_{W} \right)
\left(\lambda_{\gamma}\lambda_{\gamma}+ \bar{\lambda}_{\gamma}\bar{\lambda}_{\gamma}\right) \nonumber \\ 
&-& \frac{1}{2} \left( M \cos^{2}\theta_{W}+ M^{\prime} \sin^{2}\theta_{W} \right)
\left(\lambda_{Z}\lambda_{Z}+\bar{\lambda}_{Z}\bar{\lambda}_{Z}\right) 
- \left( M -  M^{\prime} \right) \sin({2\theta_{W}})\;
\left(\lambda_{\gamma}\lambda_{Z}+\bar{\lambda}_{\gamma}\bar{\lambda}_{Z}\right) \nonumber \\
&=&- \left\{   M_{\tilde{W}}\,\bar{\tilde{W}}\tilde{W}+
\frac{1}{2} \left[ \,  M_{\tilde{\gamma}}\; \bar{\tilde{\gamma}}\tilde{\gamma}
+ \,M_{\tilde{Z}}\,\bar{\tilde{Z}}\tilde{Z}
+\left( M_{\tilde{Z}} -  M_{\tilde{\gamma}} \right) \tan({2\theta_{W}})\bar{\tilde{\gamma}}\tilde{Z} 
\right] \right\} \,\ ,
\label{Component Field Expansion of L sub Soft prop 6}
\end{eqnarray}
where
\begin{eqnarray}
M_{\tilde{W}} &\equiv& M, \nonumber \\
M_{\tilde{\gamma}}&=& M \sin^{2}\theta_{W}+ M^{\prime}\cos^{2}\theta_{W}  \,\ , \nonumber \\
M_{\tilde{Z}}&=& M \cos^{2}\theta_{W}+  M^{\prime}\sin^{2}\theta_{W}  \,\ .
\end{eqnarray}
Therefore a priori the winos are mass eigenstate however the photinos and zino mixing to each other and their mass eigenstates are the neutralinos, 
see Sec.(\ref{subsec:neutralinos}). The unification condition give us the following result \cite{dress,Baer:2006rs,Aitchison:2005cf,mssm,charginos,neutralinos,indiano}
\begin{equation}
\frac{M^{\prime}}{M}= \frac{5}{3}\tan^{2}\theta_{W}.
\end{equation} 

\subsection{Higgs Masses}

The scalar potential in the MSSM is given by:
\begin{eqnarray}
V^{H}_{MSSM}&=&( \mu^{2}_{H}+M^{2}_{H_{1}})|H_{1}|^{2}+ 
\frac{1}{8}\left( g^{2}+g^{2}_{Y^{\prime}} \right)(|H_{1}|^{2})^{2}+
( \mu^{2}_{H}+M^{2}_{H_{2}})|H_{2}|^{2}+ \frac{1}{8}\left( g^{2}+g^{2}_{Y^{\prime}} \right)(|H_{2}|^{2})^{2} 
\nonumber \\ &-&
\frac{1}{4}\left( g^{2}+g^{2}_{Y^{\prime}} \right)|H_{1}|^{2}|H_{2}|^{2}+
\frac{g^{2}}{2}|\bar{H}_{1}H_{2}|^{2}- ( M^{2}_{12}\epsilon_{ab} H_{1}^{a}H_{2}^{b}+H.c.).
\nonumber \\
\end{eqnarray}
The $8 \times 8$ Higgs mass squared matrix breaks up diagonally into 
three set of $2 \times 2$ matrix. After the diagonalization procedure 
we finish with five physical degrees of freedom form
a neutral CP--odd, two neutral CP--even and two charged Higgs bosons
denoted by $A^{0}$, $h^{0}$, $H^{0}$, and $H^{\pm}$, respectively \cite{dress,Baer:2006rs,Aitchison:2005cf,INO82a,INO82b}.

We start our analyses in the CP--odd sector. The mass squared matrix 
in this sector is found to be
\begin{equation}
{\cal M}_{\Im(H^{0})}=\frac{M^{2}_{12}}{v_{1}v_{2}} \left(
\begin{array}{cc}
v_{2}^{2} & v_{1}v_{2} \\
v_{1}v_{2} & v_{1}^{2}
\end{array}
\right).
\end{equation}
We can in a simple way show
\begin{eqnarray}
\det[{\cal M}_{\Im(H^{0})}]&=&0, \nonumber \\
{\mbox Tr}[{\cal M}_{\Im(H^{0})}]&=&\frac{M^{2}_{12}}{v_{1}v_{2}} 
\left( v_{1}^{2}+v_{2}^{2} \right).
\end{eqnarray}

The vanishing determinant and the no vanishing trace of this matrix 
imply massless (Goldstone boson $G^{0}$) as well as massive neutral model 
($A^{0}$). The massive $A^{0}$ particle remains as a pseudoscalar 
Higgs boson and its mass is proportional to the soft SUSY breaking 
parameter $M^{2}_{12}$, therefore it should be a heavy than the Higgs 
boson defined at SM. As in general we suppose all soft parameters are 
in the TeV range, we can conclude $M_{A} \sim {\cal O}( \mbox{TeV})$. 

The mass spectrum in this sector is given by 
\begin{equation}
\left(
\begin{array}{c}
G^{0}\\
A^{0}
\end{array}
\right) = \sqrt{2} 
\left(
\begin{array}{cc}
\sin \beta & \cos \beta \\
- \cos \beta & \sin \beta
\end{array}
\right)
\left(
\begin{array}{c}
\Im(H_{2}^{0})\\
\Im(H_{1}^{0})
\end{array}
\right)  ,
\label{pseudoatmssm}
\end{equation}
and the mass of this sector is given by:
\begin{eqnarray}
M^{2}_{G^{0}}&=&0, \nonumber \\
M^{2}_{A^{0}}&=& \frac{M^{2}_{12}}{v_{1}v_{2}}\left( 
v^{2}_{1}+v^{2}_{2}\right) 
= M^{2}_{12}\left( \frac{v_{1}}{v_{2}}+ \frac{v_{2}}{v_{1}} \right)
=M^{2}_{12} \left( \tan \beta + \cot \beta \right)
=M^{2}_{12} \csc \beta \sec \beta , \nonumber \\
M^{2}_{A^{0}}&=& \frac{2M^{2}_{12}}{\sin (2 \beta )} \gg M^{2}_{Z}. \nonumber \\
\end{eqnarray}
$G^{0}$ combines with the massless $Z$ to give their mass, as in the SM. 
As we want finite mass, the above equation put some constraints in the 
$\beta$ parameter. It has to satisfy the following constraints
\begin{equation}
\beta \neq 0 \,\ \mbox{rad}, \,\ \mbox{and} \,\
\beta \neq \frac{\pi}{2} \,\ \mbox{rad}.
\end{equation}
$\Re(H)$ denotes the real and $\Im(H)$ the imaginary part of $H$.

On Fig.(\ref{fig4}), where we can see this pseudoscalar can be very 
heavy (as we have discussed above) and as we mentioned above we can 
see that it diverge when $\beta \rightarrow 0$ rad or when 
$\beta \rightarrow (\pi /2)$ rad. Of course the minium value of $A$ depend 
of the value choosen to $M_{12}$ parameter (see the differents colors 
at Fig.(\ref{fig4})), but we can say its mass can 
go from $3$ TeV until infinity.

\begin{figure}[ht]
\begin{center}
\vglue -0.009cm
\mbox{\epsfig{file=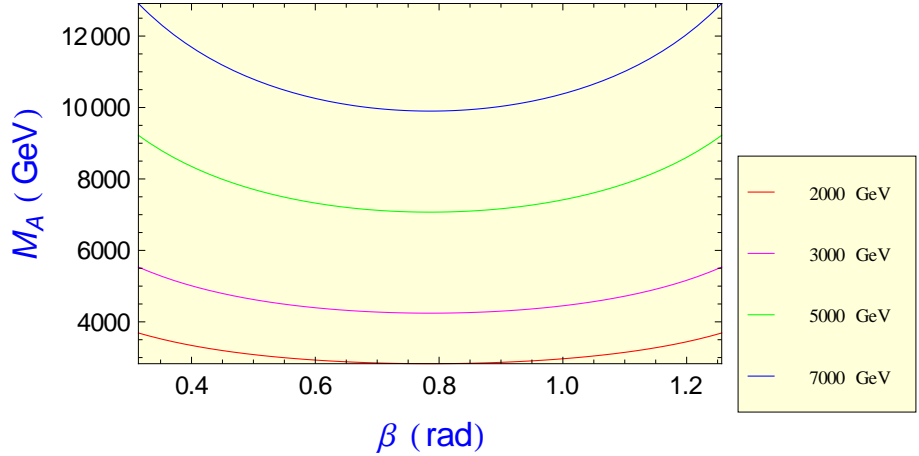,width=0.7\textwidth,angle=0}}       
\end{center}
\caption{The masses of the pseudoscalar $A$ in terms of the parameter 
$\beta$, and several values of $M_{12}$.}
\label{fig4}
\end{figure}

\begin{figure}[ht]
\begin{center}
\vglue -0.009cm
\mbox{\epsfig{file=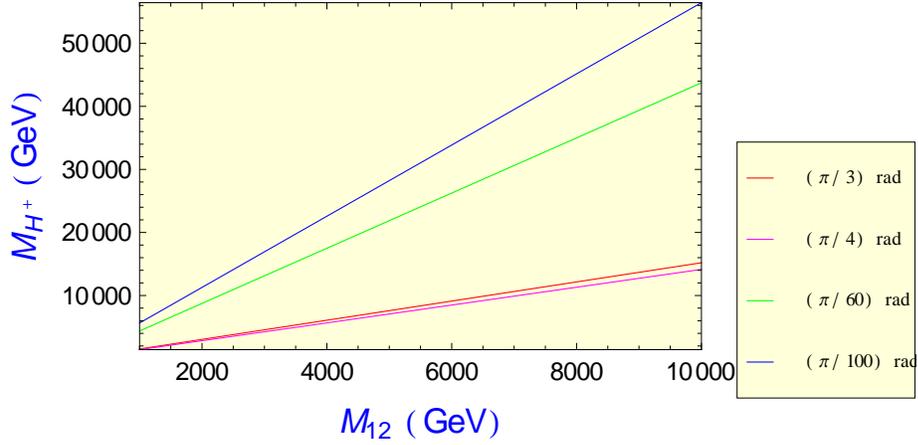,width=0.7\textwidth,angle=0}}       
\end{center}
\caption{The masses of the charged scalar $H^{+}$ in terms of 
the $M_{12}$ and several values to the $\beta$ parameter.}
\label{fig5}
\end{figure}

The charged Higgs sector is very similar to the CP--odd sector. The physical states in terms 
of the symmetry eigen states are defined in the following way
\begin{equation}
\left(
\begin{array}{c}
H^{\pm}\\
G^{\pm}
\end{array}
\right) =  
\left(
\begin{array}{cc}
\sin \beta &- \cos \beta \\
\cos \beta & \sin \beta
\end{array}
\right) \cdot
\left(
\begin{array}{c}
H_{2}^{\pm}\\
(H_{1}^{\mp})^{\dagger}
\end{array}
\right) . 
\label{chargedscalartmssm}
\end{equation}
its mass is given by
\begin{eqnarray}
M^{2}_{G^{\pm}}&=&0, \nonumber \\
M^{2}_{H^{\pm}}&=&M^{2}_{W}+M^{2}_{A^{0}} \gg M^{2}_{W}.
\label{chargedhiggsmass}
\end{eqnarray}
$G^{\pm}$ combines with the massless $W^{\pm}$ to give them mass, as in the SM. Remember $M_{W}$ is the gauge $W^{\pm}$ mass, 
see Eq.(\ref{wmass}). On Fig.(\ref{fig5}) we see that the mass of the charged Higgs is almost 
linear in terms of $M_{12}$, as we expected from 
Eq.(\ref{chargedhiggsmass}).

The mass squared matrix in the CP--even sector is given by:
\begin{equation}
{\cal M}_{\Re(H^{0})}=\left(
\begin{array}{cc}
M^{2}_{A^{0}}\sin^{2}\beta +M^{2}_{Z}\cos^{2}\beta & 
-(M^{2}_{A^{0}}+M^{2}_{Z})\sin \beta \cos \beta \\
-(M^{2}_{A^{0}}+M^{2}_{Z})\sin \beta \cos \beta & 
M^{2}_{A^{0}}\cos^{2}\beta +M^{2}_{Z}\sin^{2}\beta
\end{array}
\right).
\end{equation}
Both the determinant as well the trace of this matrix is non vanishing, 
therefore in this sector we have two massive real scalars. We can in a simple way show
\begin{eqnarray}
\det[{\cal M}_{\Re(H^{0})}]&=&M^{2}_{A^{0}}M^{2}_{Z}
\cos^{2} \left( 2 \beta \right), \nonumber \\
{\mbox Tr}[{\cal M}_{\Im(H^{0})}]&=&M^{2}_{A^{0}}+M^{2}_{Z}.
\end{eqnarray}

The eigenvalues of this mass matrix are
\begin{equation}
M^{2}_{h^{0},H^{0}}= \frac{1}{2}\left[
M^{2}_{A^{0}}+M^{2}_{Z} \mp 
\sqrt{(M^{2}_{A^{0}}+M^{2}_{Z})^{2}-4M^{2}_{A^{0}}M^{2}_{Z}\cos^{2}( 2 \beta)}
\right].
\label{cpevenscalarmass}
\end{equation}
We note that
\begin{equation}
M^{2}_{h^{0}}+M^{2}_{H^{0}}=M^{2}_{A^{0}}+M^{2}_{Z}.
\end{equation}
In equation above we have defined $H$ to be the heavier of the two, it 
means $M_{h^{0}}<M_{H^{0}}$. The corresponding mass eigenstates are
\begin{equation}
\left(
\begin{array}{c}
h^{0}\\
H^{0}
\end{array}
\right) = \sqrt{2} 
\left(
\begin{array}{cc}
\sin \alpha & \cos \alpha \\
- \cos \alpha & \sin \alpha
\end{array}
\right)
\left(
\begin{array}{c}
\Re(H_{2}^{0})- \frac{v_{2}}{\sqrt{2}}\\
\Re(H_{1}^{0})- \frac{v_{1}}{\sqrt{2}}
\end{array}
\right) . 
\label{realatmssm}
\end{equation}
The angle of rotation $\alpha$, defined in the equation above, is seen 
to obey the followings constraints \cite{dress}
\begin{eqnarray}
\sin (2 \alpha)&=&- 
\frac{M^{2}_{H^{0}}+M^{2}_{h^{0}}}{M^{2}_{H^{0}}-M^{2}_{h^{0}}}\sin (2 \beta), 
\nonumber \\
\cos (2 \alpha)&=&-
\frac{M^{2}_{A^{0}}-M^{2}_{Z}}{M^{2}_{H^{0}}-M^{2}_{h^{0}}}\cos (2 \beta), 
\nonumber \\
\tan (2 \alpha)&=& 
\frac{M^{2}_{H^{0}}+M^{2}_{h^{0}}}{M^{2}_{A^{0}}-M^{2}_{Z}}
\tan (2 \beta),
\end{eqnarray}
considerating Eq.(\ref{defbetapar}) in the first equation above 
it implies $\sin (2 \alpha) <0$\footnote{$M_{H^{0}}>M_{h^{0}}$.} while in the 
second equation we get $\cos (2 \alpha) <0$
\footnote{$M_{A^{0}}>M_{Z}$.}. Taking this information 
we can  conclude the range of the new parameter $\alpha$ is given by
\begin{equation}
-\frac{\pi}{2} \leq \alpha \leq 0 \,\ ( \mbox{rad}).
\end{equation}

\begin{figure}[ht]
\begin{center}
\vglue -0.009cm
\mbox{\epsfig{file=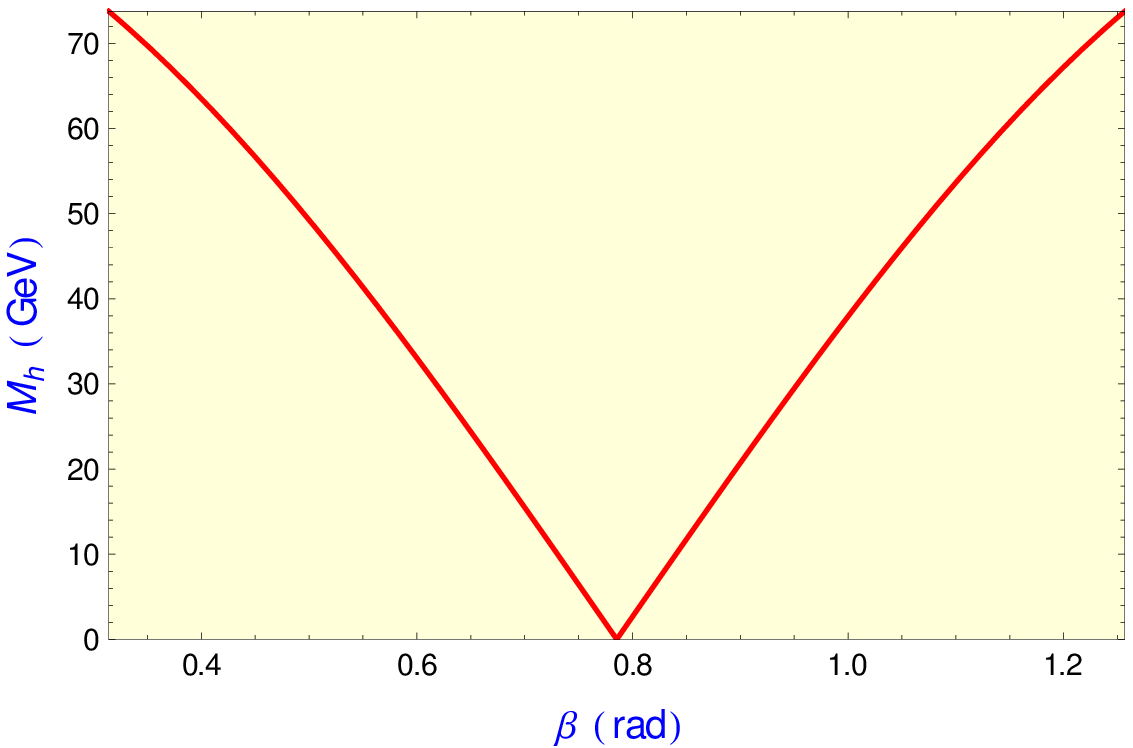,width=0.7\textwidth,angle=0}}       
\end{center}
\caption{The masses of lighest higgs even scalar $h^{0}$ in terms of 
the $\beta$ parameter considerating $M_{12}=10000$GeV.}
\label{fig6}
\end{figure}

\begin{figure}[ht]
\begin{center}
\vglue -0.009cm
\mbox{\epsfig{file=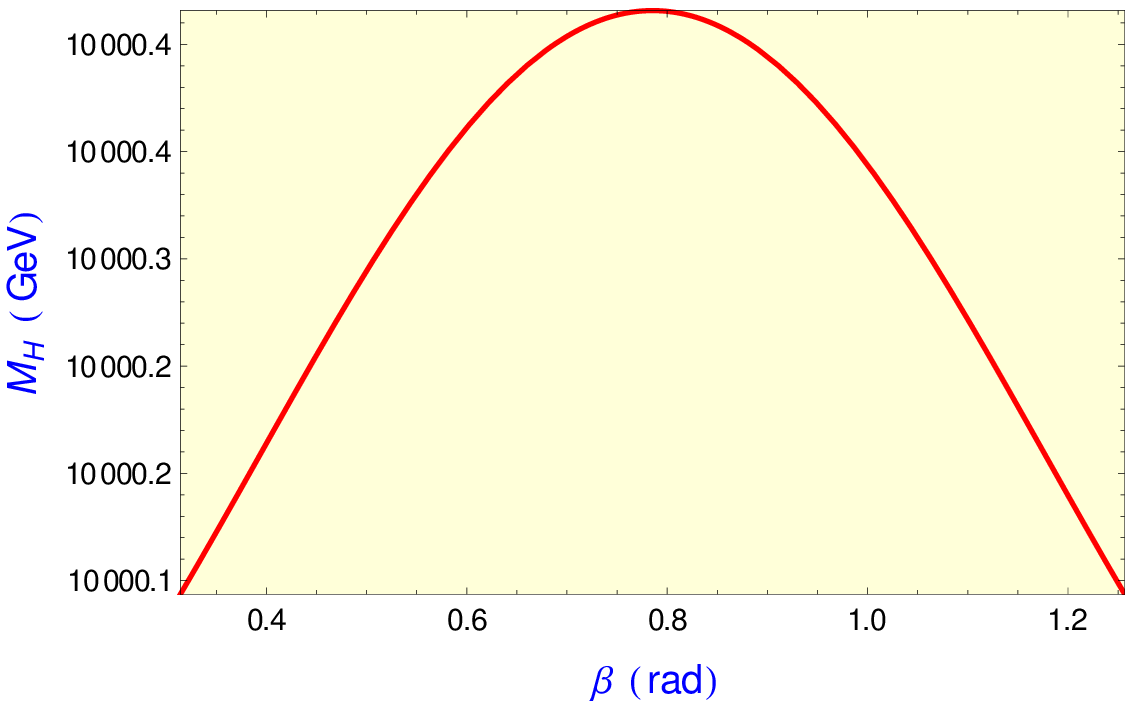,width=0.7\textwidth,angle=0}}       
\end{center}
\caption{The masses of heaviest higgs Evenn scalar $H^{0}$ in terms of 
the $\beta$ parameter considerating $M_{12}=1000$GeV.}
\label{fig7}
\end{figure}

On Fig(\ref{fig6}) we show same if we consider $M_{12}=10$TeV we 
get $m_{h^{0}}<M_{Z}$. This results is in agreement with the following 
well known constraints \cite{dress,Baer:2006rs,Aitchison:2005cf}
\begin{equation}
M_{h^{0}}\leq M_{Z}|\cos (2 \beta)|,
\end{equation}
this implies that $m_{h^{0}}=0$ if $\tan \beta =1$, this results is 
shown at Fig.(\ref{fig6}) where we shown the behaviour of the mass of 
the ligest scalar as function of $\beta$ parameter. Similar information to 
the heaviest Higgs is shown in Fig(\ref{fig7}). If we consider $M=1000$ GeV we get for several values 
of $\beta$ parameter masses to all usual scalars of this model, as we 
can see at Tab.(\ref{tab:scalars}).

\begin{table}[h]
\begin{center}
\begin{tabular}{|c|c|c|c|c|}
\hline 
$\tan \beta $ & $ m_{A^{0}} (\mbox{GeV}) $ 
& $m_{H^{\pm}} (\mbox{GeV}) $ & 
$ m_{h^{0}} (\mbox{GeV}) $ 
& $m_{H^{0}} (\mbox{GeV}) $ \\
\hline
0.10 & 3172.85 & 3173.87 & 89.35 & 1000.17 \\ 
\hline
0.20 & 2266.24 & 2267.67 & 83.94 & 1000.63  \\
\hline
0.55 & 1541.68 & 1543.78 & 49.12 & 1002.95  \\
\hline
1.03 & 1414.52 & 1416.80 & 2.65 & 1004.15 \\
\hline
1.56 & 1483.07 & 1485.25 & 37.82 & 1003.44  \\
\hline
2.57 & 1720.74 & 1722.61 & 67.11 & 1001.90 \\
\hline
3.60 & 1969.70 & 1971.34 & 78.05 & 1001.11 \\
\hline
5.80 & 2443.43 & 2444.75 & 85.88 & 1000.47 \\
\hline
\end{tabular}
\end{center}
\caption{\small Masses of all usual scalars to several values of 
$\tan \beta$.}
\label{tab:scalars}
\end{table}

Therefore, the light scalar $h^{0}$ has a mass smaller than the $Z^{0}$ gauge boson at the tree level. This implies that one has to 
consider the one-loop corrections which lead to the following result \cite{Haber:1990aw}
\begin{equation}
M^{2}_{h^{0}}\simeq M^{2}_{Z}+ 
\frac{3g^{2}M^{4}_{Z}}{16 \pi^{2}M^{2}_{W}}\left\{ 
\ln \left( \frac{M^{2}_{\tilde{t}}}{M^{2}_{t}}\right) \left[ \frac{2M^{4}_{t}-M^{2}_{t}M^{2}_{Z}}{M^{4}_{Z}}\right] +
\frac{M^{2}_{t}}{3M^{2}_{Z}}\right\}.
\end{equation}
However, radiative corrections rise it to 130 GeV~\cite{haber2}.

We can also write the following ralation \cite{dress}:
\begin{eqnarray}
\cos^{2}( \beta - \alpha )&=& 
\frac{M^{2}_{h^{0}}(M^{2}_{Z}-M^{2}_{h^{0}})}{M^{2}_{A^{0}}(M^{2}_{H^{0}}-M^{2}_{h^{0}})}. 
\end{eqnarray}
This equation together with Eq.(\ref{cpevenscalarmass}), show that 
the Higgs sector is completely controlled by two new parameters 
wchich can be taken to be $M_{A^{0}}$ and $\tan \beta$.

Here we presented only the analyses when CP is conserved. To see the 
case when CP is not conserved see \cite{macarena}.

\subsection{Charged Sleptons Masses}

The Lagrangian contains off-diagonal mass terms for the sleptons in the basis 
$(\tilde{l}_{L}, \tilde{l}^{c}_{L}\equiv \tilde{l}_{R} )$.
So, also here we have to perform a diagonalizing procedure to obtain the physical mass eigenstates, and hence we have
\begin{eqnarray}
{\cal L}_{slepton}^{mass}&=&-\mu f^{l}v_{2}\;\tilde{l}_{L}^{\dagger}\tilde{l}_{R}
-\mu f^{l}v_{2}\;\tilde{l}_{R}^{\dagger}\tilde{l}_{L}-f^{l 2}v_{1}^{2} \left( \tilde{l}_{L}^{\dagger}\tilde{l}_{L}
+\tilde{l}_{R}^{\dagger}\tilde{l}_{R} \right) 
- M_{L}^{2}\;\tilde{l}_{L}^{\dagger}\tilde{l}_{L}-M_{l}^{2}\;\tilde{l}_{R}^{\dagger}\tilde{l}_{R} \nonumber \\
&=&- \left( 
\begin{array}{cc} 
\tilde{l}_{L}^{\dagger} & \tilde{l}_{R}^{\dagger} 
\end{array}\right)
\left(
\begin{array}{cc}  
M_{L}^{2} +f^{ l2}v_{1}^{2} & \mu f^{l}v_{2} \\
\mu f^{l}v_{2} & M_{l}^{2}+f^{l 2}v_{1}^{2} 
\end{array}\right)
\left( 
\begin{array}{c} 
\tilde{l}_{L} \\ 
\tilde{l}_{R} 
\end{array}\right) .  \nonumber
\end{eqnarray}
By diagonalizing, one obtains the mass eigenstates (in the usual way)
\begin{eqnarray}
\left( 
\begin{array}{c} 
\tilde{l}_{1} \\ 
\tilde{l}_{2} 
\end{array}\right) =
\left( 
\begin{array}{cc} 
\cos{\theta_{\tilde{l}}} & \sin{\theta_{\tilde{l}}}  \\ 
\sin{\theta_{\tilde{l}}} & - \cos{\theta_{\tilde{l}}} 
\end{array}\right)
\left( 
\begin{array}{c} 
\tilde{l}_{L} \\ 
\tilde{l}_{R} 
\end{array}\right) , \nonumber
\label{physicalchargedsleptons}
\end{eqnarray}
with\footnote{Notice that $f^{l}v_{2}=f^{l}v_{1} \frac{v_{2}}{v_{1}}=m_{l}\tan\beta$, $m_{l}$ is the charged lepton mass.}
\begin{eqnarray}
\tan{2\theta_{\tilde{l}}} &=& \frac{2\mu f^{l}v_{2}}{\left( M^{2}_{L}-M^{2}_{l}\right)}
= \frac{ 2 \mu m_{l} \tan\beta }{ \left( M^{2}_{L}-M^{2}_{l}\right) },
  \nonumber
\end{eqnarray}
and masses respectively given by
\begin{eqnarray}
M_{\tilde{l}_{1},\tilde{l}_{2}}^{2}&=& m^{2}_{l}+ \frac{1}{2}\left[ \left( M^{2}_{L}-M^{2}_{l}\right)
\pm \sqrt{ \left( M^{2}_{L}-M^{2}_{l}\right)^{2}+4\mu^{2}m^{2}_{l}\tan^{2}\beta}\;\right].
\end{eqnarray}
For selectrons and smuons, the ``left'' and ``right'' states 
($\tilde{e}_{L,R}$ and $\tilde{\mu}_{L,R}$) are also the mass eigenstates. 
For staus, however, the eigenstates are $\tilde{\tau}_{1}$ and 
$\tilde{\tau}_{2}$. The production of selectrons and sneutrinos were sutied at \cite{dress,Baer:2006rs,Haber:1994pe,dreiner1,Gluck:1983zc,Baer:1993ew}. 

Let us now turn to the sneutrinos. In the case of massless neutrinos, 
there is only one sneutrino, $\tilde{\nu}_{L}$, with a mass 
\begin{equation}
m^{2}_{\tilde{\nu}_{L}}=M^{2}_{L} + \frac{M_{Z}^{2} \cos 2 \beta}{2},
\end{equation}
for each generation. Some Feynman rules to fermion, sfermion and gauge bosons will be presented at 
\ref{sec:feynmanrulesinMSSM}. The production of selectrons and sneutrinos were sutied at \cite{dress,Baer:2006rs,Haber:1994pe,dreiner1,Bartl:1987zg}. 

\subsection{Squarks}

The squarks $\tilde{q}_{L}$ and $\tilde{q}_{R}$ will mix, in a similar way as happened to the charged sleptons, to more details see 
\cite{dress,Baer:2006rs,Aitchison:2005cf}. We will donate the physical squark states as $\tilde{q}_{1},\tilde{q}_{2}$, and they are define as
\begin{eqnarray}
\left( 
\begin{array}{c} 
\tilde{q}_{1} \\ 
\tilde{q}_{2} 
\end{array}\right) =
\left( 
\begin{array}{cc} 
\cos{\theta_{\tilde{q}}} & \sin{\theta_{\tilde{q}}}  \\ 
\sin{\theta_{\tilde{q}}} & - \cos{\theta_{\tilde{q}}} 
\end{array}\right)
\left( 
\begin{array}{c} 
\tilde{q}_{L} \\ 
\tilde{q}_{R} 
\end{array}\right).
\label{physicalsquarks}
\end{eqnarray}
The Feynman rules to the squarks are presented at \cite{dress,Baer:2006rs,Aitchison:2005cf}. The squark production in nuclear collisions was presented at 
\cite{dress,Baer:2006rs,Haber:1994pe,dreiner1,Bartl:1987da,Espindola:2012vsa}. 

The ``Snowmass Points and Slopes'' (SPS), following \cite{Aitchison:2005cf,sps1,sps2}, are a set of benchmark points
and parameter lines in the MSSM parameter space corresponding to different scenarios in the search 
for Supersymmetry at present and future experiments. There is a very nice review about this convention give at \cite{{sps2}}. From this 
reference we take the Tab.(\ref{defparsps}).  The mass values of squarks and gluinos on these cenarios are shown at Tab.(\ref{tab:tmasses}).

\begin{table}[h]
\begin{center}
\begin{tabular}{|ccccccc|}
\hline\hline
SPS & \multicolumn{6}{c}{Point \hspace{3em}} \\ 
\hline\hline
mSUGRA: & $m_{0}$ & $m_{1/2}$ & $A_{0}$ & $\tan\beta$ & & \\
\hline
1a & 100  & 250 &   -100 & 10 & &   \\
1b & 200  & 400 &      0 & 30 & &  \\
2  & 1450 & 300 &      0 & 10 & & \\
3  &   90 & 400 &      0 & 10 & & \\
4  &  400 & 300 &      0 & 50 & &  \\
5  &  150 & 300 &  -1000 &  5 & &  \\
\hline\hline
mSUGRA-like: & $m_{0}$ & $m_{1/2}$ & $A_0$ & $\tan\beta$ & 
               $M_{1}$ & $M_{2} = M_{3}$  \\
\hline      
6  &  150 & 300 &      0 & 10 & 480 & 300  \\ 
\hline\hline         
GMSB: & $\Lambda/(10^{3})$ & $M_{\rm mes}/(10^{3})$ & $N_{\rm mes}$ & $\tan\beta$ & &  \\ 
\hline     
7  &   40 &  80 &      3 & 15 & &  \\    
8  &  100 & 200 &      1 & 15 & &  \\   
\hline\hline 
AMSB: & $m_{0}$ & $m_{\rm aux}/(10^{3})$ & & $\tan\beta$ & & \\ 
\hline         
9  &  450 &  60 &        & 10 & & \\
\hline\hline   
\end{tabular}
\caption{The parameters for the Snowmass Points and Slopes (SPS). On this table all the scenarios consider $\mbox{sign}\mu =+$ take from \cite{sps2}.}
\label{defparsps}
\end{center}
\end{table}

\begin{table}[htb]
\renewcommand{\arraystretch}{1.10}
\begin{center}
\normalsize
 \vspace{0.5cm}
\begin{tabular}{|c|c|c|}
\hline
\hline
Scenario & $m_{\tilde{g}}\, (GeV)$ & $M_{\tilde{q}}\, (GeV)$  \\
\hline
\hline
SPS1a & 595.2  & 539.9  \\
SPS1b & 916.1  & 836.2  \\
SPS2 & 784.4  & 1533.6  \\
SPS3 & 914.3  & 818.3  \\
SPS4 & 721.0  & 732.2  \\
SPS5 & 710.3  & 643.9  \\
SPS6 & 708.5  & 641.3  \\
SPS7 & 926.0  & 861.3  \\
SPS8 & 820.5  & 1081.6  \\
SPS9 & 1275.2  & 1219.2  \\
  \hline
\hline
\end{tabular}
\caption{Masses of gluinos, squarks, photinos and selectrons in the SPS scenarios \cite{sps1,sps2}.}
\label{tab:tmasses} 
\end{center}
\end{table}

\subsection{Charginos Masses}

The supersymmetric partners of the $W^{\pm}$ and the $H^{\pm}$
mix to mass eigenstates called charginos $\chi^{\pm}_{i}$ ($i=1,2$) which 
are four--component Dirac fermions. 
In order to deduce the properties of the latter we start with the 
basis \cite{charginos}\footnote{In this article, the authors studied the chargino production and decay in the energy region of LEP 200.} 
\begin{equation}
\psi^{+} = \left( - \imath \lambda^{+} \!,\, \tilde{H}^{+}_{2} \right)^{T},  \hspace{6mm}
\psi^{-} = \left( - \imath \lambda^{-} \!,\,  \tilde{H}^{-}_{1}\right)^{T}, 
\end{equation}
where 
\begin{equation}
\lambda^{\pm} = \frac{1}{\sqrt{2}}\,(\lambda^{1} \mp \imath \lambda^{2}),
\label{wino2comp}
\end{equation} 
see definition of $W$ boson given at Eq.(\ref{wdef}). The mass terms of the lagrangian of the charged 
gaugino--higgsino system can then be written as 
\begin{equation}
{\cal L}_{m} = - \frac{1}{2} \, \left( (\psi^{+})^{T}\!,\,(\psi^{-})^{T} \right)\,
\, Y^{\pm} \,
\left( \begin{array}{c} 
\psi^{+} \\ 
\psi^{-} 
\end{array} \right) + hc
\end{equation}
where
\begin{equation}
Y^{\pm}= \left( 
\begin{array}{cc} 
0 & X^{T} \\ 
X & 0 
\end{array} 
\right),
\label{y+}
\end{equation}
with 
\begin{equation}
X = \left( \begin{array}{cc} 
M & \sqrt{2}\, M_{W} \sin \beta \\
\sqrt{2}\, M_{W} \cos \beta & \mu
\end{array} \right).
\label{eq:chmassmat}      
\end{equation}
Its matrix has
\begin{eqnarray}
\det [X]&=&\det [X^{T}]=\mu M-2M^{2}_{W}\sin \beta \cos \beta =
\mu M-M^{2}_{W}\sin \left( 2 \beta \right), \nonumber \\
{\mbox Tr}[X]&=&{\mbox Tr}[X^{T}]=\mu +M.
\end{eqnarray}

The matrix $Y^{\pm}$ in Eq.(\ref{y+}) satisfy the following relation
\begin{eqnarray}
\det (Y^{\pm}- \lambda I)= \det \left[ 
\left( 
\begin{array}{cc}
- \lambda & X^{T} \\
X  &- \lambda 
\end{array} 
\right) 
\right]= \det( \lambda^{2}-X^{T} \cdot X),
\label{propmat1}
\end{eqnarray}
so we only have to calculate $X^{t} \cdot X$ to obtain the eigenvalues. 

From Eq.(\ref{eq:chmassmat}) we can write
\begin{eqnarray}
X \cdot X^{T} &=& \left( 
\begin{array}{cc} 
M^{2}+2M^{2}_{W}\sin^{2} \beta &\sqrt{2}M_{W}\left( M \cos \beta + \mu \sin \beta \right)  \\
\sqrt{2}M_{W}\left( M \cos \beta + \mu \sin \beta \right) &\mu^{2}+2M^{2}_{W}\cos^{2} \beta 
\end{array} 
\right), \nonumber \\
X^{T} \cdot X &=& \left( 
\begin{array}{cc} 
M^{2}+2M^{2}_{W}\cos^{2} \beta &\sqrt{2}M_{W}\left( M \sin \beta + \mu \cos \beta \right)  \\
\sqrt{2}M_{W}\left( M \sin \beta + \mu \cos \beta \right) &\mu^{2}+2M^{2}_{W}\sin^{2} \beta 
\end{array} 
\right).
\end{eqnarray}
Therefore $X \cdot X^{T} \neq X^{T} \cdot X$, however we can show the folowings results
\begin{eqnarray}
\det [X \cdot X^{T}]&=& \det [X^{T} \cdot X]= \left[ \mu M-M^{2}_{W}\sin \left( 2 \beta \right) \right]^{2}, 
\nonumber \\
{\mbox Tr}[X \cdot X^{T}]&=& {\mbox Tr}[X^{T} \cdot X]=\mu^{2} +M^{2}+2M^{2}_{W}.
\end{eqnarray} 
Since $X^{T} \cdot X$ is a symmetric matrix, $\lambda^2$ must be real, and 
positive because $Y^{\pm}$ is also symmetric.

The mass matrix $X$ is diagonalized by two $2\!\times\!2$ unitary matrices $U$ and $V$: 
\begin{eqnarray}
{\cal M}^{\mbox{diag}}_{C}=U^{*} X\, V^{-1}. 
\label{eq:mchdiagA}  
\end{eqnarray}
We can see this result from the following
\begin{eqnarray}
\left( \psi^{-} \right)^{T}X \psi^{+}&=&
\left( U^{-1}U \psi^{-} \right)^{T}XV^{-1}V \psi^{+}= 
\left( \psi^{-} \right)^{T}U^{t}U^{*}XV^{-1}V \psi^{+}
\nonumber \\ &=&
\left( \tilde{\chi}^{-} \right)^{t}{\cal M}^{\mbox{diag}}_{C}\tilde{\chi}^{+}.
\end{eqnarray}
In the last equality we have defined, $U$ and $V$ are the unitary matrices, the following rotations 
\begin{eqnarray}
\chi^{+}_{i} &=& V_{ij}\,\psi^{+}_{j}, \nonumber \\ 
\chi^{-}_{i} &=& U_{ij}\,\psi^{-}_{j}, \hspace{6mm} i,j = 1,2,
\label{eq:UVdef}
\end{eqnarray}
with 
\begin{equation}
{\cal M}^{\mbox{diag}}_{C}= \mbox{diag} \left( 
m_{\chi_{1}},m_{\chi_{2}} \right). 
\end{equation} 
with real nonnegative entries. From Eq.(\ref{eq:mchdiagA}) we see that
\begin{equation}
\left( {\cal M}^{\mbox{diag}}_{C}\right)^{2}=
V \left( X^{T} \cdot X \right) V^{-1}=
U^{*}\left( X \cdot X^{T}\right) \left( U^{*}\right)^{-1},
\end{equation} 
therefore $U$ and $V$ diagonalize the hermitian matrices $X \cdot X^{T}$ and $X^{T} \cdot X$. 

Moreover, assuming CP conservation, the CP violate case is presented at \cite{coreanos}, we choose a phase convention in which $U$ and $V$ are real. The eigenvalues of 
Eq.(\ref{eq:chmassmat}) is given by ($i=1,2$)  
\begin{eqnarray}
m^{2}_{\chi_{i}} &=& \frac{1}{2}\,\left[ 
|M^{2}|+| \mu^{2}|+2M_{W}^{2} \right. \nonumber \\ &\pm& \left.
\sqrt{(|M^{2}|-| \mu^{2}|)^{2}+4M^{4}_{W}\cos^{2}(2 \beta )+4M^{2}_{W}\left[ |M^{2}|+| \mu^{2}|
+2\Re(M \mu ) \sin (2 \beta) \right]}
\right].
\end{eqnarray} 
The mass eigenstates in Dirac notation are given by
\begin{equation}
\tilde{\chi}^{+}_{i}= \left( \begin{array}{c} 
\chi^{+}_{i} \\
\overline{\chi^{-}_{i}}
\end{array} \right), \hspace{6mm} i,j = 1,2.  
\end{equation}
We take $\tilde{\chi}^{+}_{1}$ to be the lighter chargino per definition. The charginos, 
like all the charged fermions in the SM, are Dirac fermions \cite{Dirac}.

\begin{figure}[ht]
\begin{center}
\vglue -0.009cm
\mbox{\epsfig{file=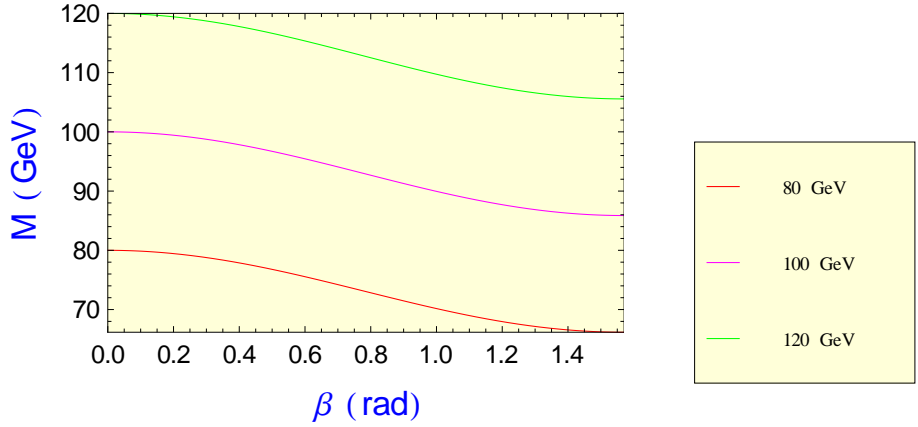,width=0.7\textwidth,angle=0}}       
\end{center}
\caption{The masses of lighest chargino in terms of the $\beta$ 
parameter considerating $M=1000$ GeV and several values of $\mu$ 
parameter.}
\label{fig8}
\end{figure}

\begin{figure}[ht]
\begin{center}
\vglue -0.009cm
\mbox{\epsfig{file=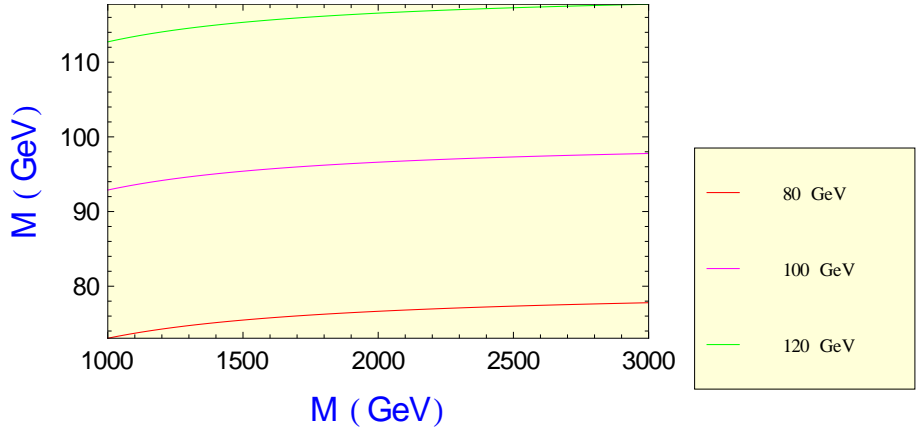,width=0.7\textwidth,angle=0}}       
\end{center}
\caption{The masses of lighest chargino in terms of the $M$ 
parameter considerating $\beta =( \pi /4)$ rad and several values of 
$\mu$ parameter.}
\label{fig9}
\end{figure}

\begin{figure}[ht]
\begin{center}
\vglue -0.009cm
\mbox{\epsfig{file=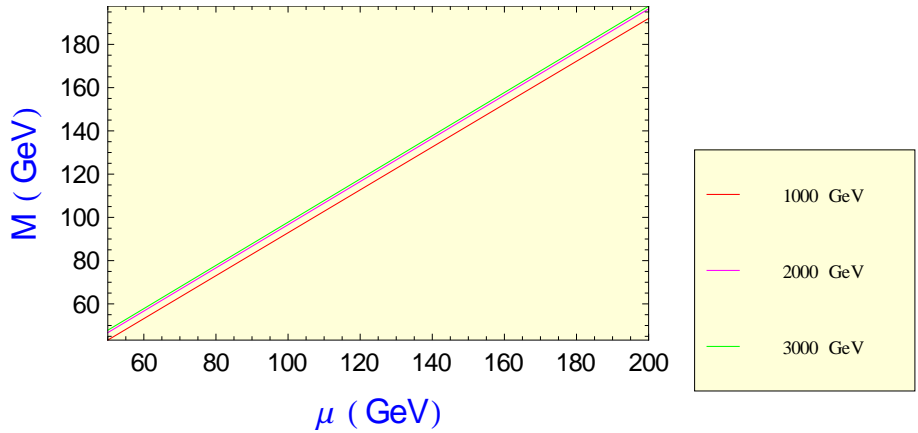,width=0.7\textwidth,angle=0}}       
\end{center}
\caption{The masses of lighest chargino in terms of the $\mu$ 
parameter.}
\label{fig10}
\end{figure}

If we consider $\mu =100$ GeV and $M=1000$ GeV we get for several values 
of $\beta$ parameter masses to the lighest charginos of 
${\cal O}(M_{Z})$ while the heaviest charginos has masses around 
TeV region, as we can see at Tab.(\ref{tab:char-neutralinos}). On Fig.(\ref{fig8}) we have fixed $M$ and we have considered several 
values to the $\mu$ parameter. On Fig.(\ref{fig9}) we have fixed $\beta$ 
and we have considered several values to the $\mu$ parameter. On 
Fig.(\ref{fig10}) we have fixed $\beta$ and we have considered several 
values to the $M$ parameter. We see from all this figure that in media 
the average mass to the lighest charginos is around 100 GeV. Tḧe couplings of charginos are presented at \cite{dress,Baer:2006rs,Aitchison:2005cf,charginos}. 
The productions of charginos was studied at \cite{dress,Baer:2006rs,Aitchison:2005cf,Haber:1994pe,dreiner1,Dawson,charginos}.

\subsection{Neutralinos Masses}
\label{subsec:neutralinos} 

The neutral gauginos, see Sec.(\ref{sec:mixphotino-zino}),  and neutral higgsinos also mix and these new state is known as neutralinos 
\cite{dress,Baer:2006rs,Aitchison:2005cf,mssm,neutralinos,indiano}. Their mass eigenstates are the neutralinos. In this review we 
choose the basis \cite{mssm,neutralinos,indiano} 
\begin{equation}
\psi^{0}_{MSSM} = \left( \begin{array}{cccc}
\imath \lambda_{\gamma} & \imath \lambda_{Z} & \tilde{H}^{0}_{1} & \tilde{H}^{0}_{2} \end{array} \right)^{\!\rm T}.
  \hspace{6mm}
\label{neutralinomssm} 
\end{equation}
The mass terms of the neutral gaugino--higgsino system 
can then be written as 
\begin{equation}
{\cal L}_{m} = - \frac{1}{2}\, 
(\psi^{0})^{\rm T}_{MSSM} \, Y^{neutralino}_{MSSMRPC} \, \psi^{0}_{MSSM}+ \,\ hc
\end{equation}
with 
\begin{equation}
Y^{neutralino}_{MSSMRPC}= \left( \begin{array}{cccc}
M_{\tilde{\gamma}} & \frac{1}{4}\left( M_{\tilde{Z}} -  M_{\tilde{\gamma}}\right) \tan({2\theta_{W}}) & 0 & 0 \\
\frac{1}{4}\left( M_{\tilde{Z}} -  M_{\tilde{\gamma}}\right) \tan({2\theta_{W}}) & M_{\tilde{Z}} & 
M_{Z}\sin \beta &- M_{Z}\cos \beta \\
0 & M_{Z}\sin \beta & 0 &- \mu \\
0 &- M_{Z}\cos \beta &- \mu & 0 
\end{array} \right) \,\ . 
\label{neutralinonaodiagonal}  
\end{equation}
Its matrix has
\begin{eqnarray}
\det [Y]&=&\mu^{2}\left\{ \left[ \left( \frac{M_{\tilde{Z}} -  M_{\tilde{\gamma}}}{4}\right) \tan (2 \theta_{W}) \right]^{2} - M_{\tilde{\gamma}}M_{\tilde{Z}} \right\} +M^{2}_{Z}M_{\tilde{\gamma}}\mu \sin (2 \beta ) 
\nonumber \\
&=&-MM^{\prime}\mu^{2}+M^{2}_{Z}\mu 
\left( M \sin^{2} \theta_{W}-M^{\prime} \cos^{2} \theta_{W} \right) 
\sin \left( 2 \beta \right) +
\frac{M^{4}_{Z}}{4}\sin \left( 2 \theta_{W} \right) 
\sin \left( 2 \beta \right) , \nonumber \\
{\mbox Tr}[Y]&=&M_{\tilde{\gamma}}+M_{\tilde{Z}}=M+M^{\prime}.
\end{eqnarray}
We wrote the matrix in the basis os photino and zino. However, it is more used it, on the basis of 
$\lambda_{3}$ and $\lambda$, on this base the matrix is given at \cite{dress,Baer:2006rs}.
 
The mass matrix $Y$ is diagonalized by a $4\!\times\!4$ unitary\footnote{To diagonalize the mass matriz of neutral fermions, we use the 
Takagi diagonalization method \cite{dreiner1,takagi,alinear}} 
matrix $N$,
\begin{equation}
N^{*}\,Y\,N^{-1} = {\cal M}_{N} 
\label{nmatrix}
\end{equation}
with ${\cal M}_{N}$ the diagonal mass matrix. The eigenvalues 
${\cal M}_{N}$ and the matrix $N$ in general are obtained numerically. 
However if all parameter in the matrix $Y$ are real, an analytical 
calculation to eigenstates and eigenvectors are possible \cite{indiano}.

The mass eigenstates in two--component notation then are
\begin{equation}
\chi^{0}_{i} = N_{ij}\,\psi^{0}_{j}, \hspace{6mm} 
i,j=1\ldots 4,
\end{equation}
and we can find them in the following way
\begin{equation}
\left(
\begin{array}{c}
\chi^{0}_{1} \\ 
\chi^{0}_{2} \\ 
\chi^{0}_{3} \\
\chi^{0}_{4}
\end{array}
\right)
= 
\left(
\begin{array}{cccc}
\eta_{1} & 0 & 0 & 0 \\  
0 & \eta_{2} & 0 & 0 \\
0 & 0 & \eta_{3} & 0 \\
0 & 0 & 0 & \eta_{4} 
\end{array}
\right) 
\left( N  \right)
\left(
\begin{array}{c}
\imath \lambda_{\gamma} \\ 
\imath \lambda_{Z} \\ 
\tilde{H}_{1}^{0} \\
\tilde{H}_{2}^{0}
\end{array}
\right)
\end{equation}
The four-by-four matrix $N$ diagonalizes the symmetric mass matrix ${\cal M}_{N}$ of the neutral Weyl spinors, 
see Eq.(\ref{nmatrix}), where the eigenvalues are arranged such that
$|m_{\chi^{0}_{1}}| < |m_{\chi^{0}_{2}}| <
|m_{\chi^{0}_{3}}| < |m_{\chi^{0}_{4}}|$. The parameter $\eta_{i}$ is introduced in order to change the 
phase of the particle whose eigenvalue becomes negative, it means it is defined as follow
\begin{equation}
\eta_{i} = \left\{
\begin{array}{c}
1,  m_{\chi^{0}_{i}}>0, \\
\imath , m_{\chi^{0}_{i}}<0,
\end{array}
\right. , \,\ i=1, \ldots 4,
\end{equation}
and 
\begin{equation}
m_{\chi^{0}_{i}} = \eta^{2}_{i} m_{\chi^{0}_{i}}.
\end{equation}
The four--component notation to the neutralinos is given as
\begin{equation}
\tilde{\chi}^{0}_{i}= 
\left( \begin{array}{c} 
\chi^{0}_{i} \\
\overline{\chi^{0}_{i}}
\end{array} \right), \hspace{6mm} i=1\ldots 4.
\end{equation}

\begin{figure}[ht]
\begin{center}
\vglue -0.009cm
\mbox{\epsfig{file=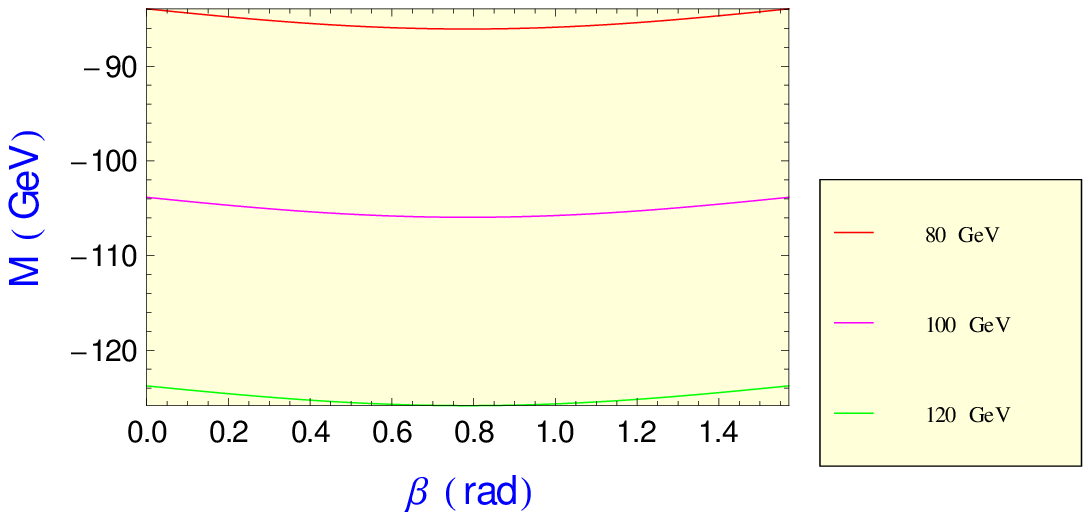,width=0.7\textwidth,angle=0}}       
\end{center}
\caption{The masses of lighest neutralino in terms of the $\beta$ 
parameter considerating $M=M^{\prime}=1000$ GeV and several values 
to $\mu$ parameter.}
\label{fig11}
\end{figure}

On Fig.(\ref{fig11}) we have fixed $M$ and we have considered several 
values to the $\mu$ parameter. We see that the mass of the lighest neutralino is around 100 GeV. If we consider 
$\mu =100$ GeV and $M=M^{\prime}=1000$ GeV we get for several values 
of $\beta$ parameter masses to the lighest charginos of 
${\cal O}(M_{Z})$ while the heaviest charginos has masses around 
TeV region, as we can see at Tab.(\ref{tab:char-neutralinos}).

\begin{table}[h]
\begin{center}
\begin{tabular}{|c|c|c|c|c|c|c|}
\hline 
$\tan \beta $ & $ m_{\tilde{\chi}^{\pm}_{1}} (\mbox{GeV}) $ 
& $m_{\tilde{\chi}^{\pm}_{2}} (\mbox{GeV}) $ & $ m_{\tilde{\chi}^{0}_{1}} (\mbox{GeV}) $ 
& $m_{\tilde{\chi}^{0}_{2}} (\mbox{GeV}) $ & 
$ m_{\tilde{\chi}^{0}_{3}} (\mbox{GeV}) $ 
& $m_{\tilde{\chi}^{0}_{4}} (\mbox{GeV}) $ \\
\hline
0.00 & 100 & 1000 & 95.51 & 103.84 & 1000.00 & 1008.33 \\
\hline
0.10 & 99.86 & 1000.14 & 96.03 & 104.27 & 1000.06 & 1008.18 \\ 
\hline
0.20 & 99.43 & 1000.57 & 96.53 & 104.68 & 1000.22 & 1007.93  \\
\hline
0.55 & 96.71 & 1003.29 & 97.68 & 105.63 & 1001.20 & 1006.74  \\
\hline
1.03 & 92.67 & 1007.33 & 98.08 & 105.95 & 1001.92 & 1005.95 \\
\hline
1.56 & 89.95 & 1010.05 & 97.85 & 105.77 & 1001.47 & 1006.45 \\
\hline
2.57 & 87.70 & 1012.30 & 97.26 & 105.28 & 1000.71 & 1007.30 \\
\hline
3.60 & 86.87 & 1013.13 & 96.86 & 104.95 & 1000.40 & 1007.69 \\
\hline
5.80 & 86.27 & 1013.73 & 96.39 & 104.56 & 1000.16 & 1008.01 \\
\hline
\end{tabular}
\end{center}
\caption{\small Masses of charginos and neutralinos to several values of $\tan \beta$.}
\label{tab:char-neutralinos}
\end{table}

Tḧe couplings of 
neutralinos are presented at \cite{dress,Baer:2006rs,Aitchison:2005cf,mssm,neutralinos} and the productions of neutralinos as well 
its decays channel was studied at \cite{dress,Baer:2006rs,Aitchison:2005cf,Haber:1994pe,dreiner1,Dawson,Espindola:2011nb,neutralinos}. 
We have presented only the analyses when we respect CP invariance. The 
calaculation of mass spectrum in the case of CP violation can be found 
at \cite{Baer:2006rs}. We want to finish this section saying that the 
chargino and neutralino mixing pattern are complex. This subject is 
discussed at \cite{dress,Baer:2006rs}.

\section{Some Feynman Rules to fermions in the MSSM.}
\label{sec:feynmanrulesinMSSM}

We can take these interactions terms from ${\cal L}_{Quarks}$ and ${\cal L}_{Leptons}$. The interaction 
between fermion-fermion with the gauge bosons, see Eqs.(\ref{kahlerpotential}), came from
\begin{eqnarray}
{\cal L}_{ffV}&=& \frac{g}{2}(\bar{Q} \bar{\sigma}^{m} \sigma^{i}Q+ 
\bar{L} \bar{\sigma}^{m} \sigma^{i}L)V^{i}_{m}+
\frac{g^{\prime}}{2}\left( Y_{Q} \bar{Q} \bar{\sigma}^{m}Q+
Y_{L} \bar{L} \bar{\sigma}^{m}L \right. \nonumber \\
&+&
\left. 
Y_{u^{c}} \overline{u^{c}} \bar{\sigma}^{m}u^{c}+
\frac{g^{\prime}}{2}Y_{d^{c}} \overline{d^{c}} \bar{\sigma}^{m}d^{c}+
\frac{g^{\prime}}{2}Y_{l^{c}} \overline{l^{c}} \bar{\sigma}^{m}l^{c} 
\right) V^{\prime}_{m} \nonumber \\
&=& {\cal L}^{charged}_{ffV}+ {\cal L}^{neutral}_{ffV}, 
\end{eqnarray}
while the interaction sfermion-sfermion with the gauge boson is get from
\begin{eqnarray}
{\cal L}_{\tilde{l}\tilde{l}V}&=& \frac{\imath g}{2} \left(
\bar{\tilde{Q}} \sigma^{i} \partial^{m}\tilde{Q}- \tilde{Q}\sigma^{i} \partial^{m}\bar{\tilde{Q}}+
\bar{\tilde{L}} \sigma^{i} \partial^{m}\tilde{L}- \tilde{L}\sigma^{i} \partial^{m}\bar{\tilde{L}}
\right) V^{i}_{m} \nonumber \\
&+&\frac{ig^{\prime}}{2}\left[
Y_{Q}\left(
\bar{\tilde{Q}} \partial^{m}\tilde{Q}- \tilde{Q} \partial^{m}\bar{\tilde{Q}}
\right) +
Y_{L}\left(
\bar{\tilde{L}} \partial^{m}\tilde{L}- \tilde{L} \partial^{m}\bar{\tilde{L}} \right) +
Y_{u^{c}}\left(
\overline{\tilde{u}^{c}} \partial^{m}\tilde{u}^{c}- \tilde{u}^{c} \partial^{m}\overline{\tilde{u}^{c}}
\right) 
\right. \nonumber \\ &+& \left.
Y_{d^{c}}\left(
\overline{\tilde{d}^{c}} \partial^{m}\tilde{d}^{c}- \tilde{d}^{c} \partial^{m}\overline{\tilde{d}^{c}}
\right) +
Y_{l^{c}}\left(
\overline{\tilde{l}^{c}} \partial^{m}\tilde{l}^{c}- \tilde{l}^{c} \partial^{m}\overline{\tilde{l}^{c}}
\right) 
\right]V^{\prime}_{m} \nonumber \\
&=& {\cal L}^{charged}_{\tilde{f}\tilde{f}V}+ {\cal L}^{neutral}_{\tilde{f}\tilde{f}V},
\end{eqnarray}
after doing some simple mathematical manipulation we arrive in the following Feynman rules see 
Eqs.(\ref{wdef},\ref{physicalchargedsleptons},\ref{physicalsquarks}), 
for charged currents \cite{dress,Baer:2006rs,Aitchison:2005cf}
\begin{eqnarray}
{\cal L}^{charged}_{ffV}&=&\frac{-g}{\sqrt{2}}\left[
\left( \bar{u}\gamma^{m}d_{L}+ \bar{\nu}\gamma^{m}e_{L}\right) W^{+}_{m}+
\left( \bar{d}\gamma^{m}u_{L}+ \bar{e}\gamma^{m}\nu_{L}\right) W^{-}_{m}\right], \nonumber \\
{\cal L}^{charged}_{\tilde{f}\tilde{f}V}&=&\frac{- \imath g}{\sqrt{2}}\left[
\left(
\tilde{u}^{\star}_{L}\stackrel{\leftrightarrow}{\partial}^{m} \tilde{d}_{L}+
\tilde{\nu}^{\star}_{L}\stackrel{\leftrightarrow}{\partial}^{m} \tilde{e}_{L} \right) W^{+}_{m}+
\left(
\tilde{d}^{\star}_{L}\stackrel{\leftrightarrow}{\partial}^{m} \tilde{u}_{L}+
\tilde{e}^{\star}_{L}\stackrel{\leftrightarrow}{\partial}^{m} \tilde{\nu}_{L} 
\right) W^{-}_{m} \right], \nonumber \\
\end{eqnarray}
the chiral Dirac matrix $\gamma_{5}$ is defined as \cite{wb}
\begin{equation}
\gamma^{5} \equiv  \left(
\begin{array}{cc}
1& 0 \\ 
0& -1
\end{array}
\right) \,\ ,
\label{gamma5}
\end{equation}
so the right-handed projector ($R$) and left-handed projector ($L$) are given as
\begin{eqnarray}
R &=& \frac{1}{2} \left( 1+ \gamma^{5} \right) = 
\left( 
\begin{array}{cc}   
1 & 0 \\ 
0 & 0 
\end{array} 
\right) \,\ . \nonumber \\
L &=& \frac{1}{2} \left( 1- \gamma^{5} \right) = 
\left( 
\begin{array}{cc}   
0 & 0 \\ 
0 & 1 
\end{array} 
\right) \,\ ,
\label{projetores}
\end{eqnarray}
and as usual, we have
\begin{eqnarray}
f_{L}\equiv Lf, \,\
f_{R}\equiv Rf.
\end{eqnarray}
where we have defined
\begin{eqnarray}
\bar{\phi}\stackrel{\leftrightarrow}{\partial}\Phi \equiv 
\bar{\phi}\left( \partial\Phi \right)- \left( \partial\bar{\phi}\right) \Phi ,
\label{ucraniano}
\end{eqnarray}
while the Feynman Rules for neutral currents see Eqs.(\ref{boson5},\ref{physicalchargedsleptons},\ref{physicalsquarks}) \cite{dress,Baer:2006rs,Aitchison:2005cf}
\begin{eqnarray}
{\cal L}^{neutral}_{ffV}&=&\frac{-g}{\cos \theta_{W}}\sum_{f}
\bar{f}\left[ 
\left( T_{3f}- \sin^{2}\theta_{W}Q_{f}\right)L- \sin^{2}\theta_{W}Q_{f}R \right] fZ_{m}- 
e \sum_{f}
\bar{f}\gamma^{m}fA_{m}, \nonumber \\
{\cal L}^{neutral}_{\tilde{f}\tilde{f}V}&=&\frac{-\imath g}{\cos \theta_{W}}\sum_{f}
\left[ 
\tilde{f}^{\star}_{L}\stackrel{\leftrightarrow}{\partial}^{m}\left( T_{3f}- \sin^{2}\theta_{W}Q_{f}\right) \tilde{f}_{L}- 
\tilde{f}^{\star}_{R}\stackrel{\leftrightarrow}{\partial}^{m}\left( \sin^{2}\theta_{W}Q_{f}\right) \tilde{f}_{R}
\right] Z_{m} \nonumber \\ &-& 
\imath e \sum_{f}Q_{f} \left( 
\tilde{f}^{\star}_{L}\stackrel{\leftrightarrow}{\partial}^{m} \tilde{f}_{L}+
\tilde{f}^{\star}_{R}\stackrel{\leftrightarrow}{\partial}^{m} \tilde{f}_{R} \right)A_{m},
\end{eqnarray}
the summation $f$ is taken over the fermion species. 

The Feynman Rules to the coupling Fermion-Fermion-Higgs, for the usual scalars see 
Eqs.(\ref{pseudoatmssm},\ref{chargedscalartmssm},\ref{realatmssm}) \cite{dress,Baer:2006rs,Aitchison:2005cf}
\begin{eqnarray}
{\cal L}_{ffH}&=&- f^{u} \epsilon_{ij} \left( 
H^{i}_{2}Q^{j}u^{c}+h.c. \right)
- f^{d} \epsilon_{ij} \left( 
H^{i}_{1}Q^{j}d^{c}+h.c. \right)
- f^{l} \epsilon_{ij} \left( 
H^{i}_{1}L^{j}l^{c}+h.c. \right)  ,
\label{eeH} 
\end{eqnarray}
using the physical fields we get
\begin{eqnarray}
{\cal L}_{ffH^{0}_{i}}&=&\frac{-gm_{u}}{2M_{W} \sin \beta}\left( \cos \alpha \bar{u}uh^{0}- \sin \alpha \bar{u}uH^{0} - 
\imath \cos \beta \bar{u}\gamma^{5}uA^{0}\right) \nonumber \\ &-& 
\frac{gm_{d}}{2M_{W} \cos \beta}\left( \sin \alpha \bar{d}dh^{0}+ \cos \alpha \bar{d}dH^{0} - 
\imath \sin \alpha \bar{d}\gamma^{5}dA^{0}\right) \nonumber \\
&-&
\frac{gm_{e}}{2M_{W}\cos \beta}\left( \sin \alpha \bar{e}eh^{0}+ \cos \alpha \bar{e}eH^{0} - 
\imath \sin \beta \bar{e}\gamma^{5}eA^{0}\right) \nonumber \\ &+&
\frac{g}{2 \sqrt{2}M_{W}} \left[ 
\left( m_{u} \cot \beta +m_{d} \tan \beta \right)\bar{u}d+
\left( m_{d} \tan \beta -m_{u} \cot \beta \right)\bar{u}\gamma^{5}d 
\right. \nonumber \\ &+& \left.
m_{e}\tan \beta \bar{\nu}_{e}\left( 1+ \gamma^{5}\right)e \right]H^{+},
\end{eqnarray}
where to the usual scalar see Tab.(\ref{tab:scalars}). 

\section{The Next to Minimal Supersymmetric Standard Model (NMSSM).}

It is possible to extend the Higgs sector in such way that the 
$SU(3)_{C}\otimes SU(2)_{L}\otimes U(1)_{Y}$ gauge symmetry is spontaneously broken at tree level, even in the supersymmetric limit. The simplest extension 
is to include a complex scalar field wchich is an $SU(3)_{C}\otimes SU(2)_{L}\otimes U(1)_{Y}$ gauge singlet, this model is known as the 
Next to Minimal Supersymmetric Standard Model (NMSSM) \cite{Hooper:2009gm,Wang:2012ry,dress,Drees:1988fc,Ellis:1988er} . 

The NMSSM is characterized by the following new singlet superfieldfield 
\begin{eqnarray}
\hat{S} \sim( {\bf 1},{\bf 1}, 0),
\end{eqnarray}
again the the numbers in parenthesis refers to the $(SU(3)_{C}, SU(2)_{L},
U(1)_{Y}$) quantum numbers. This new superfield is introduced in the
following chiral superfield \cite{wb,dress,Drees:1988fc} 
\begin{eqnarray}
\hat{S}(y, \theta )&=&S(y)+ \sqrt{2}\theta \tilde{S}(y)+ \theta \theta
F_{S}(y),
\label{singletatnmssm}
\end{eqnarray}
and $S$ is the scalar field and its vacuum expectation value (vev) is given by 
\begin{eqnarray}
\langle S \rangle \equiv \frac{x}{\sqrt{2}}.  
\label{vevsinglet}
\end{eqnarray}
The fermionic field $\tilde{S}$, defined at Eq.(\ref{singletatnmssm}), is known as singlino. Due the fact we
introduce this new superfield, we get as consequence that the mass bounds for the
Higgs bosons and neutralinos are weakened as we want to show next. The goal of this review is to
present both sector in this kind of model in details, the main motivation to
study this kind of model can be found at \cite{Hooper:2009gm,Wang:2012ry}.

The supersymetric Lagrangian of the NMSSM is given by 
\begin{equation}
{\cal L}^{NMSSM}_{SUSY} = {\cal L}^{chiral}_{SUSY} + {\cal L}^{Gauge}_{SUSY} .  
\label{SUSY-Lagrangian1}
\end{equation}
The Lagrangian defined in the equation (\ref{SUSY-Lagrangian1}) contains
the lagrangian ${\cal L}_{quarks}$ and ${\cal L}_{leptons}$, and these terms are 
the same as presented at MSSM see Eqs.(\ref{allsusyterms}). 

In this model, We  need to modify only ${\cal L}_{Higgs}$, and we get 
\begin{eqnarray}
{\cal L}_{Higgs}&=&  \int d^{4}\theta\;\left[\,
K \left( \hat{ \bar{H}}_{1}e^{2g\hat{V}+g^{\prime}\left( - \frac{1}{2}\right) \hat{V}^{\prime}}, \hat{H}_{1} \right) +
K \left( \hat{ \bar{H}}_{2}e^{2g\hat{V}+g^{\prime}\left( \frac{1}{2}\right) \hat{V}^{\prime}}, \hat{H}_{2} \right) +
K \left( \hat{ \bar{S}}, \hat{S} \right) \right].  
\label{lsupgsemssm}
\end{eqnarray}
The terms $\left( {\cal D}_{m}H_{1} \right)$ and 
$\left( {\cal D}_{m}H_{2} \right)$ give the mass to the gauge bosons $W^{\pm}$ and $Z^{0}$
in the same way as happen in the MSSM, see Eq.(\ref{originmassgaugebosons}).

The most general superpotential of NMSSM is defined as 
\begin{eqnarray}
W_{NMSSM}&=& W^{MSSM}_{3RC} + \lambda 
\left( \hat{H}_{1}\hat{H}_{2} \right) \hat{S}- 
\frac{\kappa}{3}\left( \hat{S} \right)^{3}.  
\label{suppotNMSSM}
\end{eqnarray}
where $W^{MSSM}_{3RC}$ is given by Eq.(\ref{suppotMSSM}). 

In this case, the parameter $\mu$ of the MSSM is generated as
\begin{equation}
\mu \equiv \lambda \frac{x}{\sqrt{2}}.
\end{equation}
and $x$ is expected to be ${\cal O}(v_{1},v_{2})$ in most theories. Since 
$\lambda$ is also in the perturbative domain, we have a natural explanation for keeping $\mu \ll M_{Pl}$, given that 
$v_{1},v_{2}\ll M_{Pl}$.

We want to stress that this superpotential has no bilinear terms, 
remember they lead to naturalness problems and, furthermore, do not 
appear in a large class of superstring models 
\cite{Gunion:1989we,Drees:1988fc,Ellis:1988er}. The sign of $\kappa$ 
coupling has been chosen for later convenience.

We can add the following soft supersymmetry breaking
terms to the NMSSM 
\begin{eqnarray}
{\cal L}^{NMSSM}_{Soft} &=& {\cal L}^{MSSM}_{SMT} + 
{\cal L}^{MSSM}_{GMT}+ {\cal L}^{NMSSM}_{INT} \,\ ,
\label{The Soft SUSY-BreakingNMSSM}
\end{eqnarray}
where ${\cal L}^{MSSM}_{SMT}$ and ${\cal L}^{MSSM}_{GMT}$ are introduced at Eqs.(\ref{burro},\ref{The Soft SUSY-Breaking Term prop 3}), 
respectively. There is an interaction term ${\cal L}^{NMSSM}_{INT}$ of the form \cite{Gunion:1989we,Drees:1988fc,Ellis:1988er}
\begin{eqnarray}
{\cal L}^{NMSSM}_{INT} &=&{\cal L}^{MSSM}_{INT}+ \left(
\lambda A_{\lambda}H_{1}H_{2}S+
\frac{\kappa A_{\kappa}}{3}S^{3}+h.c.
\right) \,\ .
\label{softtermintNMSSM}
\end{eqnarray}
The term ${\cal L}^{MSSM}_{INT}$ is given at Eq.(\ref{burroint}), in this model the first term at this equation is absent. 

The soft Supersymmetry breaking terms are given by 
\begin{eqnarray}
V_{soft}^{NMSSM}&=& \tilde{m}_{1}^{2} | H_{1} |^{2} + \tilde{m}_{2}^{2} |
H_{2} |^{2} + \tilde{m}_{s}^{2} | S |^{2}- \left[ \lambda A_{\lambda} \left(
\epsilon_{\alpha \beta} H_{1}^{\alpha} H_{2}^{\beta} \right) S + 
\frac{\left( \kappa A_{\kappa}\right)}{3} \left( S \right)^{3} + h.c. \right] .
\label{soft}
\end{eqnarray}

In the presence of soft supersymmetry breaking, one would expect
\begin{equation}
x= {\cal O}(|m_{i}|)={\cal O}(M_{W}),
\end{equation}
and hence $\mu = {\cal O}(M_{W}) \ll M_{Pl}$.

\section{Scalar Potential at NMSSM}

We assume that squarks and sleptons fields have zero vaccum expectation
value (VEVEs). After the scalar fields $H_{1}$,$H_{2}$ and $S$ develop their
VEVs $v_{1}$, $v_{2}$ and $x$, respectively, they can be expanded in the
usual way as 
\begin{eqnarray}
H_{1}= \left ( 
\begin{array}{c}
\frac{1}{\sqrt{2}} \left( v_{1} + \phi_{1} + \imath \varphi_{1} \right) \\ 
H_{1}^{-}
\end{array}
\right), \,\ H_{2}= \left ( 
\begin{array}{c}
H_{2}^{+} \\ 
\frac{1}{\sqrt{2}} \left( v_{2} + \phi_{2} + \imath \varphi_{2} \right)
\end{array}
\right), \,\ S= \frac{1}{\sqrt{2}} \left( x + \sigma + \imath \xi \right) \, .
\end{eqnarray}

The scalar potential, as usual in supersymmetric models, is written as 
\begin{eqnarray}
V^{NMSSM}&=& M_{H_{1}}^{2} | H_{1} |^{2} + M_{H_{2}}^{2} | H_{2} |^{2}+
M_{S}^{2} | S |^{2}- \left[ \lambda A_{\lambda} \left( \epsilon_{\alpha
\beta} H_{1}^{\alpha} H_{2}^{\beta} \right) S + \frac{\left( \kappa
A_{\kappa}\right)}{3} \left( S \right)^{3} + h.c. \right]  
\nonumber \\
&+& \left( |H_{1}|^{2}+ |H_{2}|^{2}\right) \left| \lambda S\right|^{2}+
\left| \lambda \left( H_{1}H_{2}\right) S+ \kappa S^{2} \right|^{2} 
\nonumber \\
&+& \frac{g^{2}}{8} \left( \bar{H}_{1}\sigma^{i}H_{1}+
\bar{H}_{2}\sigma^{i}H_{2} \right)^{2}+ \frac{g^{\prime 2}}{8} 
\left( \bar{H}_{1}H_{1}- \bar{H}_{2}H_{2} \right)^{2}.
\end{eqnarray}
$\lambda$, $\kappa$, $A_{\lambda}$ and $A_{\kappa}$ can be complex number.

There are various limiting cases in wchich the scalar Higgs masses and 
mixing angles can be evaluated perturbatively 
\begin{enumerate}
\item[1-)] $x \gg v_{1},v_{2}$ with $\lambda$ and $\kappa$ fixed;
\item[2-)] $x \gg v_{1},v_{2}$ with $\lambda x$ and $\kappa x$ fixed;
\end{enumerate}
in the last limit the MSSM with two Higgs doublets and no Higgs singlets 
is obtained \cite{Ellis:1988er}.

\subsection{Constraints}

We can use the minimization condition to re-express the soft supersymmetry
breaking terms $M_{H_{1}}^{2}$, $M_{H_{2}}^{2}$ and $M_{S}^{2}$ in terms of
the vevs and of the remaind parameters $\lambda$, $\kappa$, $A_{\lambda}$
and $A_{\kappa}$ 
\begin{eqnarray}
M_{H_{1}}^{2}&=& \lambda A_{\lambda} \frac{v_{2}x}{v_{1}}- \lambda^{2}\left(
x^{2}+v_{2}^{2} \right) + \lambda \kappa \frac{v_{2}x^{2}}{v_{1}}+ \left( 
\frac{g^{2}+g^{\prime 2}}{4} \right) \left( v_{2}^{2}-v_{1}^{2} \right),  \\
M_{H_{2}}^{2}&=& \lambda A_{\lambda} \frac{v_{1}x}{v_{2}}- \lambda^{2}\left(
x^{2}+v_{1}^{2} \right) + \lambda \kappa \frac{v_{1}x^{2}}{v_{2}}+ \left( 
\frac{g^{2}+g^{\prime 2}}{4} \right) \left( v_{1}^{2}-v_{2}^{2} \right),  \\
M_{S}^{2}&=& \lambda A_{\lambda} \frac{v_{1}v_{2}}{x}+ \kappa A_{\kappa}x-
\lambda^{2}\left( v_{1}^{2}+v_{2}^{2} \right) - 2 \kappa^{2}x^{2}+2 \lambda
\kappa v_{1}v_{2}.
\end{eqnarray}
Therefore, the mass terms for the Higgs fields, can be expressed in terms of
the six parameters $\lambda$, $\kappa$, $A_{\lambda}$, $A_{\kappa}$, $x$ and 
$\tan \beta$.

\subsection{Charged Higgs}

In the Charged Higgs sector we get an unphysical Goldstone boson and the
physical charged Higgs field defined at Eq.(\ref{chargedscalartmssm}) and
therefore the mass eigenvector are the same in those models MSSM and NMSSM.

The mass squared matrix in this sector is found to be 
\begin{equation}
\left( {\cal M}^{NMSSM}_{H^{\pm}} \right)^{2}= \left[ 
\frac{v_{1}v_{2}}{2}(g-2 \lambda^{2}) + \frac{\lambda x}{g}( \kappa x+A_{\lambda}) \right] 
{\cal M}^{2}_{H^{\pm}} ,
\end{equation}
and we have defined 
\begin{equation}
{\cal M}^{2}_{H^{\pm}}= \left( 
\begin{array}{cc}
\tan \beta & 1 \\ 
1 & \cot \beta
\end{array}
\right),
\end{equation}
where $\tan \beta$ is defined at Eq.(\ref{defbetapar}), it is simple to show 
\begin{equation}
\det \left( {\cal M}^{2}_{H^{\pm}} \right)= \tan \beta \cot \beta
-1=1-1=0,
\end{equation}
therefore we have one charged Goldstone boson on this sector and 
\begin{eqnarray}
{\mbox Tr} \left[ {\cal M}^{NMSSM}_{H^{\pm}} \right]&=&\tan \beta + \cot \beta = 
\frac{v_{2}}{v_{1}}+ \frac{v_{1}}{v_{2}}= \frac{v^{2}_{1}+v^{2}_{2}}{v_{1}v_{2}}= 
\frac{2M^{2}_{W}}{gv_{1}v_{2}},
\end{eqnarray}
and at lst step we used Eq.(\ref{wmass}). As conclusion, we have one massive
state $H^{\pm}$ and its mass is given by 
\begin{equation}
\left( M^{NMSSM}_{H^{\pm}}\right)^{2}= \left[ \frac{v_{1}v_{2}}{2}(g-2 \lambda^{2}) 
+ \frac{\lambda x}{g}( \kappa x+A_{\lambda}) \right] \cdot 
\frac{2M^{2}_{W}}{gv_{1}v_{2}},
\label{chargedscalaratnmssm}
\end{equation}
the last term can be rewriten as 
\begin{equation}
\frac{2 \lambda x M^{2}_{W}}{(gv_{1})(gv_{2})}( \kappa x+A_{\lambda}),
\end{equation}
using Eq.(\ref{wmass}) we can write 
\begin{eqnarray}
gv_{1}&=& \sqrt{2}M_{W}\cos \beta,   \\
gv_{2}&=& \sqrt{2}M_{W}\sin \beta,   \\
\frac{M^{2}_{W}}{(gv_{1})(gv_{2})}&=& \frac{1}{\sin \left( 2 \beta \right)},
\label{v1v2MZmssm}
\end{eqnarray}
and the NMSSM, the squared mass of the charged boson is given by 
\begin{equation}
\left( M^{NMSSM}_{H^{\pm}}\right)^{2}=M^{2}_{W} \left( 1- \frac{2 \lambda^{2}}{g}\right) 
+ \frac{2 \lambda x}{\sin \left( 2 \beta \right)} \left(
A_{\lambda}+ \kappa x \right) ,  
\label{masschargedHiggsNMSSM}
\end{equation}
$M^{2}_{H^{\pm}}$ may be less or greater than $M^{2}_{W}$, depending upon
the relative size of the last two terms at Eq.(\ref{masschargedHiggsNMSSM}).
This result can be seen in our Figs.(\ref{fig1},\ref{fig2}), where we take 
$\lambda =0.87$, $\kappa =0.63$ as used at \cite{Ellis:1988er,Gunion:1989we}. In pratice, 
however, the parameter space given by $\beta$, $\lambda$ and $\kappa$ parameters for which 
the charged scalar is smaller than $M_{W}$ are not favored by the renormalization group analysis.

We also can
get heavier then $W$-boson see our Figs.(\ref{fig3},\ref{fig4}).

\begin{figure}[ht]
\begin{center}
\vglue -0.009cm 
\mbox{\epsfig{file=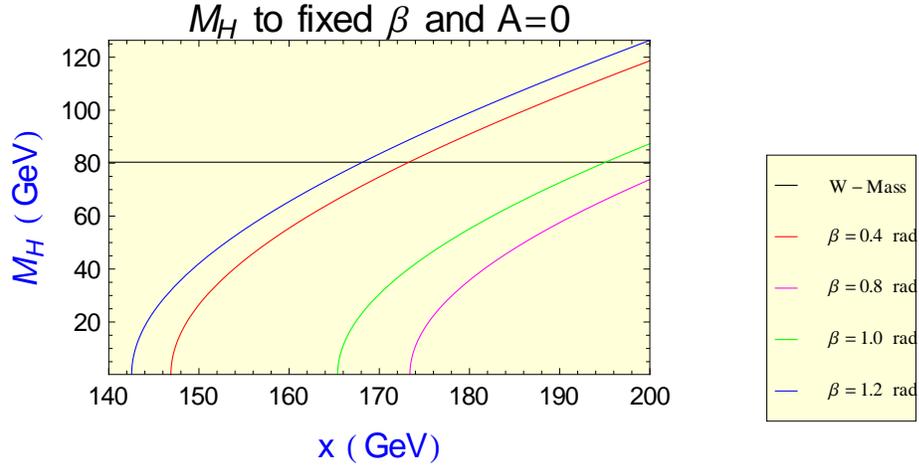,width=0.7\textwidth,angle=0}}
\end{center}
\caption{The masses of $H^{\pm}$ to several values of the $x$ to some fixed $\beta$ parameter here we used $A_{\lambda}=0$, the black line means 
the experimental values of $M_{W}$.}
\label{fig1}
\end{figure}

\begin{figure}[ht]
\begin{center}
\vglue -0.009cm \mbox{\epsfig{file=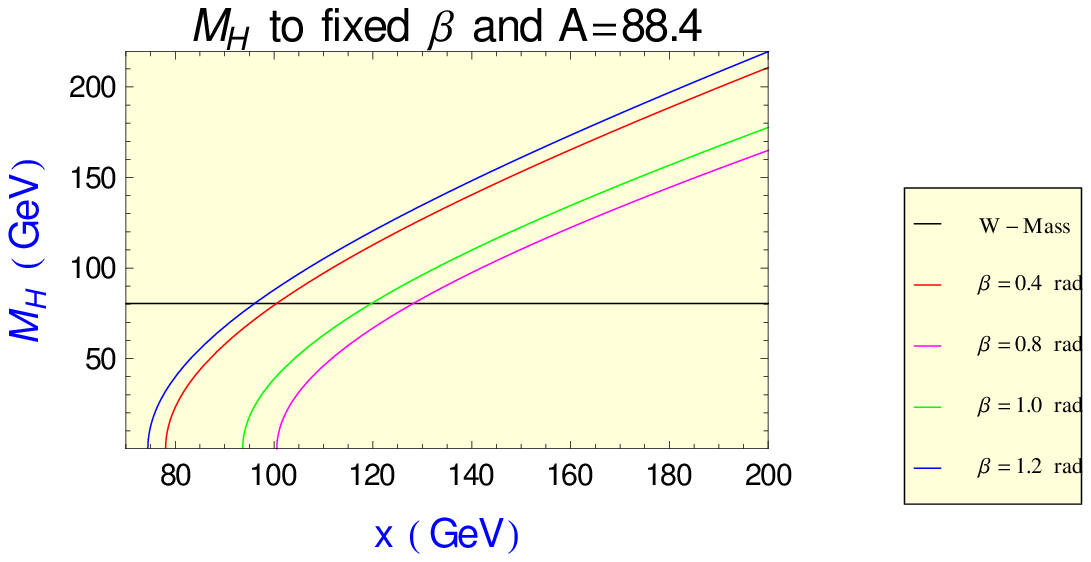,width=0.7\textwidth,angle=0}}
\end{center}
\caption{The masses of $H^{\pm}$ to several values of the $\beta$ to
some fixed $x$ parameter here we used $A_{\lambda}=0$, the black line means 
the experimental values of $M_{W}$.}
\label{fig2}
\end{figure}

\begin{figure}[ht]
\begin{center}
\vglue -0.009cm \mbox{\epsfig{file=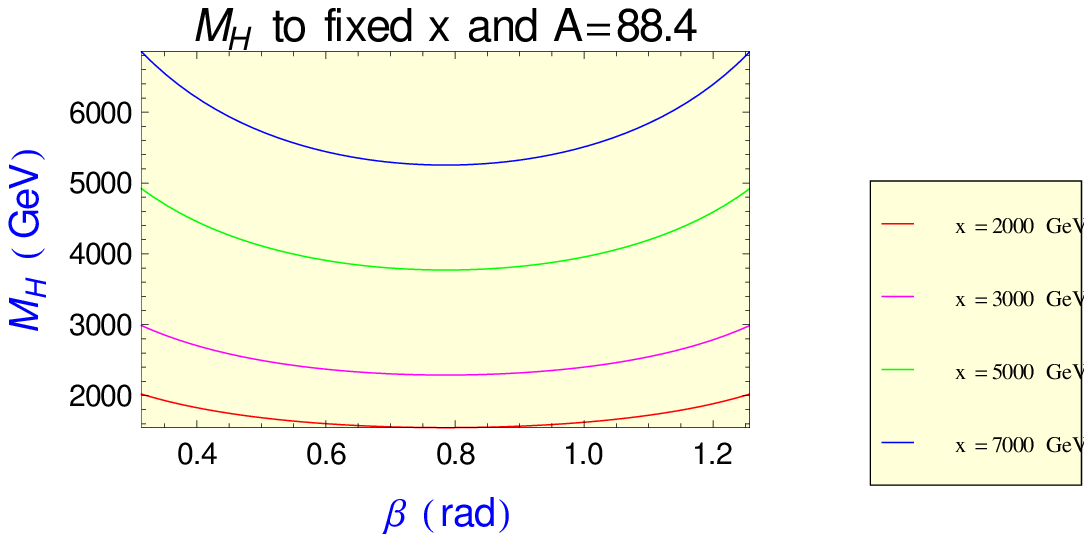,width=0.7\textwidth,angle=0}}
\end{center}
\caption{The masses of $H^{\pm}$ to several values of the $x$ to some fixed $\beta$ parameter here we used $A_{\lambda}=88.4$.}
\label{fig3}
\end{figure}

\begin{figure}[ht]
\begin{center}
\vglue -0.009cm \mbox{\epsfig{file=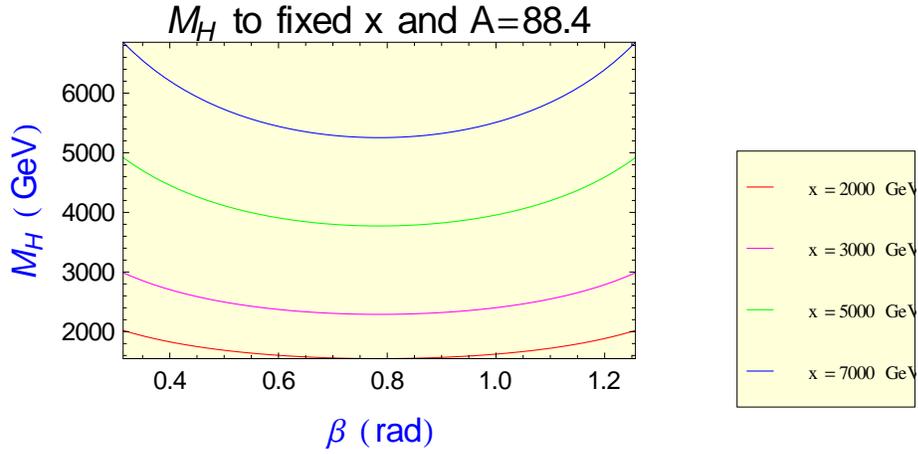,width=0.7\textwidth,angle=0}}
\end{center}
\caption{The masses of $H^{\pm}$ to several values of the $\beta$ to
some fixed $x$ parameter here we used $A_{\protect\lambda}=88.4$.}
\label{fig4}
\end{figure}

We notice that, when $\lambda =0$ then $\left(
M^{NMSSM}_{H^{\pm}}\right)^{2}=M^{2}_{W}$. Remember that in DHM and in the
MSSM the charged Higgses is always havier the $W^{\pm}$ gauge boson, see
Eq.(\ref{wmass}) and this result is agreement as presented at \cite{Ellis:1988er}.

\subsection{Pseudoscalar}

We can get the mass matrix at the basis 
\begin{equation}
\left( 
\begin{array}{ccc}
\varphi_{1} & \varphi_{2} & \xi
\end{array}
\right)^{T}
\end{equation}
and we get the matrix ${\cal M}^{2}_{CP-odd}$ and we will not write their
elements here. Our first analytical result is 
\begin{eqnarray}
\det[{\cal M}^{NMSSM}_{CP-odd}]=0,
\end{eqnarray}
and therefore we have Goldstone boson in this sector and this result is
agreement as presented at \cite{Ellis:1988er}.

As we want to compare our result to the mass values presented at \cite{Ellis:1988er} 
we need to do the following rotation 
\begin{equation}
\left( 
\begin{array}{c}
G^{0} \\ 
P_{1} \\ 
P_{2}
\end{array}
\right) = \left( 
\begin{array}{ccc}
\cos \beta & - \sin \beta & 0 \\ 
\cos \gamma \sin \beta & \cos \gamma \cos \beta & \sin \gamma \\ 
- \sin \gamma \sin \beta & - \sin \gamma \sin \beta & \cos \gamma
\end{array}
\right) \left( 
\begin{array}{c}
\varphi_{1} \\ 
\varphi_{2} \\ 
\xi
\end{array}
\right) ,  
\label{pseudoatnmssm}
\end{equation}
where the $\tan \beta$ is defined at Eq.(\ref{defbetapar}) and the new
mixing angle is found to be 
\begin{eqnarray}
\sin \left( 2 \gamma \right)&=&- \frac{2S}{\sqrt{(T-R)^{2}+4S^{2}}},  \\
\cos \left( 2 \gamma \right)&=&\frac{(T-R)}{\sqrt{(T-R)^{2}+4S^{2}}}.
\end{eqnarray}
We have defined 
\begin{eqnarray}
R&=&\lambda A_{\Sigma}\frac{xv^{2}}{v_{1}v_{2}},  \\
S&=&\lambda v \left( A_{\Sigma}-3 \kappa x \right),  \\
T&=&\lambda A_{\Sigma}\frac{v_{1}v_{2}}{x}+3 \kappa A_{\kappa}x+3 \lambda
\kappa v_{1}v_{2},
\end{eqnarray}
where we have defined 
\begin{eqnarray}
A_{\Sigma}&=&A_{\lambda}+\kappa x,  \nonumber \\
v&=&\sqrt{v^{2}_{1}+v^{2}_{2}}.
\label{usefulparameters1}
\end{eqnarray}
The angle $\gamma$ may be chosen as $0 < \gamma < \pi$ according to 
\cite{Ellis:1988er}
\begin{equation}
\gamma \in \left\{
\begin{array}{c}
\left[ 0, \frac{\pi}{4} \right] \,\ S<0, \,\ T>R, \\
\left[ \frac{\pi}{4}, \frac{\pi}{2} \right] \,\ S<0, \,\ T<R, \\
\left[ \frac{\pi}{2}, \frac{3 \pi}{4} \right] \,\ S>0, \,\ T<R, \\
\left[ \frac{3 \pi}{4}, \pi \right] \,\ S>0, \,\ T>R.
\end{array}
\right. 
\end{equation} 

\begin{figure}[ht]
\begin{center}
\vglue -0.009cm 
\mbox{\epsfig{file=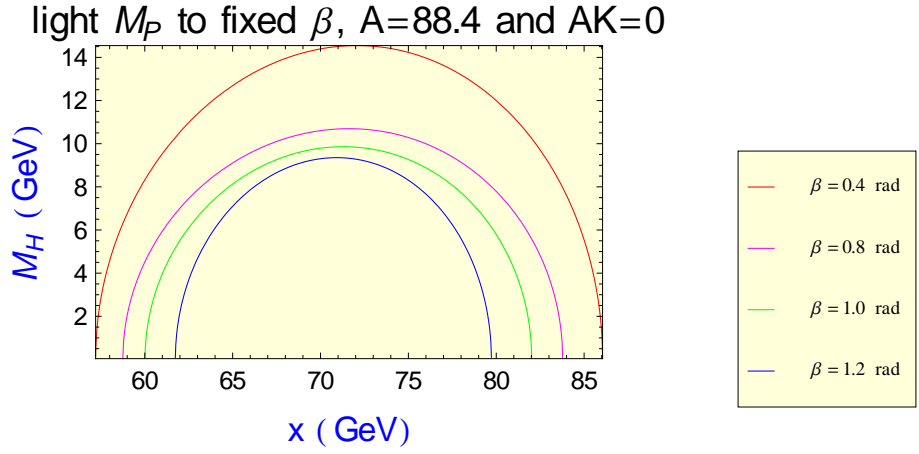,width=0.7\textwidth,angle=0}}
\end{center}
\caption{The masses of $P_{1}$ to several values of the $x$ to some fixed $\beta$ parameter here we used $A_{\kappa}=0$ and $A_{\lambda}=0$.}
\label{figpseudoleve1}
\end{figure}

\begin{figure}[ht]
\begin{center}
\vglue -0.009cm 
\mbox{\epsfig{file=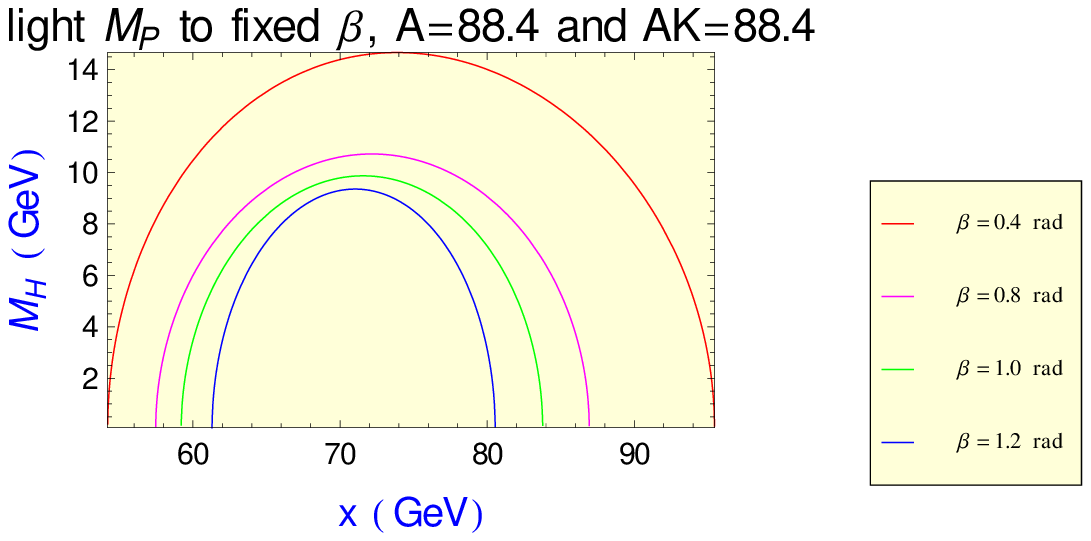,width=0.7\textwidth,angle=0}}
\end{center}
\caption{The masses of $P_{1}$ to several values of the $x$ to some fixed $\beta$ parameter here we used $A_{\kappa}=0$ and $A_{\lambda}=88.4$.}
\label{figpseudoleve2}
\end{figure}

The eigenvalues of ${\cal M}^{2}_{CP-odd}$ are given by: 
\begin{equation}
M^{2}_{P_{1},P_{2}}=\frac{1}{2}\left[ \left( T+R \right) \mp 
\sqrt{(T-R)^{2}+4S^{2}} \right],
\label{masspseudoHiggsNMSSM}
\end{equation}
where we choose $M^{2}_{P_{1}}<M^{2}_{P_{2}}$ and the eigenvectors are
defined at Eq.(\ref{pseudoatnmssm}) and $G^{0}$ is Goldstone boson. We show at 
Figs.(\ref{figpseudoleve1},\ref{figpseudoleve2}) we can get very light pseudoscalars as required 
by cosmological analyses presented at \cite{Hooper:2009gm}.

\begin{figure}[ht]
\begin{center}
\vglue -0.009cm 
\mbox{\epsfig{file=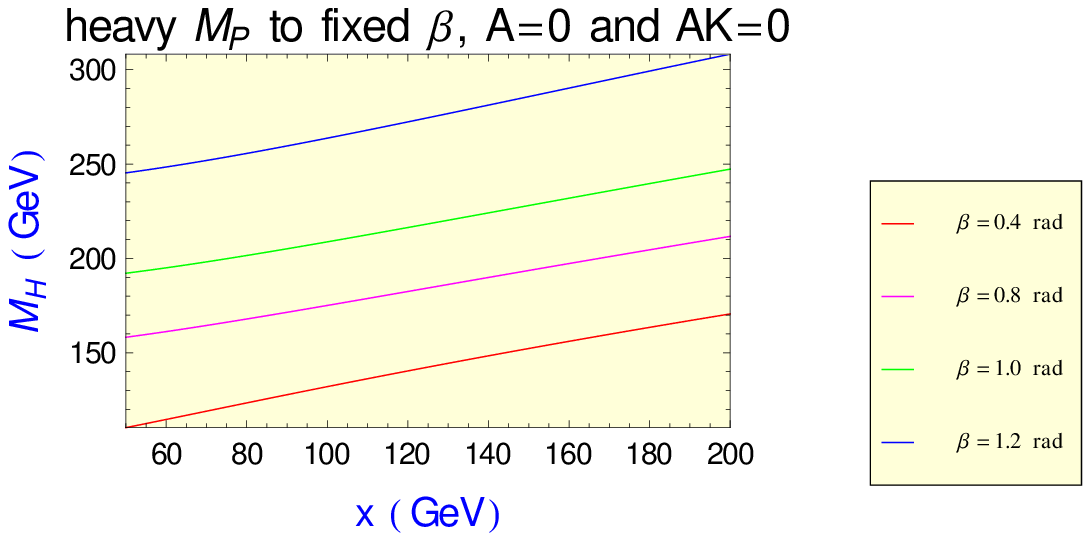,width=0.7\textwidth,angle=0}}
\end{center}
\caption{The masses of $P_{2}$ to several values of the $x$ to some fixed $\beta$ parameter here we used $A_{\kappa}=0$ and $A_{\lambda}=0$.}
\label{figpseudo1}
\end{figure}

\begin{figure}[ht]
\begin{center}
\vglue -0.009cm 
\mbox{\epsfig{file=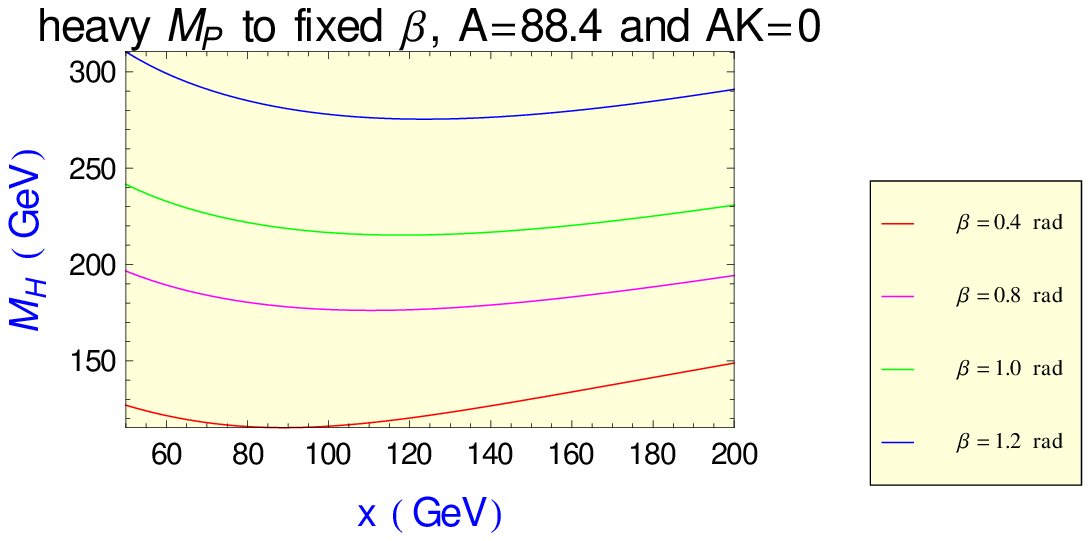,width=0.7\textwidth,angle=0}}
\end{center}
\caption{The masses of $P_{2}$ to several values of the $x$ to some fixed $\beta$ parameter here we used $A_{\kappa}=0$ and $A_{\lambda}=88.4$.}
\label{figpseudo2}
\end{figure}

We can also showed, thet $P_{2}$ are heavier states than $P_{1}$, as 
we shown at Figs.(\ref{figpseudo1},\ref{figpseudo2},\ref{figpseudo3}).

\begin{figure}[ht]
\begin{center}
\vglue -0.009cm 
\mbox{\epsfig{file=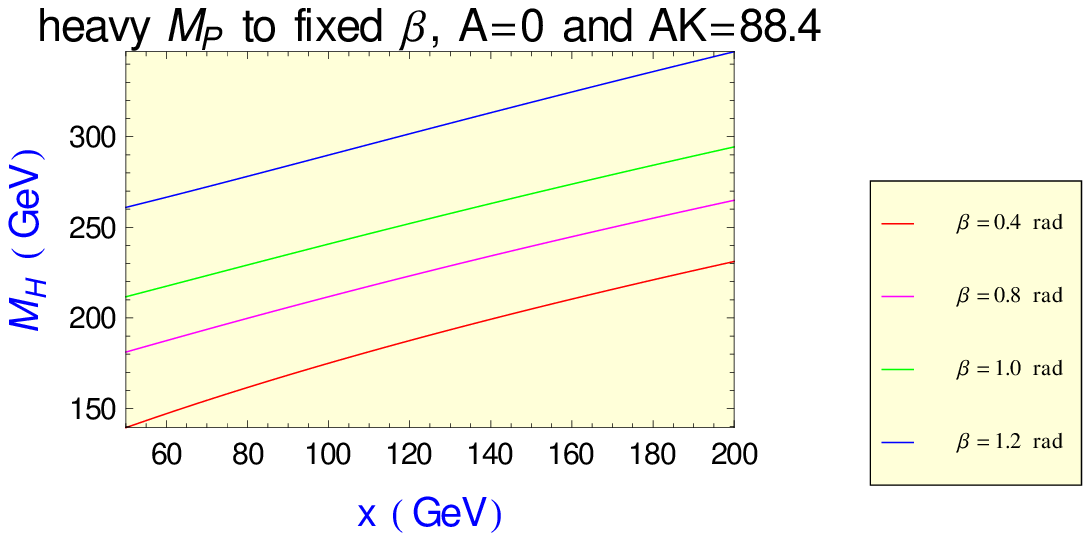,width=0.7\textwidth,angle=0}}
\end{center}
\caption{The masses of $P_{2}$ to several values of the $x$ to some fixed $\beta$ parameter here we used $A_{\kappa}=88.4$ and $A_{\lambda}=0$.}
\label{figpseudo3}
\end{figure}

\begin{figure}[ht]
\begin{center}
\vglue -0.009cm 
\mbox{\epsfig{file=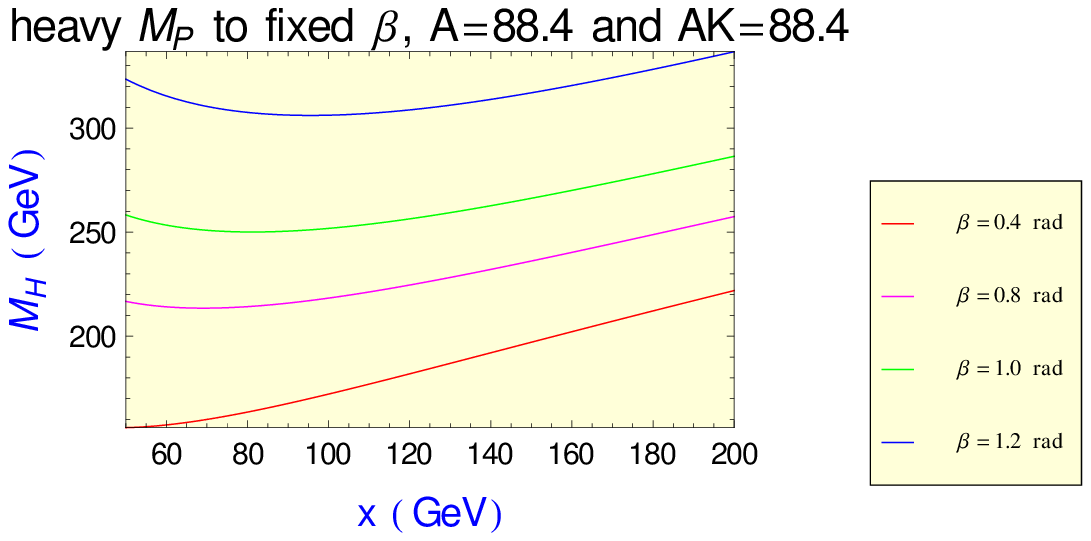,width=0.7\textwidth,angle=0}}
\end{center}
\caption{The masses of $P_{2}$ to several values of the $x$ to some fixed $\beta$ parameter here we used $A_{\kappa}=0$ and $A_{\lambda}=88.4$.}
\label{figpseudo3}
\end{figure}

\subsection{The Neutral Scalars}

The squared mass matrix for the ``scalar" neutral in the basis 
\begin{equation}
\left(
\begin{array}{ccc} 
\phi_{1}& \phi_{2}& \sigma 
\end{array}
\right)^{T},
\end{equation} 
takes the form \cite{Ellis:1988er}
\begin{equation}
\lambda 
\left( 
\begin{array}{ccc}
\frac{\bar{g}^{2}v^{2}_{1}}{\lambda}+ 
\frac{A_{\Sigma}xv_{2}}{v_{1}} & 
-A_{\Sigma}x+ \frac{v_{1}v_{2}}{x}(2 \lambda^{2}-\bar{g}^{2}) & 
v_{2}\left[ \frac{2 \lambda xv_{1}}{v_{2}}- \kappa x-A_{\Sigma} 
\right] \\
-A_{\Sigma}x+ \frac{v_{1}v_{2}}{x}(2 \lambda^{2}-\bar{g}^{2}) & 
\frac{\bar{g}^{2}v^{2}_{2}}{\lambda}+ 
\frac{A_{\Sigma}xv_{1}}{v_{2}} & 
v_{1}\left[ \frac{2 \lambda xv_{2}}{v_{1}}- \kappa x-A_{\Sigma} 
\right] \\ 
v_{2}\left[ \frac{2 \lambda xv_{1}}{v_{2}}- \kappa x-A_{\Sigma} 
\right] &
v_{1}\left[ \frac{2 \lambda xv_{2}}{v_{1}}- \kappa x-A_{\Sigma} 
\right] &
\frac{4 \kappa^{2}x^{2}- \kappa A_{\kappa}x}{\lambda}+ 
\frac{A_{\Sigma}v_{1}v_{2}}{x}
\end{array}
\right)
\end{equation}
where we have defined
\begin{eqnarray}
r &\equiv& \frac{x}{v}, \nonumber \\
\bar{g}&\equiv& \frac{1}{\sqrt{2}}\sqrt{g^{2}+g^{\prime 2}},
\end{eqnarray}
and $v$ and some others parameters of this Equation as defined at Eq.(\ref{usefulparameters1}). It is not useful to present analytic results 
for the diagonalization of this matrix.

\section{Neutralinos at NMSSM}

The diagonal contribution to neutralinos came from the gaugino mass term
given by ${\cal L}^{MSSM}_{GMT}$, while Higgsinos mass term came 
from the superpotential throught 
\begin{eqnarray}
{\cal L}^{NMSSM}_{HMT}&=&- \lambda \left[ \left( \tilde{H}_{1}
\tilde{H}_{2} \right)S+ \left( \tilde{H}_{1}H_{2}+ H_{1}\tilde{H}_{2} \right) \tilde{S} \right] + 
\kappa \tilde{S}\tilde{S}S   \\
&=&- \left[ \lambda x \tilde{H}^{0}_{1}\tilde{H}^{0}_{2}+ 
\lambda v_{1} \tilde{H}^{0}_{2}\tilde{S}+ \lambda v_{2}\tilde{H}^{0}_{1}\tilde{S}+ \kappa
x \tilde{S}\tilde{S} \right] + \ldots.
\end{eqnarray}
where $\ldots$ means the terms to charged higgsinos. Using the 
expression to $M_{W}$ in the MSSM, we can get 
\begin{eqnarray}
{\cal L}^{NMSSM}_{HMT}&=&- \left[ \lambda x \tilde{H}^{0}_{1}
\tilde{H}^{0}_{2}+ \frac{\sqrt{2}M_{W}}{g}\lambda \left( \cos \beta\tilde{H}^{0}_{2}+
\sin \beta \tilde{H}^{0}_{1} \right) \tilde{S}+ \kappa x \tilde{S}\tilde{S} \right] + \ldots.
\end{eqnarray}

The mixing between higgsinos and gauginos came from Eq.(\ref{lsupgsemssm}),
as the singlinos are singlet under $SU(2)_{L}\otimes U(1)_{Y}$ 
they can not mix with the gauginos. However the mixing between the gauginos and higgsinos, as in the MSSM, came from 
\begin{eqnarray}
{\cal L}^{mixing}_{\tilde{H} \tilde{V}}&=& \sqrt{2} \imath \;\bar{H}_{1}
\left( gT^{i}\lambda^{i}- \frac{g^{\prime}}{2} \lambda^{\prime} \right) 
\tilde{H}_{1} - \sqrt{2} \imath \;\bar{\tilde{H}}_{1} \left(gT^{i}
\bar{\lambda}^{i}- \frac{g^{\prime}}{2} \overline{\lambda^{\prime}} \right) H_{1}
\notag \\
&+&\sqrt{2}i\;\bar{H}_{2} \left( gT^{i}\lambda^{i}+ \frac{g^{\prime}}{2}
\lambda^{\prime} \right) \tilde{H}_{2} - \sqrt{2}i\;\bar{\tilde{H}}_{2}
\left( gT^{i}\bar{\lambda}^{i}+ \frac{g^{\prime}}{2} 
\overline{\lambda^{\prime}} \right) H_{2},   \\
&=& \frac{\imath g}{\sqrt{2}}\left( v_{1}\tilde{H}^{0}_{1}\lambda^{3}+ 
v_{2} \tilde{H}^{0}_{2}\lambda^{3} + h.c. \right)+ 
\frac{\imath g^{\prime}}{\sqrt{2}}\left( v_{1}\tilde{H}^{0}_{1}\lambda^{\prime}+ v_{2}
\tilde{H}^{0}_{2}\lambda^{\prime} + h.c. \right).
\end{eqnarray}
From the MSSM is so simply to show the Eq.(\ref{v1v2MZmssm}) and in similar way we can write the following expression
\begin{eqnarray}
\frac{g^{\prime}v_{1}}{\sqrt{2}}&=&\tan \theta_{W}\frac{gv_{1}}{\sqrt{2}}=
\tan \theta_{W} M_{W}\cos \beta = \tan \theta_{W} M_{Z}\cos \theta_{W}\cos \beta = 
M_{Z}\sin \theta_{W}\cos \beta ,  \nonumber \\
\frac{g^{\prime}v_{2}}{\sqrt{2}}&=&\tan \theta_{W}\frac{gv_{2}}{\sqrt{2}}=
\tan \theta_{W} M_{W}\sin \beta = \tan \theta_{W} M_{Z}\cos \theta_{W}\sin \beta = 
M_{Z}\sin \theta_{W}\sin \beta .
\end{eqnarray}
then we get 
\begin{eqnarray}
{\cal L}^{\mbox{mixing}}_{\tilde{H} \tilde{V}}&=&\imath M_{Z}\left[ \left( \cos \theta_{W}\cos \beta \tilde{H}^{0}_{1}+ \cos \theta_{W}
\sin \beta \tilde{H}^{0}_{2} \right) \lambda^{3} + 
\left( \sin \theta_{W}\cos \beta \tilde{H}^{0}_{1}+ \sin \theta_{W}
\sin \beta \tilde{H}^{0}_{2} \right) \lambda^{\prime} + h.c. \right].  
\end{eqnarray}

It generate a symmetric $5 \times 5$ mass matrix ${\cal M}_{0}$. In the
basis 
\begin{equation}
\psi^{0} = \left(
\begin{array}{ccccc}
- \imath \lambda^{3} & - \imath \lambda^{\prime} & 
\tilde{H}^{0}_{1} & \tilde{H}^{0}_{2} & \tilde{S}
\end{array}
\right)^{T},
\label{neutralinonmssm}
\end{equation} 
the resulting mass terms in the Lagrangian read 
\begin{equation}
{\cal L} = - \frac{1}{2} (\psi^{0})^{T} {\cal M}_{0} (\psi^{0}) + h.c.
\label{2.31e}
\end{equation}
where 
\begin{equation}
{\cal M}_{0} = \left( 
\begin{array}{ccccc}
M_{1} & 0 & M_{Z} \sin \beta \cos \theta_{W} & -M_{Z} \cos \beta \cos
\theta_{W} & 0 \\ 
0 & M_{2} & M_{Z} \sin \beta \cos \theta_{W} & -M_{Z} \cos \beta \cos
\theta_{W} & 0 \\ 
M_{Z} \sin \beta \cos \theta_{W} & M_{Z} \sin \beta \cos \theta_{W} & 0 & -
\frac{\lambda x}{\sqrt{2}} & - \frac{\lambda v_{2}}{\sqrt{2}} \\ 
-M_{Z} \cos \beta \cos \theta_{W} & -M_{Z} \cos \beta \cos \theta_{W} & -
\frac{\lambda x}{\sqrt{2}} & 0 & - \frac{\lambda v_{1}}{\sqrt{2}} \\ 
0 & 0 & - \frac{\lambda v_{2}}{\sqrt{2}} & - \frac{\lambda v_{1}}{\sqrt{2}} & \sqrt{2} \kappa x
\end{array}
\right) . 
\label{2.32e}
\end{equation}
using Eq.(\ref{v1v2MZmssm}) we can rewrite the elements $\left( {\cal M}_{0}\right)_{3,5}$ and 
$\left( {\cal M}_{0}\right)_{4,5}$ in the following way
\begin{eqnarray}
\left( {\cal M}_{0}\right)_{3,5}&=&\frac{\lambda v_{2}}{\sqrt{2}}= \frac{\lambda M_{Z}\cos \theta_{W}\sin \beta}{g},
\nonumber \\
\left( {\cal M}_{0}\right)_{4,5}&=&\frac{\lambda v_{1}}{\sqrt{2}}= \frac{\lambda M_{Z}\cos \theta_{W}\cos \beta}{g},
\end{eqnarray}
using the expressions above we can write
\begin{equation}
{\cal M}_{0} = \left( 
\begin{array}{ccccc}
M_{1} & 0 & M_{Z} \sin \beta \cos \theta_{W} & -M_{Z} \cos \beta \cos
\theta_{W} & 0 \\ 
0 & M_{2} & M_{Z} \sin \beta \cos \theta_{W} & -M_{Z} \cos \beta \cos
\theta_{W} & 0 \\ 
M_{Z} \sin \beta \cos \theta_{W} & M_{Z} \sin \beta \cos \theta_{W} & 0 & -
\frac{\lambda x}{\sqrt{2}} & - \frac{\lambda M_{Z}\cos \theta_{W}\sin \beta}{g} \\ 
-M_{Z} \cos \beta \cos \theta_{W} & -M_{Z} \cos \beta \cos \theta_{W} & -
\frac{\lambda x}{\sqrt{2}} & 0 & - \frac{\lambda M_{Z}\cos \theta_{W}\cos \beta}{g} \\ 
0 & 0 & - \frac{\lambda M_{Z}\cos \theta_{W}\sin \beta}{g} & - \frac{\lambda M_{Z}\cos \theta_{W}\cos \beta}{g} & \sqrt{2} \kappa x
\end{array}
\right) . 
\label{massneutralinoMZbeta}
\end{equation}
We want to stress the following \cite{dress}
\begin{itemize}
\item[1-)] The singlino, $\tilde{S}$, does not mix directly with the gauginos, see $\left( {\cal M}_{0}\right)_{1,5}$ and 
$\left( {\cal M}_{0}\right)_{2,5}$;
\item[2-)] The singlino, $\tilde{S}$, mix directly with the higgsinos $\tilde{H}^{0}_{1}$ 
and $\tilde{H}^{0}_{2}$see $\left( {\cal M}_{0}\right)_{3,5}$ and $\left( {\cal M}_{0}\right)_{4,5}$;
\item[3-)] If $|x| \gg v_{1,2}$ the singlino decouples from the other four neutralinos, wchich will 
be MSSM-like;
\item[4-)] If $| \kappa |$ is very small this singlinolike state will become the LSP \cite{Ellwanger:1996gw}.
\end{itemize}

The five-by-five matrix $N$ diagonalizes, in the following way, as in the MSSM
\begin{equation}
{\cal M}_{0}=N^{*}\,Y\,N^{-1},  
\label{nmatrix}
\end{equation}
the symmetric mass matrix ${\cal M}_{0}$ of the neutral Weyl spinors, 
see Eq.(\ref{nmatrix}), where the eigenvalues are arranged such that
$|m_{\chi^{0}_{1}}| < |m_{\chi^{0}_{2}}| <
|m_{\chi^{0}_{3}}| < |m_{\chi^{0}_{4}}|< |m_{\chi^{0}_{5}}|$. The parameter $\eta_{i}$ is introduced in order to change the 
phase of the particle whose eigenvalue becomes negative, it means it is defined as follow
\begin{equation}
\eta_{i} = \left\{
\begin{array}{c}
1,  m_{\chi^{0}_{i}}>0, \\
\imath , m_{\chi^{0}_{i}}<0,
\end{array}
\right. , \,\ i=1, \ldots 5,
\end{equation}
and 
\begin{equation}
m_{\chi^{0}_{i}} = \eta^{2}_{i} m_{\chi^{0}_{i}}.
\end{equation}
The four--component notation to the neutralinos is given as
\begin{equation}
\tilde{\chi}^{0}_{i}= 
\left( \begin{array}{c} 
\chi^{0}_{i} \\
\overline{\chi^{0}_{i}}
\end{array} \right), \hspace{6mm} i=1\ldots 5.
\end{equation}

\begin{figure}[ht]
\begin{center}
\vglue -0.009cm 
\mbox{\epsfig{file=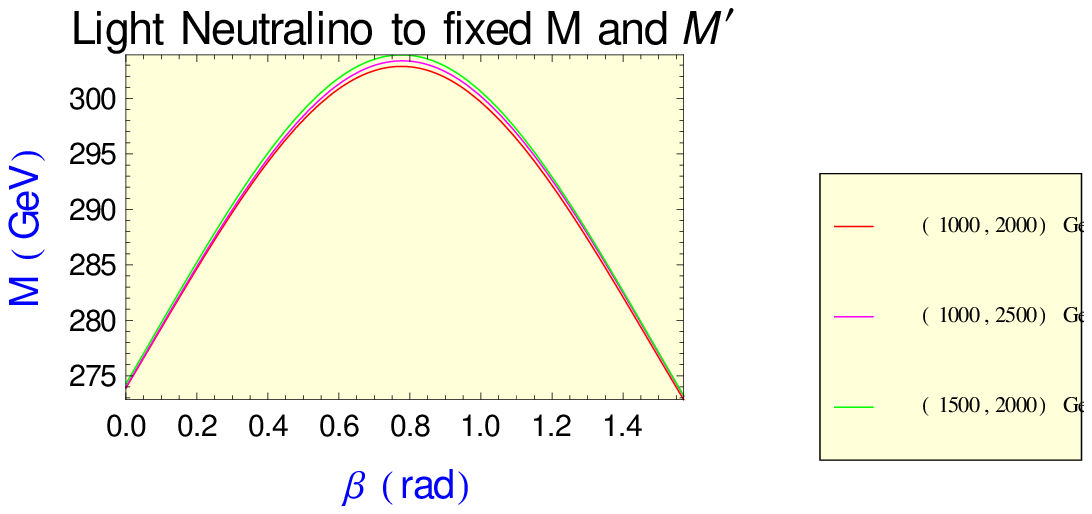,width=0.7\textwidth,angle=0}}
\end{center}
\caption{The masses of $\tilde{\chi}^{0}_{1}$ to several values of the $x$ to $M=1000$ Gev and $M^{\prime}=2000$ Gev as function of $\beta$ parameter.}
\label{figlspnmssm}
\end{figure}

We take $\lambda =0.87$, $\kappa =0.63$ as used at \cite{Ellis:1988er} and $x=500$ GeV and used $M=1000$ Gev and $M^{\prime}=2000$ Gev, under 
these parameters we get the masses of LSP is of order of $270$ GeV as shown at Fig.(\ref{figlspnmssm}). This results is in agreement with we 
need to get some nice resuts in cosmological analyses as
presented at \cite{Hooper:2009gm}.  

\section{Motivation to study some minimal modification}

The existence of a ``light" chiral gauge singlet superfield in the observable sector can cause other difficulties \cite{dress}
\begin{itemize}
\item[1-)]The stability of gauge hierarchy;
\item[2-)]The superpotential of NMSSM, defined at Eq.(\ref{suppotNMSSM}), 
possesses a discrete ${\cal Z}_{3}$ symmetry;
\end{itemize}
however within the framework of gravity mediated SUSY breaking terms, the required amount of violation of the ${\cal Z}_{3}$ symmetry can be introduced through nonrenormalizable operators. In this context we 
can defined the General Singlet Extensions of the MSSM (GSEMSSM). 

The most general superpotential to Singlet extension of MSSM, it can be get from super-GUT models or from super-string $E(6)$ models, 
is given by \cite{Drees:1988fc}
\begin{eqnarray}
W_{GSEMSSM}&=&\mu\; \left( \hat{H}_{1}\hat{H}_{2} \right) +
\lambda \left( \hat{H}_{1}\hat{H}_{2} \right) \hat{S}+
\left( \xi_{F} M_{n}^{2}\right) \hat{S}+ \frac{\mu_{2}}{2}\left( \hat{S} \right)^{2}+
\frac{\kappa}{3}\left( \hat{S} \right)^{3} \nonumber \\
&+& \sum_{i,j=1}^{3}\left[\,
f^{l}_{ab}\left( \hat{H}_{1}\hat{L}_{a}\right) \hat{l}^{c}_{b}+
f^{d}_{ij}\left( \hat{H}_{1}\hat{Q}_{i}\right) \hat{d}^{c}_{j}+
f^{u}_{ij}\left( \hat{H}_{2}\hat{Q}_{i}\right) \hat{u}^{c}_{j}\,\right].
\label{suppotGSEMSSM}
\end{eqnarray}
The parameters 
$\lambda , \kappa$ and $\xi_{F}$ are dimensionless coefficients while the parameters $\mu_{2}$ and $M_{n}$ have mass dimension. 
Before we continue, is useful stress the following, a term of the form 
$M \hat{S}$ can be absorved by a shift in $\hat{S}$ \cite{Drees:1988fc}.

We can get, from Eq.(\ref{suppotGSEMSSM}), the the Next-to-the-Minimal Supersymmetric Standard-Model (NMSSM) \cite{R,nmssm,barr,nilsredwy,derendinger,Drees:1988fc,Ananthanarayan:1996zv,Ellwanger:1998jk}, we need 
by setting 
\begin{eqnarray}
\mu = \xi_{F}= \mu_{2}=0, 
\label{condtogetNMSSMfromGEMSSM}
\end{eqnarray}
we get the following superpotential 
\begin{eqnarray}
W_{NMSSM}&=&\lambda \left( \hat{H}_{1}\hat{H}_{2} \right) \hat{S}+
\frac{\kappa}{3}\left( \hat{S} \right)^{3}+
\sum_{i,j=1}^{3}\left[\,
f^{l}_{ab}\left( \hat{H}_{1}\hat{L}_{a}\right) \hat{l}^{c}_{b}+
f^{d}_{ij}\left( \hat{H}_{1}\hat{Q}_{i}\right) \hat{d}^{c}_{j}+
f^{u}_{ij}\left( \hat{H}_{2}\hat{Q}_{i}\right) \hat{u}^{c}_{j}\,\right] , \nonumber \\
\label{suppotNMSSM}
\end{eqnarray}
and the nearly Minimal Supersymmetric Model (nMSM) \cite{xnMSSM1,xnMSSM2}, by setting 
\begin{eqnarray}
\mu = \mu_{2}= \kappa =0, 
\label{condtogetnMSSMfromGEMSSM}
\end{eqnarray}
\begin{eqnarray}
W_{nMSM}&=&\lambda \left( \hat{H}_{1}\hat{H}_{2} \right) \hat{S}+
\left( \xi_{F} M_{n}^{2}\right) \hat{S}+
\sum_{i,j=1}^{3}\left[\,
f^{l}_{ab}\left( \hat{H}_{1}\hat{L}_{a}\right) \hat{l}^{c}_{b}+
f^{d}_{ij}\left( \hat{H}_{1}\hat{Q}_{i}\right) \hat{d}^{c}_{j}+
f^{u}_{ij}\left( \hat{H}_{2}\hat{Q}_{i}\right) \hat{u}^{c}_{j}\,\right]. \nonumber \\
\label{suppotnMSSM}
\end{eqnarray}
Note that the nMSM differs from the NMSSM in the last term with the trilinear singlet term $\kappa$ of the NMSSM replaced by the tadpole term 
$\xi_{F}$ and both models have nice cosmological consequences, see for example \cite{Rodriguez:2016esw,Hooper:2009gm,Wang:2012ry}.

\subsection{Scalar Potential}

The scalar potential is defined as \cite{Drees:1988fc}
\begin{eqnarray}
V^{Dress}&=& M_{H_{1}}^{2} | H_{1} |^{2} + M_{H_{2}}^{2} | H_{2} |^{2}+ M_{S}^{2} | S |^{2}+
\left[ \left( B \mu \right) \left( H_{1} H_{2} \right) +
\lambda A_{\lambda} \left( H_{1} H_{2} \right) S \right. \nonumber \\
&+& \left. \frac{\left( B_{s} \mu_{2}\right)}{2} \left( S \right)^2 + 
\frac{\left( \kappa A_{\kappa}\right)}{3} \left( S \right)^{3} + h.c. \right] \nonumber \\
&+&
\left( |H_{1}|^{2}+ |H_{2}|^{2}\right) \left| \mu + \lambda S\right|^{2}+  
\left| \lambda \left( H_{1}H_{2}\right) + \mu_{2}S+ \kappa S^{2} \right|^{2} \nonumber \\
&+&
\frac{g^{2}}{8} 
\left( \bar{H}_{1}\sigma^{i}H_{1}+\bar{H}_{2}\sigma^{i}H_{2}
\right)^{2}+ 
\frac{g^{\prime 2}}{8} \left( \bar{H}_{1}H_{1}- \bar{H}_{2}H_{2} \right)^{2}.
\end{eqnarray}
In this general case the mass of all scalars is given by \cite{Drees:1988fc}
\begin{eqnarray}
M^{2}_{H^{\pm}}&=&M^{2}_{W}\left( 1- \frac{2 \lambda^{2}}{g^{2}}\right) - 
\frac{2C_{1}}{\sin \left( 2 \beta \right)}, \nonumber \\
\label{chargedscalaratgsemssm}
\end{eqnarray}
compare this equation with Eq.(\ref{chargedscalaratnmssm}). We see that both the 
results at NMSSM and at GSEMSSM are similar. In the case of pseudoscalar we get
\begin{eqnarray}
M^{2}_{H^{PS}_{1,2}}&=&\frac{1}{2}\left[ - 
\frac{2C_{1}}{\sin \left( 2 \beta \right)}+C_{2} \pm 
\sqrt{\left( \frac{2C_{1}}{\sin \left( 2 \beta \right)}+C_{2} \right)^{2}+
\frac{16M^{2}_{W}}{g^{2}}C^{2}_{3}}\,\ \right], \nonumber \\
\label{masspseudoHiggsgsemssm}
\end{eqnarray}
the pseudoscalar at NMSSM is given at Eq.(\ref{masspseudoHiggsNMSSM}).

In Eqs.(\ref{chargedscalaratgsemssm},\ref{masspseudoHiggsgsemssm}), we have defined the following coefficients
\begin{eqnarray}
C_{1}&=&\lambda x \left( A_{\lambda}+ \mu_{2}\right) +B \mu + \frac{\lambda \kappa}{2}x^{2}, \nonumber \\
C_{2}&=&- \frac{\lambda M^{2}_{W}}{g^{2}x}\left[ \sin \left( 2 \beta \right) \left( A_{\lambda}+ \mu_{2}\right) +2 
\mu \right] - \frac{\kappa x}{2}\left( 3A_{\kappa}+ \mu_{2}\right) -2B_{s}\mu_{2}- 
\frac{2 \lambda \kappa}{g^{2}}M^{2}_{W}\sin \left( 2 \beta \right) , \nonumber \\
C_{3}&=&\lambda \left( \kappa x+ \mu_{2}-A_{\lambda}\right).
\end{eqnarray}
As happen at NMSSM in this case the charged Higgs bosons can also be lighter than the $W$ boson. Note that the 
condition $M^{2}_{H^{PS}_{1}} \geq 0$\footnote{$H^{PS}_{1}$ denotes the ligther eigenstate} implies $C_{2} \geq 0$, 
whereas $C_{3}$ can have either sign. The interesting fact is that no absolute bound on the masses of the physical 
pseudoscalars can be given; in particular, $H^{PS}_{1,2}$ can both be very light. These results are in agreement 
with the results we presented to NMSSM, and we can in this case reproduce the masses to pseudoscalars, see 
Figs.(\ref{figpseudoleve1},\ref{figpseudoleve2}) in such way that the 
GSEMSSM can be useful in explore cosmological analyses as presented at \cite{Hooper:2009gm,Wang:2012ry}.

On this case we can get an upper bound on the mass of the lighest neutral scalar $H^{0}_{1}$ and it is given 
by \cite{Gunion:1989we,Drees:1988fc,dress}
\begin{equation}
M^{2}_{H^{0}_{1}}\leq M^{2}_{Z}\left[
\cos^{2}\left( 2 \beta \right) + \frac{2 \lambda^{2}\cos^{2} \theta_{W}}{g^{2}}\sin^{2} \left( 2 \beta \right)
\right].
\end{equation}
We can show that in this model, we get the following upper bound \cite{dress}
\begin{equation}
M^{2}_{H^{0}_{1}} < 145 \,\ \mbox{GeV},
\end{equation}
where $H^{0}_{1}$ is the lighest $CP$-even physical scalar and one can therefore say that supersymmetric theories 
always contain one neutral scalar Higgs boson with mass 
proportional to $M_{Z}$ \cite{Drees:1988fc}. To our knowledge, $200$ GeV is the absolute limit to wchich the 
upper bound on the lightest Higgs mass can be raised in any perturbatively treatable model with weak 
scale supersymmetry \cite{dress}.

\section{Neutralino in GSEMSSM}

In the base defined at Eq.(\ref{neutralinonmssm}), the mass matrix to the neutralino masses is
\begin{equation}
{\cal M}_{0} = \left( 
\begin{array}{ccccc}
M_{1} & 0 & M_{Z} \sin \beta \cos \theta_{W} & -M_{Z} \cos \beta \cos
\theta_{W} & 0 \\ 
0 & M_{2} & M_{Z} \sin \beta \cos \theta_{W} & -M_{Z} \cos \beta \cos
\theta_{W} & 0 \\ 
M_{Z} \sin \beta \cos \theta_{W} & M_{Z} \sin \beta \cos \theta_{W} & 0 & -
\left( \mu + \frac{\lambda x}{\sqrt{2}}\right) & - \frac{\lambda M_{Z}\cos \theta_{W}\sin \beta}{g} \\ 
-M_{Z} \cos \beta \cos \theta_{W} & -M_{Z} \cos \beta \cos \theta_{W} & -
\left( \mu + \frac{\lambda x}{\sqrt{2}}\right) & 0 & - \frac{\lambda M_{Z}\cos \theta_{W}\cos \beta}{g} \\ 
0 & 0 & - \frac{\lambda M_{Z}\cos \theta_{W}\sin \beta}{g} & - \frac{\lambda M_{Z}\cos \theta_{W}\cos \beta}{g} & 
\sqrt{2} \kappa x + \mu_{2}
\end{array}
\right) . 
\label{massneutralinoMZbetaGSEMSSM}
\end{equation}
It is similar to Eq.(\ref{massneutralinoMZbeta}), then we can reproduce the results presented at LSP, see 
Fig.(\ref{figlspnmssm}).

\section{Conclusions}
\label{sec:conclusion}

In this article we have presented the MSSM and NMSSM lagrangian in terms of superfields. Then we presented the mass spectrum of those models. We shown 
that the masses of lighest chargino and neutralino have their masses 
of ${\cal O}(M_{Z})$, while the gluinos are the heavier ones, because its 
mass comes from SUSY soft breaking terms, and it mass is ${\cal O}(TeV)$. 

We show that all the Higgs sector in the MSSM can be described in terms of 
$M_{A^{0}}$ and $\tan \beta$. We also showed how to get some Feynman Rules of this sector. 

We have, also, presented the NMSSM and also the GSEMSSM models. We, also, show some choose of free 
parameter that can get the masses to pseudoscalars, see Figs.(\ref{figpseudoleve1},\ref{figpseudoleve2}), and LSP, see 
Fig.(\ref{figlspnmssm}), necessary to get the nice results in some cosmological analyses as presented at \cite{Hooper:2009gm,Wang:2012ry}.

We hope this review can be useful to all the people wants to learn about Supersymmetry.

\begin{center}
{\bf Acknowledgments} 
\end{center}
The author would like to thanks to Instituto de F\'\i sica Te\'orica (IFT-Unesp) for their nice 
hospitality during the period I developed this review about SUSY and from the nice workshop about Dark Matter.

The author would like to thanks to Instituto de F\'\i sica Te\'orica (IFT-Unesp), Laboratoire 
de Physique Math\'ematique et Th\'eorique at Universit\'e  Montpellier II, Laborat\'{o}rio de 
F\'{\i}sica Experimental at Centro Brasileiro de Pesquisas F\'\i sicas (LAFEX-CBPF), Instituto de 
F\'\i sica da Universidade Federal do Rio Grande do Sul (IF-UFRGS) and Institut of Physics at Vietnam Academy of Science and Technology (IOP-VAST) 
for their nice hospitality during the period I developed some works on SUSY. Special thanks to Professores V. Pleitez, J. C. Montero, N. Berkovitz 
(to teach me superfields language), M. Capdequi-Peyran\`ere, 
M. Manna, G. Moultaka, A. Djouadi, Jean-Loïc Kneur, Pierre Fayet\footnote{To send me the originals articles about MSSM},  H. N. Long, P. V. Dong, D. T. Huong, 
C. M. Maekawa, C. B. Mariotto, J.A. Helay\"{e}l-Neto and A. J. Accioly to several discussion about 
SUSY and also E. V. Gorbar, A. Belyaev (classes about COMPHEP) to give me the originals articles of SUSY in Russian. I would like to say thanks to 
Funda\c c\~ao de Amparo \`a Pesquisa do Estado de S\~ao Paulo (FAPESP), under contract number 96/10046-0 and 00/14221-9, Brazilian funding 
agency Conselho Nacional de Desenvolvimento Cient\'\i fico e Tecnol\'ogico (CNPq), under contract number 309564/2006-9 and 
Funda\c c\~ao de Amparo \`a Pesquisa do Estado do Rio Grande do Sul (FAPERGS), under contract number 02/1266-6, for financial support.



\begin{thebibliography}{99}
\bibitem{Susskind:1978ms}L. Susskind, 
{\sl Phys. Rev.}{\bf D20}, 2619, (1979). 


\bibitem{Gunion:1989we} J. F. Gunion, H. E. Haber, G. L. Kane and S. Dawson, 
{\it The Higgs Hunter's Guide}, 
{\sl Front. Phys.}{\bf 80}, 1, (2000).
\bibitem{Deshpande:1977rw}N. G. Deshpande and E. Ma, 
{\sl Phys. Rev.}{\bf D18}, 2574, (1978). 
\bibitem{Georgi:1978wr}H. Georgi, 
{\sl Hadronic J.}{\bf 1}, 1227, (1978).
\bibitem{Donoghue:1978cj}J. F. Donoghue and L. F. Li, 
{\sl Phys. Rev.}{\bf D19}, 945, (1979); 
\bibitem{Abbott:1979dt}L. F. Abbott, P. Sikivie and M. B. Wise, 
{\sl Phys. Rev.}{\bf D21}, 1393, (1980); 
\bibitem{McWilliams:1980kj}B. McWilliams and L. F. Li, 
{\sl Nucl. Phys.}{\bf B179}, 62, (1981); 
\bibitem{Haber:1978jt}H. E. Haber, G. L. Kane and T. Sterling, 
{\sl Nucl. Phys.}{\bf B161}, 493,  (1979).
\bibitem{Gunion:1984yn}J. F. Gunion and H. E. Haber, 
{\sl Nucl. Phys.} B {\bf B272}, 1, (1986); Erratum: [ {\sl Nucl. Phys.}{\bf B402}, 567, (1993)].


\bibitem{Hooper:2009gm}D. Hooper and T. M. P. Tait, 
{\sl Phys. Rev.}{\bf D80}, 055028, (2009); [arXiv:0906.0362 [hep-ph]].

\bibitem{Wang:2012ry}W. Wang, 
{\sl Adv. High Energy Phys.}{\bf 2012}, 216941, (2012); [arXiv:1205.5081 [hep-ph]].

\bibitem{Langacker:1984nf}P. Langacker {\it et al.}, {\it Nonstandard Higgs Bosons}, {\sl Snowmass Summer Study 1984:771}.
\bibitem{Ellis:1986ij}J. R. Ellis, {\it Supersymmetry at the SSC}, {\sl Snowmass Summer Study 1984:782}.

\bibitem{Salam:1974ig}A. Salam and J. A. Strathdee, 
{\sl Phys. Lett.}{\bf B51}, 353, (1974).

\bibitem{ogievetskivi}V. I. Ogievetski\v{i} and L. Mezincescu, 
{\sl Sov. Phys. Usp}{\bf 18}, 960, (1976).
\bibitem{wb}J. Wess and J. Bagger, {\it Supersymmetry and Supergravity},
2nd edition, Princeton University Press, Princeton NJ, (1992).
\bibitem{MullerKirsten:1986cw}H. J. W. M\"uller-Kirsten and A. Wiedemann, 
{\it SUPERSYMMETRY: AN INTRODUCTION WITH CONCEPTUAL AND CALCULATIONAL DETAILS}, 
Second Edition, World Scientific Publishing Co. Pte. Ltd., Singapore, (2010).

\bibitem{Steane:2013wra}A. M. Steane, {\it An introduction to spinors}, arXiv:1312.3824 [math-ph].
\bibitem{Willenbrock:2004hu}S. Willenbrock, {\sl Symmetries of the standard model}, hep-ph/0410370.

\bibitem{R}P. Fayet, {\sl Nucl. Phys.}{\bf B90}, 104, (1975).
\bibitem{ssm}P. Fayet, {\sl Phys. Lett.}{\bf B64}, 159, (1976); {\bf B69}, 489, (1977).
\bibitem{grav}P. Fayet, {\sl Phys. Lett.}{\bf B70}, 461, (1977).

\bibitem{INO82a}K. Inoue, A. Komatsu and S. Takeshita, {\sl Prog. Theor. Phys.} {\bf 68}, 927, (1982).
\bibitem{INO82b}K. Inoue, A. Komatsu and S. Takeshita, {\sl Prog. Theor. Phys.} {\bf 70}, 330, (1983).

\bibitem{r1} A. Salam and J. Strathdee, {\sl Nucl. Phys.} {\bf B87}, 85, (1975).

\bibitem{gl} Yu. A. Gol'fand and E.P. Likhtman, 
{\sl ZhETF Pis. Red.}{\bf 13}, 452, (1971) [{\sl JETP Lett.}{\bf 13}, 323, (1971)].
\bibitem{va} D.V. Volkov and V.P. Akulov, 
{\sl JETP Lett.}{\bf 16}, 438 (1972) [{\sl Pisma Zh. Eksp. Teor. Fiz.}{\bf 16}, 621, (1972)];
{\sl Phys. Lett.}{\bf B46}, 109, (1973);
{\sl Theor. Math. Phys.} {\bf 18}, 28 (1974) [Teor.\ Mat.\ Fiz.\  {\bf 18}, 39 (1974)].
\bibitem{wz} J. Wess and B. Zumino, 
{\sl Nucl. Phys.}{\bf B70}, 39, (1974);
{\sl Phys. Lett.}{\bf B49}, 52, (1974); 
{\sl Nucl. Phys.}{\bf  B78}, 1, (1974).
\bibitem{Ferrara:1974ac}S. Ferrara, J. Wess and B. Zumino,
{\sl Phys. Lett.}{\bf 51B}, 239, (1974).


\bibitem{volkov1}D. V. Volkov, 
Talk given at International Conference on the History of Original Ideas and Basic Discoveries in Particle Physics, Erice, Italy, 29 Jul - 4 Aug 1994, 
e-Print: hep-th/9410024.
\bibitem{shifman}G. Kane and M. Shifman,  {\it Supersymmetric World, The Beginning of the Theory}, 
1st edition, World Scientific Publishing Company, Singapore, (2000).
\bibitem{shifman1} M. Shifman,  {\it The Many Faces of the Superworld: Yuri Golfand Memorial Volume}, 1st edition, 
World Scientific Publishing Company, Singapore, (2000).



\bibitem{Chung:2003fi}D. J. H. Chung, L. L. Everett, G. L. Kane, S. F. King, J. D. Lykken and L. T. Wang, 
{\sl Phys.Rept.}{\bf 407}, 1, (2005).

\bibitem{dress}M. Drees, R. M. Godbole and P. Royr, {\it Theory and Phenomenology of Sparticles} 
First Edition, World Scientific Publishing Co. Pte. Ltd., Singapore, (2004).
\bibitem{Baer:2006rs} H.~Baer and X.~Tata, {\it Weak scale supersymmetry: From superfields to scattering events}
First Edition, Cambridge University Press, Cambridge, UK, (2006).
\bibitem{Aitchison:2005cf}I. J. R. Aitchison, {\it Supersymmetry and the MSSM: An Elementary introduction}, hep-ph/0505105.
\bibitem{mssm}H. E. Haber and G. L. Kane, {\sl Phys. Rep.}{\bf 117}, 75, (1985).
\bibitem{Simonsen:1995cf}I. Simonsen, 
hep-ph/9506369.
\bibitem{Kuroda:1999ks}M. Kuroda, 
hep-ph/9902340.
\bibitem{kraml}S. Kraml, 
hep-ph/9903257.

\bibitem{sugra} P. Nath and R. Arnowitt, {\sl Phys. Lett.} {\bf B56}, 177, (1975);
D. Z. Freedman, P. van Nieuwenhuizen and S. Ferrara, {\sl Phys. Rev.} {\bf D13}, 3214, (1976); 
S. Deser and B. Zumino, {\sl Phys. Lett.} {\bf B62}, 335, (1976).

\bibitem{ABF}U. Amaldi, W. de Boer, H. F\"{u}rstenau, {\sl Phys. Lett.}{\bf B260}, 447, (1991).

\bibitem{running}V. Barger, M. S. Berger and P. Ohmann, {\sl Phys. Rev.} {\bf D47}, 1093, (1993);\\
W. de Boer, R. Ehret and D. Kazakov, {\sl Z. Phys.} {\bf C67}, 647, (1995);\\ 
W. de Boer et al., {\em Z. Phys.} {\bf C71}, 415, (1996).

\bibitem{Ibanez:wd}L. E. Iba\~nez and G. G. Ross, 
{\sl Phys. Lett.}{\bf B131}, 335, (1983).

\bibitem{Pendleton:as}B. Pendleton and G. G. Ross, 
{\sl Phys. Lett.}{\bf B98}, 291, (1981).
\bibitem{Sirunyan:2018gqx}A. M. Sirunyan {\it et al.} [CMS Collaboration],
arXiv:1805.01428 [hep-ex].

\bibitem{Dimopoulos:1981yj}S. Dimopoulos, S. Raby and F. Wilczek, 
{\sl Phys. Rev.}{\bf D24}, 1681, (1981).

\bibitem{Dimopoulos:1981zb}S. Dimopoulos and H. Georgi, 
{\sl Nucl. Phys.}{\bf B193}, 150, (1981).

\bibitem{Ibanez:yh}L. E. Iba\~nez and G. G. Ross, 
{\sl Phys. Lett.}{\bf B105}, 439, (1981).

\bibitem{Einhorn:1981sx}M. B. Einhorn and D. R. T. Jones,
{\sl Nucl. Phys.}{\bf B196}, 475, (1982).

\bibitem{Kane:1992kq}G. L. Kane, C. F. Kolda and J. D. Wells,
{\sl Phys. Rev. Lett.}{\bf 70}, 2686, (1993).

\bibitem{Espinosa:1992hp}J. R. Espinosa and M. Quiros,
{\sl Phys. Lett.}{\bf B302}, 51, (1993).

\bibitem{haber2} H. E. Haber, {\sl Eur. Phys. J.}{\bf C15}, 817, (2000).

\bibitem{Djouadi:2005gj}A. Djouadi, 
{\sl Phys. Rept.}{\bf 459}, 1, (2008).

\bibitem{Casas:1994us}J. A. Casas, J. R. Espinosa, M. Quiros and A.Riotto,
{\sl Nucl. Phys.}{\bf B436}, 3, (1995); Erratum: [ {\sl Nucl. Phys.}{\bf B439}, 466, (1995)].

\bibitem{Aad:2015zhl}G. Aad {\it et al.} [ATLAS and CMS Collaborations],
{\sl Phys. Rev. Lett.}{\bf 114}, 191803, (2015).

\bibitem{Fayet:2001xk}P. Fayet, 
{\sl Nucl. Phys. Proc. Suppl.}{\bf 101}, 81, (2001) (Also in *Minneapolis 2000, 30 years of supersymmetry* 81-98). 
\bibitem{Rodriguez:2009cd} M. C. Rodriguez, 
{\sl Int. J. Mod. Phys.}{\bf A25}, 1091, (2010).

\bibitem{Ellis:2010kf}J. Ellis and K. A. Olive, 
In *Bertone, G. (ed.): Particle dark matter* 142-163; [arXiv:1001.3651 [astro-ph.CO]].

\bibitem{nmssm}U. Ellwanger, M. Rausch de Traubenberg and C. A. Savoy, 
{\sl Phys. Lett.}{\bf B315}, 331, (1993), [arXiv:hep-ph/9307322].
\bibitem{barr}S.M. Barr, {\sl Phys. Lett.}{\bf B112}, 219, (1982).
\bibitem{nilsredwy}H. P. Nilles, M. Srednicki and D. Wyler, {\sl Phys. Lett.}{\bf B120}, 346, (1983).
\bibitem{derendinger}J.-P. Derendinger and C. A. Savoy, {\sl Nucl. Phys.}{\bf B 237}, 307, (1984).
\bibitem{Drees:1988fc}M. Drees, 
{\sl Int. J. Mod. Phys.}{\bf A4}, 3635, (1989).
doi:10.1142/S0217751X89001448  
\bibitem{Ellis:1988er}J. R. Ellis, J. F. Gunion, H. E. Haber, L. Roszkowski and F. Zwirner, 
{\sl Phys. Rev.}{\bf D39}, 844, (1989).  
\bibitem{Ananthanarayan:1996zv}B. Ananthanarayan and P. N. Pandita, 
{\sl Int. J. Mod. Phys.}{\bf A12}, 2321, (1997); [hep-ph/9601372].  
\bibitem{Ellwanger:1998jk}U. Ellwanger and C. Hugonie, 
hep-ph/9901309.  
\bibitem{Maniatis:2009re}M. Maniatis, 
{\sl Int. J. Mod. Phys.}{\bf A25}, 3505, (2010); [arXiv:0906.0777 [hep-ph]].
\bibitem{Ellwanger:2009dp}U. Ellwanger, C. Hugonie and A. M. Teixeira, 
{\sl Phys. Rept.}{\bf 496}, 1, (2010); [arXiv:0910.1785 [hep-ph]].

\bibitem{superspcae1} A. Salam e J. Strathdee, 
{\it Supergauge Transformations} em {\sl  Nucl. Phys.}{\bf B76}, 477, (1974).
\bibitem{superspcae2} S. Ferrara, J. Wess e B. Zumino, 
{\it Supergauge Multiplets and Superfields} em {\sl Phys. Lett.}{\bf 51B}, 239, (1974).
\bibitem{kane1}G. Kane,  {\it Supersymmetry Squarks, Photinos, and the Unveiling of the Ultimate Laws of Nature}, 
Primeira Edi\c c\~ao, Helix Books, Cambridge, Massachusetts, (2000).


\bibitem{vdWaerden1} B.L. van der Waerden, {\sl Nachrichten Akad. Wiss. G\"ottingen, Math.-Physik. Kl.}, 100, (1929).

\bibitem{Haber:1994pe}H.E.Haber, 
arXiv:hep-ph/9405376.
\bibitem{Martin:tasi}S. P. Martin, 
arXiv:1205.4076.
\bibitem{dreiner1}H. K. Dreiner, H. E. Haber, S. P. Martin,
{\sl Phys. Rept.}{\bf 494}, 1, (2010).

\bibitem{Lykken:1996xt}J. D. Lykken, 
hep-th/9612114.

\bibitem{Majorana} E. Majorana, {\sl Nuovo Cim.}{\bf 14}, 171, (1937).
\bibitem{Dirac} P.A.M. Dirac, {\sl Proc. Royal Soc. A}{\bf 117}, 610, (1928);
{\bf 118}, 351, (1928).

\bibitem{Dong:2006vk}P. V. Dong, D. T. Huong, M. C. Rodriguez and H. N. Long,
{\sl Eur. Phys. J.}{\bf C48}, 229, (2006).
\bibitem{barbier} R. Barbier {\it et al.},
{\sl Phys. Rept.}{\bf 420}, 1, (2005).

\bibitem{cmmc} C.M. Maekawa and M. C. Rodriguez, {\sl JHEP}{\bf 04}, 031, (2006).
\bibitem{cmmc1} C.M.Maekawa and M.C.Rodriguez,  {\sl JHEP}{\bf 0801}, 072, (2008).

\bibitem{neutral-LSP} J. Ellis, J. S. Hagelin, D. V. Nanopoulos, K. Olive 
and M. Srednicki, {\sl Nucl. Phys.}{\bf B238}, 453, (1984). 

\bibitem{hillwalkers}J. Ellis, hep-ph/9812235.


\bibitem{banks}T. Banks, {\sl Nucl. Phys.}{\bf B303}, 172, (1988).
\bibitem{hall}L. J. Hall and M. Suzuki, {\sl Nucl. Phys.}{\bf B231}, 419, (1984).
\bibitem{rv1}M. A. Diaz, J. C. Rom\~ao and J. W. F. Valle, {\sl Nucl. Phys.}{\bf B524}, 23, (1998).
\bibitem{fb}F. Borzumati and Y. Nomura, {\sl Phys. Rev.}{\bf D64}, 053005, (2001);
F. Borzumati, K. Hamaguchi and T. Yanagida, {\sl Phys. Lett.}{\bf B497}, 259, (2001);
F. Borzumati, K. Hamaguchi, Y. Nomura and T. Yanagida, hep-ph/0012118. 
\bibitem{rnm} R. N. Mohapatra, {\sl Phys. Rev.}{\bf D34}, 3457, (1986).
\bibitem{rv2} J. C. Rom\~ao and J. W. F. Valle, {\sl Nucl. Phys.}{\bf B381}, 87, (1992).
\bibitem{marta}S. Davison and M. Losada, hep-ph/0010325.
\bibitem{Montero:2001ch}J. C. Montero, V. Pleitez and M. C. Rodriguez, 
{\sl Phys. Rev.}{\bf D65}, 095008, (2002).

\bibitem{10}L. Girardello and  M. T. Grisaru, {\sl Nucl.  Phys.} {\bf B194}, 65, (1982).

\bibitem{Srednicki:2004hg}M. Srednicki, {\it Quantum field theory}, Fourth Edition, Cambridge University Press, United Kindom, (2010); 
and also avaliable at arXiv:hep-th/0409035 and arXiv:hep-th/0409036.

\bibitem{moultaka}J. L. Kneur and G. Moultaka, {\sl Phys. Rev.}{\bf D59}, 015005, 1999.

\bibitem{Mariotto:2008zt}C. B. Mariotto and M. C. Rodriguez,
arXiv:0805.2094 [hep-ph].

\bibitem{Dawson}S. Dawson, E. Eichten and C. Quigg, 
{\sl Phys. Rev.}{\bf D31}, 1581, (1985).
 
\bibitem{Espindola:2011nb} D. B. Espindola, M. C. Rodriguez and C. B. Mariotto,
{\sl Braz. J. Phys.}{\bf 43}, 105, (2013).
\bibitem{BrennerMariotto:2011wm}C. Brenner Mariotto, D. B. Espindola and M. C. Rodriguez,
{\sl Phys. Rev.}{\bf C83}, 064902, (2011).

\bibitem{Espindola:2010zz}D. B. Espindola, C. Brenner Mariotto and M. C. Rodriguez,
{\sl AIP Conf. Proc.}{\bf 1296}, 262, (2010).

\bibitem{Heinemeyer:2011ab}S. Heinemeyer and C. Schappacher,
{\sl Eur. Phys. J.}{\bf C72}, 1905, (2012).

\bibitem{ma}E. Ma, {\sl Phys. Rev.}{\bf D39}, 1922, (1989).

\bibitem{charginos} A. Bartl, H. Fraas, W. Majerotto and B. M\"o{\ss}lacher, {\sl Z. Phys.}{\bf C55}, 257, (1992).
\bibitem{neutralinos} A. Bartl, H. Fraas and W. Majerotto, {\sl Nucl. Phys.}{\bf B278}, 1, (1986).
\bibitem{indiano}M. Guchait, {\sl Z. Phys.}{\bf C57}, 157, (1993); Erratum {\bf C67}, 178, (1994).

\bibitem{Haber:1990aw}H. E. Haber and R. Hempfling, 
{\sl Phys. Rev. Lett.}{\bf 66},1815, (1991).

\bibitem{macarena}M. Carena, J. Ellis, J. S. Lee, A. Pilaftsis and C. E. M. Wagner, arXiv:1512.00437.

\bibitem{Gluck:1983zc}M. Gl\"uck and E. Reya, 
{\sl Phys. Rev.}{\bf D31}, 620, (1985).
\bibitem{Baer:1993ew}H. Baer, C. h. Chen, F. Paige and X. Tata, 
{\sl Phys. Rev.}{\bf D49}, 3283, (1994).
\bibitem{Bartl:1987zg}A. Bartl, H. Fraas and W. Majerotto, 
{\sl Z. Phys.}{\bf C34}, 411, (1987).

\bibitem{Bartl:1987da}A. Bartl, H. Fraas and W. Majerotto, 
{\sl Nucl. Phys.}{\bf B297}, 479, (1988).
\bibitem{Espindola:2012vsa} D. B. Espindola, C. B. Mariotto and M. C. Rodriguez,
{\sl AIP Conf. Proc.}{\bf 1520}, 273, (2013).

\bibitem{sps1}B.C. Allanach {\it et al}, {\sl Eur.Phys.J.}{\bf C25}, 113, (2002).
\bibitem{sps2}Nabil Ghodbane and Hans-Ulrich Martyn, hep-ph/0201233.


\bibitem{coreanos} S. Y. Choi, Y. S. Shim, H. S. Song and W. Y. Song, hep-ph/9808227.
\bibitem{takagi}T. Takagi, {\sl Japan J. Math}{\bf 1}, 83, (1925).
\bibitem{alinear}R. A. Horn and C. R. Johnson, {\it Matrix Analysis}, first edition, Cambridge University Press, Cambridge, UK, (1990). 

\bibitem{Rodriguez:2016esw}M. C. Rodriguez and I. V. Vancea, 
arXiv:1603.07979 [hep-ph].

\bibitem{Ellwanger:1996gw}U. Ellwanger, M. Rausch de Traubenberg and C. A. Savoy, 
{\sl Nucl. Phys.}{\bf B492}, 21, (1997). 

\bibitem{xnMSSM1}C. Panagiotakopoulos, K. Tamvakis,
{\sl Phys. Lett.}{\bf B446}, 224, (1999); 
{\sl Phys. Lett.}{\bf B469}, 145, (1999);
C. Panagiotakopoulos, A. Pilaftsis, 
{\sl Phys. Rev.}{\bf D63}, 055003, (2001);
A. Dedes, C. Hugonie, S. Moretti and K. Tamvakis, 
{\sl Phys. Rev.}{\bf D63}, 055009, (2001);
A. Menon, D.E. Morrissey, C.E.M. Wagner, 
{\sl Phys. Rev.}{\bf D70}, 035005, (2004);
V. Barger, P. Langacker and H. S. Lee, 
{\sl Phys. Lett.}{\bf B630}, 85, (2005);
C. Balazs, M. Carena, A. Freitas and C. E. M. Wagner, 
{\sl JHEP}{\bf 0706}, 066, (2007);
J. Cao, Z. Heng and J. M. Yang, 
{\sl JHEP}{\bf 1011}, 110, (2010).

\bibitem{xnMSSM2}J. Cao, H. E. Logan, J. M. Yang, 
{\sl Phys. Rev.}{\bf D79}, 091701, (2009).
    

\end{thebibliography}
\end{document}